\documentclass[amssymb, nobibnotes, aps, prfluids]{revtex4-1}

\usepackage{graphicx}

\usepackage{array,mathtools,amssymb,booktabs}
\newcolumntype{C}{>{$}c<{$}}
\AtBeginDocument{
\heavyrulewidth=.08em
\lightrulewidth=.05em
\cmidrulewidth=.03em
\belowrulesep=.65ex
\belowbottomsep=0pt
\aboverulesep=.4ex
\abovetopsep=0pt
\cmidrulesep=\doublerulesep
\cmidrulekern=.5em
\defaultaddspace=.5em
}

\usepackage[numbered,framed]{matlab-prettifier}

\lstMakeShortInline"

\usepackage{tikz}
	\usetikzlibrary{automata,topaths}
\usepackage{physics}
\usepackage{mathtools}
\usepackage{enumitem}
\usepackage{hhline} 
\usepackage{gensymb}
\usepackage{textcomp}
\usepackage{siunitx}
\usepackage{bm}
\DeclareSIUnit\molar{\mole\per\cubic\deci\metre}
\DeclareSIUnit\Molar{\textsc{m}}
\usepackage[caption=false]{subfig}

\usetikzlibrary{arrows.meta}

\makeatletter
\DeclareFontFamily{OMX}{MnSymbolE}{}
\DeclareSymbolFont{MnLargeSymbols}{OMX}{MnSymbolE}{m}{n}
\SetSymbolFont{MnLargeSymbols}{bold}{OMX}{MnSymbolE}{b}{n}
\DeclareFontShape{OMX}{MnSymbolE}{m}{n}{
	<-6>  MnSymbolE5
	<6-7>  MnSymbolE6
	<7-8>  MnSymbolE7
	<8-9>  MnSymbolE8
	<9-10> MnSymbolE9
	<10-12> MnSymbolE10
	<12->   MnSymbolE12
}{}
\DeclareFontShape{OMX}{MnSymbolE}{b}{n}{
	<-6>  MnSymbolE-Bold5
	<6-7>  MnSymbolE-Bold6
	<7-8>  MnSymbolE-Bold7
	<8-9>  MnSymbolE-Bold8
	<9-10> MnSymbolE-Bold9
	<10-12> MnSymbolE-Bold10
	<12->   MnSymbolE-Bold12
}{}

\let\llangle\@undefined
\let\rrangle\@undefined
\DeclareMathDelimiter{\llangle}{\mathopen}%
{MnLargeSymbols}{'164}{MnLargeSymbols}{'164}
\DeclareMathDelimiter{\rrangle}{\mathclose}%
{MnLargeSymbols}{'171}{MnLargeSymbols}{'171}
\makeatother

\usepackage{hyperref}

\usepackage{placeins}

\usepackage{natbib}

\makeindex

\begin{document}


\title{Regularized Stokeslet rings -- an efficient method for axisymmetric Stokes flow, with application to the growing pollen tube}
\author{J. Tyrrell}
\author{D.J. Smith}
\author{R.J. Dyson}
\email[Corresponding author:	]{r.j.dyson@bham.ac.uk}
\affiliation{University of Birmingham, School of Mathematics, Edgbaston, Birmingham, B15 2TT, United Kingdom}

\begin{abstract}
The method of regularized Stokeslets, based on the divergence-free exact solution to the equations of highly viscous flow due to a spatially-smoothed concentrated force, is widely employed in biological fluid mechanics. Many problems of interest are axisymmetric, motivating the study of the azimuthally-integrated form of the Stokeslet which physically corresponds to a ring of smoothed forces. The regularized fundamental solution for the velocity (single layer potential) and stress (double layer potential) due to an axisymmetric ring of smoothed point forces, the `regularized ringlet', is derived in terms of complete elliptic integrals of the first and second kind. The relative errors in the total drag and surrounding fluid velocity for the resistance problem on the translating, rotating unit sphere, as well as the condition number of the underlying resistance matrix, are calculated; the regularized method is also compared to 3D regularized Stokeslets, and the singular method of fundamental solutions. The velocity of Purcell's toroidal swimmer is calculated; regularized ringlets enable accurate evaluation of surface forces and propulsion speeds for non-slender tori. The benefits of regularization are illustrated by a model of the internal cytosolic fluid velocity profile in the rapidly-growing pollen tube. Actomyosin transport of vesicles in the tube is modelled using forces immersed in the fluid, from which it is found that transport along the central actin bundle is essential for experimentally-observed flow speeds to be attained. The effect of tube growth speed on the internal cytosolic velocity is also considered. For axisymmetric problems, the regularized ringlet method exhibits a comparable accuracy to the method of fundamental solutions whilst also allowing for the placement of forces inside of the fluid domain and having more satisfactory convergence properties. 
\end{abstract}

\maketitle

\section{Introduction}
The Stokes equations for incompressible flow at zero Reynolds number are used extensively to model the viscous-dominated regime of microscale flows, particularly biological flows associated with cilia-driven transport, and the motility and feeding of flagellated cells such as bacteria, spermatozoa, algae and choanoflagellates. For an overview, see Lauga and Powers \cite{lauga2009hydrodynamics}. The fundamental solution of the Stokes flow equation, which corresponds to the flow driven by a single spatially-concentrated force is often referred to as the Oseen tensor or Stokeslet. The linearity of the Stokes flow equations enables the construction of solutions to problems involving moving boundaries with complex geometry through integral sums of Stokeslets, forming the basis for the method of fundamental solutions, slender body theory and boundary integral methods. The latter numerical method has the principle major advantage of avoiding the need to mesh the fluid volume, which has enabled highly accurate and efficient simulation of biological flow systems for several decades \cite{phan1987boundary,ramia1993role,shum2010modelling}. Indeed more approximate methods based on line distributions of Stokeslets and higher order singularities also enabled major progress in this area before the present era of computationally-intensive research. For review see the earlier work of Chwang, Wu and co-authors
\cite{chwang1975hydromechanicsP2}, who also explore a wide range of applications as part of a series of papers on low-Reynolds number flow \cite{chwang1974hydromechanicsP1,chwang1975hydromechanicsP3,chwang1976hydromechanicsP4,johnson1979hydromechanicsP5,huang1986hydromechanicsP6}.

Nevertheless, two implementational issues arise with methods based on singular solutions. The first is that boundary integrals of solutions with a \(1/r\)-type singularity can be technically complex to evaluate on or near the boundary. Line integrals associated with models of slender bodies such as cilia and flagellar are `more singular', and can require careful distinction between the inside and outside of the body. Moreover, there are cases in which immersed forces due to, e.g.\ many suspended moving particles, are desired to be modelled by an immersed volumetric force. Cortez \emph{et al.} developed the method of regularized Stokeslets \cite{cortez2001method,cortez2005method} based on the exact, divergence-free solution to the Stokes flow equations due to a concentrated but spatially-smoothed (regularized) force. This approach has enabled the use of Stokeslet methods in a wider range of applications, such as those in which an inducing force is present in  the interior of the fluid domain (as either a point in $\mathbb{R}^2$ or a point/curve in $\mathbb{R}^3$). 

While conceptually elegant, the standard implementation of the method of regularized Stokeslets is computationally expensive, motivating the development of boundary element discretization \cite{smith2009boundary}, line integration \cite{cortez2018regularized}, and meshless interpolation \cite{smith2018nearest} among other approaches. Many diverse biological flow problems of interest exhibit rotational symmetry, examples including spherical `squirmer' swimmers \cite{blake1971spherical}, the conceptual toroidal swimmer of Purcell \cite{purcell1977life}, and cytosolic flow in elongating pollen tubes \cite{chebli2013transport}. Thus motivated, in this paper we study axisymmetric Stokes flows in which the singular and regularized Stokeslets can be integrated azimuthally to yield an axisymmetric ring of point forces. The singular solution to this problem is already known \cite{pozrikidis1992boundary}; the regularized solution, which we term the `regularized ringlet', is newly derived. This solution forms the basis for an efficient axisymmetric method of regularized Stokeslets.

We begin in Section \ref{sec:singularRegularizedStokeslets} by introducing the singular and regularized Stokeslets, and review their application in solving the resistance problem for a rigid body translating in a viscous fluid. In Section \ref{sec:regStokesRing} we derive the regularized ringlet, whose analytical solution is given in Section \ref{sec:analyticalRinglet} and is applied to the resistance problem for the translating and rotating sphere in Sections \ref{sec:singSolComp} -- \ref{sec:rotSphere}. The double layer potential (DLP), relevant to bodies undergoing volume-changing deformation, is considered in Section \ref{sec:mainDLP}, with analytical evaluation of the azimuthal integral in the DLP given in Appendix \ref{sec:DLP}. In Section \ref{sec:toroidalSwimmer} we consider the case of Purcell's toroidal swimmer \cite{purcell1977life,leshansky2008surface}, in which the method of regularized ringlets enables the calculation of propulsion speeds which are in excellent agreement with analytical results for both slender and non-slender tori. We then present a study of fluid flow in the angiosperm pollen tube in Section \ref{sec:pollenTube}, an illustrative example highlighting the benefits of being able to apply smoothed forces inside the fluid (two prominent features of regularization). In Section \ref{sec:conclusion} we conclude with a summary of results and a discussion of future work.

\FloatBarrier
\section{Singular and regularized Stokeslet solutions} \label{sec:singularRegularizedStokeslets}
For the viscous--dominated very low Reynolds number flow associated with microscopic length scales and slow velocities, incompressible Newtonian flow is well--approximated by the steady Stokes flow equations
\begin{align}
  \label{stokes1} \mu\nabla^2\mathbf{u} &= \nabla p - \mathbf{F}, \\
  \nabla \cdot \mathbf{u} &= 0, \label{stokes2}
\end{align}
where $\mu$ is dynamic viscosity, $p$ the pressure, $\mathbf{u}$ the velocity, and $\mathbf{F}$ the applied force per unit volume. In the case of a singular force of the form $\mathbf{F}(\mathbf{x}_0)=\mathbf{g}^p\delta(\mathbf{x}_0-\mathbf{x})$ for arbitrary point force $\mathbf{g}^p$, arbitrary point $\mathbf{x}$ at which the singularity is located, and where $\delta$ is the Dirac delta function, the fundamental solution \cite{fairweather1998method,young2006method} for $\mathbf{u}$ is given by
\begin{equation}
  u_i(\mathbf{x}_0)=\frac{1}{8\pi\mu}S_{ij}(\mathbf{x}_0,\mathbf{x})g^p_j,
\end{equation}
where
\begin{equation}
  S_{ij}(\mathbf{x}_0,\mathbf{x})=\frac{\delta_{ij}}{|\mathbf{x}_0-\mathbf{x}|} + \frac{(x_{0,i}-x_i)(x_{0,j}-x_j)}{|\mathbf{x}_0-\mathbf{x}|^3},
\end{equation}
is known as the Stokeslet. 

The singularity in the Stokeslet solution can be eliminated without loss of incompressibility by regularization of the force $\mathbf{F}$, as described by Cortez and colleagues \cite{cortez2001method, cortez2005method}. The Dirac delta function is replaced with $\mathbf{F}(\mathbf{x}_0)=\mathbf{g}^p\phi_\varepsilon(\mathbf{x}_0-\mathbf{x})$ where $\phi_\varepsilon$ is a radially symmetric, smooth `cutoff' function with the property $\int_{\mathbb{R}^3}\phi_\varepsilon(\mathbf{x})\,d\mathbf{x}=1$. This is in essence applying the force over a small ball, varying smoothly from a maximum at its centre to $\approx 0$ sufficiently far away, instead of using an infinite point force as in the classical Stokeslet solution. The numerical parameter $\varepsilon$ dictates the radius of support of the force, and as $\varepsilon\rightarrow 0$ the classical solution is recovered. Solutions for $\mathbf{u}$ using regularized Stokeslets differ from those found using the singular Stokeslet only near the point where the force is applied. Following Cortez \emph{et al.} \cite{cortez2005method} we take
\begin{equation} \label{eq:cutoff}
  \phi_\varepsilon(\mathbf{x}_0-\mathbf{x})=\frac{15\varepsilon^4}{8\pi(|\mathbf{x}_0-\mathbf{x}|^2+\varepsilon^2)^{7/2}},
\end{equation}
which yields
\begin{equation}
  S_{ij}^\varepsilon(\mathbf{x}_0,\mathbf{x})=\delta_{ij}\frac{|\mathbf{x}_0-\mathbf{x}|^2+2\varepsilon^2}{(|\mathbf{x}_0-\mathbf{x}|^2+\varepsilon^2)^{3/2}}+\frac{(x_{0,i}-x_i)(x_{0,j}-x_j)}{(|\mathbf{x}_0-\mathbf{x}|^2+\varepsilon^2)^{3/2}}.
\end{equation}

By considering a solid body $D$ moving through the fluid, it can be shown that
\begin{equation} \label{cortez_integrals}
  \int_{\mathbb{R}^3} u_i(\mathbf{x})\phi_\varepsilon(\mathbf{x}_0-\mathbf{x}) \, dV(\mathbf{x}) =  \frac{1}{8\pi\mu} \int_{\partial D}  S_{ij}^\varepsilon (\mathbf{x}_0,\mathbf{x}) g_j^a \, dS(\mathbf{x}),
\end{equation} 
where $\mathbf{g}^a$ is the force per unit area exerted by the body surface (denoted $\partial D$) on the surrounding fluid \cite{cortez2005method}. The above equation is exact; replacing the left hand side with the velocity \(u_i(\mathbf{x}_0)\) introduces an error \(\mathcal{O}(\epsilon^p)\) where \(p=1\) on or near the body surface, and \(p=2\) sufficiently far away.

Discretising Equation \eqref{cortez_integrals} using $N$ Stokeslets on the surface of the solid body $D$ enables the approximation of the fluid velocity at any point $\mathbf{x}_0$ via a numerical quadrature formula
\begin{equation} u_i(\mathbf{x}_0) = \frac {1}{8\pi\mu} \sum_{n=1}^N \sum_{i=1}^3 S_{ij}^\varepsilon (\mathbf{x}_0,\mathbf{x}_n)g^a_{n,j}W_n,
\end{equation} where $g^a_{n,j}$ denotes the $j$th component of the force per unit area applied at the point $\mathbf{x}_n$ (a Stokeslet location) and $W_n$ is the quadrature weight associated with the $n$th particle. The value of $W_n$ is dependent on the geometry of the body surface $\partial D$ and in Cortez and colleagues' work has units of area.

\subsection{Derivation of the regularized ringlet} \label{sec:regStokesRing}
\begin{figure}[ht]  \centering \begin{tikzpicture} [scale=.5]
	\draw[step=1cm,gray!0,very thin] (-8,-5) grid (8,8);
	
	\draw [ultra thick] (0,0)  circle (5);
	
	\draw [ultra thick, ->] (0,0) -- (8,0);
	\draw [ultra thick, ->] (0,0) -- (0,8);
	\node [left] at (9,0) {$x$};
	\node [right] at (-1,8) {$y$};
	\node at (-.5,-.5) {$0$};
	
	\draw [dashed, ultra thick] (0,0) -- (3,4); 
	\draw [ultra thick, dashed] (3,0) -- (3,4);
	\draw [ultra thick, dashed] (0,4) -- (3,4);
	\draw (1.75,0) arc [radius=1.75, start angle=0, end angle =50];
	\node [below] at (1,1) {$\theta$};
	\node at (1.4,2.6) {$r$};
	\node at (3,-.5) {$r\cos\theta$};
	\node at (-1.35,3.9) {$r\sin\theta$};
	\node at (4,4.75) {$(r,\theta,z)$};
	\node at (7,-.75) {$(r_0,0,z_0)$};
	\fill (3,4) circle (0.25);
	\fill [white] (7,0) circle (0.25);
	\draw (7,0) circle (0.25);
	\end{tikzpicture} \caption{Stokeslet ring in $x,y$ view. Note that the fluid point (open circle) and points on the ring (closed circle) do not necessarily both lie in the plane of the page (i.e. $z=z_0$ is not required).} \label{ringfig} \end{figure}
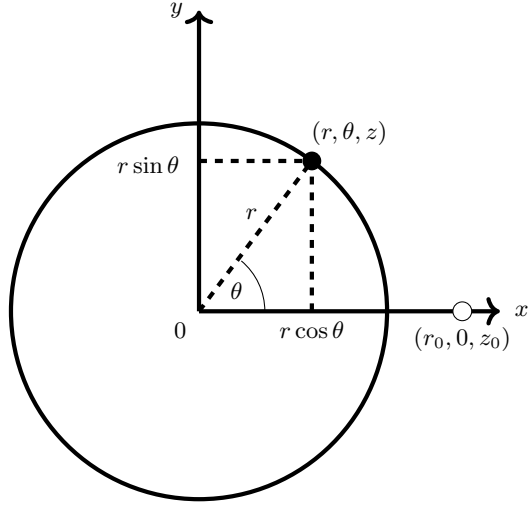

Consider a specific case of Equation \eqref{cortez_integrals} using a cylindrical $(r,\theta,z)$ coordinate system in which the body $D$ exhibits rotational symmetry about the $z$ axis. This symmetry enables analytical integration azimuthally, reducing the surface discretisation to a line discretisation and increasing accuracy. In doing so, we are effectively placing `rings' of regularized Stokeslets at positions  $\mathbf{x}_n = (r_n,\theta,z_n)$ for $n=1,...,N,$ and $\theta\in[0,2\pi)$ (see Figure \ref{ringfig} for a diagram of a single ring). This is analogous to covering the surface of the body in `strips' instead of the patches used in a standard $3$D Cartesian discretisation. With a surface parameterization $\mathbf{x}(s,\theta)$ where $0\le s\le\ell$ denotes arclength and $0\le\theta\le 2\pi$, the boundary integral equation \eqref{cortez_integrals} reads
\begin{equation} \label{ringsol}
u_i(\mathbf{x}_0) = \frac{1}{8\pi\mu} \int_{\partial D}  S_{ij}^\varepsilon (\mathbf{x}_0,\mathbf{x}) g^a_j(\mathbf{x}) \, dS(\mathbf{x}) = \frac{1}{8\pi\mu}\int_{s=0}^\ell \left(\int_{\theta=0}^{2\pi} S_{ij}^\varepsilon (\mathbf{x}_0,\mathbf{x}(s,\theta)) g^a_j(s,\theta) \, r(s) \, d\theta\right)  ds. \end{equation} 

Converting to cylindrical polar coordinates, we introduce the transformation matrix 
\begin{equation}\Theta(\theta) = \left(\begin{array}{ccc} \cos\theta & -\sin\theta & 0 \\ \sin\theta & \cos\theta & 0 \\ 0 & 0 & 1 \end{array} \right). \end{equation}
Letting indices $i,j,k$ and $\alpha,\beta,\gamma$ correspond to Cartesian and cylindrical polar bases respectively (with Einstein summation convention employed for both sets), it follows that $g_j^a = \Theta_{j\alpha}(\theta)g_\alpha^a$. Assuming that velocity is evaluated at fluid point $\mathbf{x}_0=(r_0,\theta_0,z_0)$ in cylindrical polars, it further follows that $u_i = \Theta_{i\alpha}(\theta_0)u_\alpha.$ Recognizing that $\Theta^{-1}=\Theta^T$, substitution of the cylindrical forms of the velocity and force vectors into Equation \eqref{ringsol} thus yields
\begin{align} \nonumber
u_\alpha(r_0,\theta_0,z_0) & = \Theta_{\alpha i}(\theta_0)u_i(x_0,y_0,z_0) \\ & = \frac{1}{8\pi\mu}\Theta_{\alpha i}(\theta_0)\int_0^\ell \int_0^{2\pi} S_{ij}^\varepsilon (\mathbf{x}_0,\mathbf{x}(s,\theta)) \Theta_{j \beta }(\theta)g^a_\beta(s) \, r(s) \, d\theta \, ds. \label{ringsol3} \end{align}
Under the assumption of axisymmetry, it is sufficient to only consider cases $y_0=\theta_0=0$. This results in the Cartesian $x-z$ and the cylindrical polar $r-z$ planes coinciding, such that the transformation matrix with $\theta_0=0$ simply yields the identity matrix and Equation \eqref{ringsol3} reduces to
\begin{align} \label{ringsol4}
u_\alpha(r_0,z_0) &= \frac{1}{8\pi\mu}\int_0^\ell  g_\beta^a(s)  \left[\delta_{\alpha i}r(s)\int_0^{2\pi} S_{ij}^\varepsilon (\mathbf{x}_0,\mathbf{x}(s,\theta)) \Theta_{j \beta }(\theta) \, d\theta\right]ds, \\ &= \frac{1}{8\pi\mu}\int_0^\ell  g_\beta^a(s) R^\varepsilon_{\alpha\beta}(\mathbf{x}_0,\mathbf{x}(s)) \, ds, \label{ringsol5} \end{align}
in which the `ringlet' kernel
\begin{equation} \label{eq:ringDef} R^\varepsilon_{\alpha\beta} (\mathbf{x}_0,\mathbf{x}(s)) := \delta_{\alpha i}r(s)\int_0^{2\pi} S^\varepsilon_{ij}(\mathbf{x}_0,\mathbf{x}(s,\theta))\Theta_{j\beta}(\theta) \, d\theta, \end{equation} is implicitly defined. Here, $\delta_{\alpha i}$ is the Kronecker delta function in which $\delta_{\alpha i}=1$ if $\alpha = i$ and $0$ otherwise. Unlike the Stokeslet which has the symmetric property $S_{ij}^\varepsilon(\mathbf{x}_0,\mathbf{x}) \equiv S_{ij}^\varepsilon(\mathbf{x},\mathbf{x}_0)$, ordering of arguments in the ringlet is important; the first and second arguments in $R_{\alpha\beta}^\varepsilon(\mathbf{x}_0,\mathbf{x})$ denote the fluid point and ring location respectively, with the ring radius $r(s)$ being the crucial non-symmetric term.

The various terms in the Stokeslet $S_{ij}^\varepsilon$ can be evaluated in cylindrical polar coordinates via
\begin{align} (x_{0,1}-x_1) & = r_0-r\cos\theta, \\ (x_{0,2}-x_2) &= -r\sin\theta, \\ (x_{0,3}-x_3) &= z_0-z, \\ |\mathbf{x}_0-\mathbf{x}|^2 &= (r_0-r\cos\theta)^2+(r\sin\theta)^2 + (z_0-z)^2.
\end{align} 
The resulting form of each $S^\varepsilon_{ij}$ is given in Appendix \ref{sec:IntKer} and used in the evaluation of $R_{\alpha\beta}^\varepsilon$ in Equation \eqref{eq:ringDef}. These integrals $R_{\alpha\beta}^\varepsilon$ yield the regularized fundamental solution for an axisymmetric ring of concentrated force (the regularized ringlet). Analytical evaluation reveals that $R_{r\theta}^\varepsilon=R_{z\theta}^\varepsilon=R_{\theta r}^\varepsilon=R_{\theta z}^\varepsilon=0$ such that the rotational problem for $g_{\theta}^a$ decouples from $g_r^a$ and $g_z^a$. The remaining nonzero $R_{\alpha\beta}^\varepsilon$ yield the equations
\begin{equation} \label{eq:rzForce}
  \left(\begin{array}{c}u_r(\mathbf{x}_0) \\ u_z(\mathbf{x}_0) \end{array}\right) =\frac{1}{8\pi\mu} \int_0^\ell \left[\begin{array}{cc}R_{rr}^\varepsilon(\mathbf{x}_0,\mathbf{x}(s)) & R_{rz}^\varepsilon(\mathbf{x}_0,\mathbf{x}(s)) \\ R_{zr}^\varepsilon(\mathbf{x}_0,\mathbf{x}(s)) & R_{zz}^\varepsilon(\mathbf{x}_0,\mathbf{x}(s)) \end{array}\right]\left(\begin{array}{c} g^a_r(s) \\  g^a_z(s)\end{array}\right) \, ds,\end{equation} and \begin{equation} \label{eq:aziForce} u_\theta(\mathbf{x}_0) = \frac{1}{8\pi\mu}\int_0^\ell R^\varepsilon_{\theta\theta}(\mathbf{x}_0,\mathbf{x}(s)) g^a_\theta(s) \, ds.
\end{equation}
Utilising Equations \eqref{eq:rzForce} -- \eqref{eq:aziForce} in tandem models axisymmetric problems with or without azimuthal rotation, in which the fluid experiences a constant force in each principal direction $(\hat{\mathbf{r}},\hat{\boldsymbol{{\theta}}},\hat{\mathbf{z}})$ at points along which the ringlet is located. This could be used, for example, to model the flow around a mobile axisymmetric body rotating about an axis defined by its direction of translation. 

Approximating the integrals in Equations \eqref{eq:rzForce} -- \eqref{eq:aziForce} numerically using a series of $N$ rings yields a system of equations of the form
\begin{align} \label{eq:stokes_summation} \left(\begin{array}{c}u_r(\mathbf{x}_0) \\ u_z(\mathbf{x}_0) \end{array}\right) &= \frac{1}{8\pi\mu} \sum_{n=1}^N \left[\begin{array}{cc}R_{rr}^\varepsilon(\mathbf{x}_0,\mathbf{x}_n) & R_{rz}^\varepsilon(\mathbf{x}_0,\mathbf{x}_n) \\ R_{zr}^\varepsilon(\mathbf{x}_0,\mathbf{x}_n) & R_{zz}^\varepsilon(\mathbf{x}_0,\mathbf{x}_n) \end{array}\right]\left(\begin{array}{c} g^a_r (\mathbf{x}_n) \\ g^a_z (\mathbf{x}_n)\end{array}\right)w_n, \\ u_\theta(\mathbf{x}_0) &= \frac{1}{8\pi\mu}\sum_{n=1}^N R^\varepsilon_{\theta\theta}(\mathbf{x}_0,\mathbf{x}_n) g^a_\theta(\mathbf{x}_n)w_n, \label{eq:stokes_summation2}\end{align} where $g^a_r(\mathbf{x}_n),g^a_\theta(\mathbf{x}_n),g^a_z(\mathbf{x}_n)$ are the radial, azimuthal, and axial components of the forces per unit area applied at ringlet location $\mathbf{x}_n$ respectively, and $w_n$ is the quadrature weight associated with $\mathbf{x}_n$ for numerical integration over $s$. The quantity $w_n$ has units of length unlike its counterpart $W_n$ in Cortez and colleagues' work (units of length squared). It is also possible to recompile the force per unit area $\mathbf g^a$ and quadrature weight $w_n$ into a force per unit length $\mathbf g^l$, such that Equation \eqref{eq:stokes_summation} can alternatively represent the fluid velocity induced by a series of rings. 

By considering the fluid velocity at each individual ringlet location, an invertible system can be produced. In the irrotational case ($u_\theta\equiv 0$), this takes the block matrix form
\begin{equation} \label{eq:invertStart} \mathsf{G}^l = 8\pi\mu {\underbrace{\left[ \begin{array}{cc} \mathsf R_{rr}^\varepsilon & \mathsf R_{rz}^\varepsilon\\ \mathsf R_{zr}^\varepsilon & \mathsf R_{zz}^\varepsilon \end{array}\right]}_{\mathsf{R}^\varepsilon}}^{-1} \mathsf{U},
\end{equation} where
\begin{equation} {\mathsf{G}^l} = \left[\begin{array}{c} g^l_r (\mathbf{x}_1)\\ \vdots \\ g^l_r (\mathbf{x}_N)\\ g^l_z (\mathbf{x}_1) \\ \vdots \\ g^l_z (\mathbf{x}_N)\end{array} \right],  \quad \mathsf{U} =  \left[ \begin{array}{c}
u_r (\mathbf{x}_1) \\ \vdots \\ u_r (\mathbf{x}_N) \\ u_z (\mathbf{x}_1) \\ \vdots \\ u_z (\mathbf{x}_N)\end{array} \right], \end{equation} and
\begin{equation}  \label{eq:invertEnd} \mathsf R_{\alpha\beta}^\varepsilon=\left[\begin{array}{cccc} R_{\alpha\beta}^\varepsilon(\mathbf{x}_1,\mathbf{x}_1) & R_{\alpha\beta}^\varepsilon(\mathbf{x}_1,\mathbf{x}_2) & \dots & R_{\alpha\beta}^\varepsilon(\mathbf{x}_1,\mathbf{x}_N) \\ R_{\alpha\beta}^\varepsilon(\mathbf{x}_2,\mathbf{x}_1) & R_{\alpha\beta}^\varepsilon(\mathbf{x}_2,\mathbf{x}_2) & \dots & R_{\alpha\beta}^\varepsilon(\mathbf{x}_2,\mathbf{x}_N) \\ \vdots & \vdots & \ddots & \vdots \\ R_{\alpha\beta}^\varepsilon(\mathbf{x}_N,\mathbf{x}_1) & R_{\alpha\beta}^\varepsilon(\mathbf{x}_N,\mathbf{x}_2) & \dots & R_{\alpha\beta}^\varepsilon(\mathbf{x}_N,\mathbf{x}_N) \end{array} \right]. \end{equation} 
Hence the forces needed to induce a given prescribed velocity in the fluid may be found (the resistance problem). This works for both a series of translating rings and a translating axisymmetric body, with the force per unit length $\mathbf g^l$ essentially `absorbing' both the force per unit area $\mathbf g^a$ and the quadrature weight $w_n$ in the latter case. Inclusion of rotation involves the formulation of a similar invertible system for Equation \eqref{eq:stokes_summation2} which can be solved separately.

\subsection{Analytical evaluation of the regularized ringlet} \label{sec:analyticalRinglet}
The nonzero elements of the regularized ringlet can be expressed in the form
\begin{align} \label{eq:R_start}
  R_{rr}^\varepsilon(\mathbf{x}_0,\mathbf{x}_n) &= r_n(-r_0r_nI_0+(2\tau-(z_0-z_n)^2)I_1 - 3r_0r_nI_2), \\ R_{rz}^\varepsilon(\mathbf{x}_0,\mathbf{x}_n) &= r_n(z_0-z_n)(r_0I_0-r_nI_1), \\ R_{zr}^\varepsilon(\mathbf{x}_0,\mathbf{x}_n) &= r_n(z_0-z_n)(-r_nI_0+r_0I_1), \\ R_{zz}^\varepsilon(\mathbf{x}_0,\mathbf{x}_n) &= r_n((\tau + (z_0-z_n)^2 + \varepsilon^2)I_0 - 2r_0r_nI_1), \\ R_{\theta\theta}^\varepsilon(\mathbf{x}_0,\mathbf{x}_n) &= r_n(r_0r_nI_0 + (\tau+\varepsilon^2)I_1 - 3r_0r_nI_2), \label{eq:R_end} \end{align} in which $\tau:=r_0^2+r_n^2+(z_0-z_n)^2+\varepsilon^2,$ and \begin{align} \label{eq:I_n} I_n &:= \int_0^{2\pi} \frac{\cos^n\theta}{(\tau-2r_0r_n\cos\theta)^{3/2}} \, d\theta, \\ &\;= \frac{4k^3}{(4r_0r_n)^{3/2}} \int_0^{\frac{\pi}{2}}  \frac{(2\cos^2\theta-1)^n}{(1-k^2\cos^2\theta)^{3/2}} \,d\theta, \label{eq:I_n2}
\end{align}
with $k^2 := 4r_0r_n/(\tau+2r_0r_n)$. Equation \eqref{eq:I_n2} is found by using the double angle formula for $\cos\theta$ as well as symmetry arguments about $\pi/2$. Following the example of Pozrikidis \cite{pozrikidis1992boundary}, the integrals $I_n$ can be computed by first expanding the numerator of the integrand in Equation \eqref{eq:I_n2} to obtain a series of polynomial integrals with respect to $\cos\theta$. Letting
\begin{equation}
  I_n' := 2^n\left(\frac{4k^3}{(4r_0r_n)^{3/2}}\right) \int_0^{\frac{\pi}{2}}  \frac{\cos^{2n}\theta}{(1-k^2\cos^2\theta)^{3/2}} \,d\theta,
\end{equation} 
it follows that \begin{equation} I_0 = I_0', \quad I_1 = I_1' - I_0, \quad I_2 = I_2'-2I_1'+I_0. \end{equation} The individual integrals $I_n'$ can be expressed in terms of complete elliptic integrals of the first and second kind, which are respectively defined
\begin{equation} F = F(k) := \int_0^{\frac{\pi}{2}} \frac{d\theta}{(1-k^2\sin^2\theta)^{1/2}} \quad  \text{and} \quad E = E(k) := \int_0^{\frac{\pi}{2}}(1-k^2\sin^2\theta)^{1/2} \, d\theta. \end{equation} The solutions for each $I_n'$ (as can be found in Section 2.58 of Gradshteyn and Ryzhik \cite{gradshteyn2014table}) are given by
\begin{align}
  I_0' &= \frac{4k^3}{(4r_0r_n)^{3/2}}\left(\frac{1}{1-k^2}E\right) , \\
  I_1' &= \frac{8k^3}{(4r_0r_n)^{3/2}}\left(\frac{1}{k^2(1-k^2)}E - \frac{1}{k^2}F\right), \\
  I_2' &= \frac{16k^3}{(4r_0r_n)^{3/2}}\left(\frac{2-k^2}{k^4(1-k^2)}E-\frac{2}{k^4}F\right),
\end{align}
from which it follows that
\begin{align}\label{eq:I_start}
  I_0 &= \frac{4k^3}{(4r_0r_n)^{3/2}}\left(\frac{1}{1-k^2}E\right) , \\
  I_1 &= \frac{4k^3}{(4r_0r_n)^{3/2}}\left(\frac{2-k^2}{k^2(1-k^2)}E - \frac{2}{k^2}F\right), \\
  I_2 &= \frac{4k^3}{(4r_0r_n)^{3/2}}\left(\frac{k^4-8k^2+8}{k^4(1-k^2)}E-\frac{4(2-k^2)}{k^4}F\right). \label{eq:I_end}
\end{align}
Substitution of Equations \eqref{eq:I_start} -- \eqref{eq:I_end} back into Equations \eqref{eq:R_start} -- \eqref{eq:R_end} yields the complete solution for the regularized ringlet
\begin{align} \label{eq:finalSolStart}
R_{rr}^\varepsilon(\mathbf{x}_0,\mathbf{x}_n) & = \frac{k}{r_0r_n}\left(\frac{r_n}{r_0}\right)^{\frac{1}{2}}\left[(\tau+(z_0-z_n)^2)F + \frac{4r_0^2r_n^2 - \tau(\tau+(z_0-z_n)^2)}{\tau-2r_0r_n}E\right], \\
R_{rz}^\varepsilon(\mathbf{x}_0,\mathbf{x}_n) & = k\frac{(z_0-z_n)}{r_0}\left(\frac{r_n}{r_0}\right)^{\frac{1}{2}}\left[F + \frac{2r_0^2-\tau}{\tau-2r_0r_n}E\right], \\ R_{zr}^\varepsilon(\mathbf{x}_0,\mathbf{x}_n) & = -k\frac{(z_0-z_n)}{(r_0r_n)^{\frac{1}{2}}}\left[F + \frac{2r_n^2-\tau}{\tau-2r_0r_n}E\right], \\ R_{zz}^\varepsilon(\mathbf{x}_0,\mathbf{x}_n) & = 2k\left(\frac{r_n}{r_0}\right)^{\frac{1}{2}}\left[F + \frac{(z_0-z_n)^2+\varepsilon^2}{\tau-2r_0r_n}E\right], \\ R_{\theta\theta}^\varepsilon(\mathbf{x}_0,\mathbf{x}_n) & = \frac{k}{r_0r_n}\left(\frac{r_n}{r_0}\right)^{\frac{1}{2}}\left[(2\tau-\varepsilon^2)F + \frac{8r_0^2r_n^2+\tau(\varepsilon^2-2\tau)}{\tau-2r_0r_n}E\right]. \label{eq:finalSolEnd}
\end{align}
The solutions given by Equations \eqref{eq:finalSolStart} -- \eqref{eq:finalSolEnd} can be readily evaluated except when $r_0=0$ or $r_n=0$. In the limit as $r_n\rightarrow 0$, all $R_{\alpha\beta}^\varepsilon \rightarrow 0$. In the limit as $r_0 \rightarrow 0$, both $R_{zr}^\varepsilon$ and $R_{zz}^\varepsilon$ tend to finite values while $R_{rr}^\varepsilon,R_{rz}^\varepsilon,R_{\theta\theta}^\varepsilon \rightarrow 0.$ This behaviour is described in detail in Appendix \ref{app:limits}. Streamlines for the flow induced by the regularized ringlet with associated unit forces in both $\hat{\mathbf{r}}$ and $\hat{\mathbf{z}}$ directions are given in Appendix \ref{sec:streamlines}.

The form of the ringlet solutions $R_{\alpha\beta}^\varepsilon$ is similar to those for the ring of singular Stokeslets (as detailed by Pozrikidis in \cite{pozrikidis1992boundary} with the exception of the newly derived $R_{\theta\theta}^0$)
\begin{align} \label{eq:poz1} R_{rr}^{0}(\mathbf{x}_0,\mathbf{x}_n)  & =  \frac{k}{r_0r_n}\left(\frac{r_n}{r_0}\right)^{\frac{1}{2}}\bigg[(r_0^2+r_n^2+2(z_0-z_n)^2)F - \frac{2(z_0-z_n)^4 +3(z_0-z_n)^2(r_0^2+r_n^2) + (r_0^2-r_n^2)^2}{(z_0-z_n)^2+(r_0-r_n)^2}E\bigg], \\ R_{rz}^0(\mathbf{x}_0,\mathbf{x}_n) &=  k\frac{(z_0-z_n)}{r_0}\left(\frac{r_n}{r_0}\right)^{\frac{1}{2}}\left[F + \frac{r_0^2 - r_n^2 - (z_0-z_n)^2}{(z_0-z_n)^2 + (r_0-r_n)^2}E\right], \\ R_{zr}^0(\mathbf{x}_0,\mathbf{x}_n) &=  -k\frac{(z_0-z_n)}{(r_0r_n)^{\frac{1}{2}}}\left[F - \frac{r_0^2 - r_n^2 + (z_0-z_n)^2}{(z_0-z_n)^2 + (r_0-r_n)^2}E\right], \\ R_{zz}^0(\mathbf{x}_0,\mathbf{x}_n) &= 2k\left(\frac{r_n}{r_0}\right)^{\frac{1}{2}}\left[F+\frac{(z_0-z_n)^2}{(z_0-z_n)^2 + (r_0-r_n)^2}E\right], \\ R_{\theta\theta}^0(\mathbf{x}_0,\mathbf{x}_n) & = \frac{k}{r_0r_n}\left(\frac{r_n}{r_0}\right)^{\frac{1}{2}}\bigg[2(r_0^2+r_n^2+(z_0-z_n)^2)F  - \frac{4(z_0-z_n)^4 + 4(z_0-z_n)^2(r_0^2+r_n^2) + 2(r_0^2-r_n^2)^2}{(z_0-z_n)^2 + (r_0-r_n)^2}E\bigg], \label{eq:poz2} \end{align} and in the limit as $\varepsilon \rightarrow 0$ our solutions are equivalent to their singular counterparts. This can be verified by substitution of $\varepsilon=0$ into $R_{\alpha\beta}^\varepsilon$ and is a result of the cutoff function $\phi_{\varepsilon}$ approaching a delta distribution as $\varepsilon \rightarrow 0$. 

Equations \eqref{eq:rzForce} -- \eqref{eq:aziForce} and \eqref{eq:finalSolStart} -- \eqref{eq:finalSolEnd} provide the exact analytical solution for the fluid velocity at any point due to the drag force per unit area on the surface of a generalized axisymmetric body. Using Equations \eqref{eq:stokes_summation} -- \eqref{eq:stokes_summation2} in place of \eqref{eq:rzForce} -- \eqref{eq:aziForce} yields the numerical solution based on discretization over the arclength $s$. In the case of a single ring, removing the integral over $s$ and replacing the force per unit area $\mathbf{g}^a$ with a force per unit length $\mathbf{g}^\ell$ yields the exact analytical solution for the fluid velocity induced by the force acting along the ring in $3$D space.
\FloatBarrier
\subsection{On the double layer potential} \label{sec:mainDLP}
A more complete formulation of Equation \eqref{cortez_integrals} for the fluid velocity induced by a translating body $D$ is given by
\begin{equation} \int_{\mathbb{R}^3} u_j(\mathbf{x})\phi_\varepsilon(\mathbf{x}_0-\mathbf{x}) \, dV(\mathbf{x}) = \frac{1}{8\pi\mu} \int_{\partial D}  S_{ij}^\varepsilon (\mathbf{x}_0,\mathbf{x}) g_i^a(\mathbf{x}) \, dS(\mathbf{x}) + \frac{1}{8\pi}\int_{\partial D} u_i(\mathbf{x})T_{ijk}^\varepsilon(\mathbf{x}_0,\mathbf{x})n_k(\mathbf{x}) \, dS(\mathbf{x}),\end{equation} 
where the first and second integrals on the right hand side are known as the single layer potential (SLP) and the double layer potential (DLP) respectively. The stress tensor $T_{ijk}^\varepsilon$ present in the DLP is given by
\begin{equation} T_{ijk}^\varepsilon(\mathbf{x}_0,\mathbf{x}) = -6\frac{(x_{0,i}-x_i)(x_{0,j}-x_j)(x_{0,k}-x_k)}{(|\mathbf{x}_0-\mathbf{x}|^2 + \varepsilon^2)^{5/2}} - 3\varepsilon^2 \frac{(x_{0,i}-x_i)\delta_{jk} + (x_{0,j}-x_j)\delta_{ik} + (x_{0,k}-x_k)\delta_{ij}}{(|\mathbf{x}_0-\mathbf{x}|^2 + \varepsilon^2)^{5/2}}.\end{equation}

The DLP can be neglected for problems in which the condition $\int_{\partial D} \mathbf{u}\cdot\hat{\mathbf{n}} \, dS =0$ is satisfied \cite[Chapter~2.3]{pozrikidis1992boundary}, which is the case throughout this manuscript. However we provide the expressions for the DLP in Appendix \ref{sec:DLP}, which may be of value for future studies of systems in which the condition $\int_{\partial D} \mathbf{u}\cdot\hat{\mathbf{n}} \, dS =0$ is violated (e.g. bubbles).

\FloatBarrier
\section{Simple examples and test cases} \label{sec:examples}
In Section \ref{sec:singularRegularizedStokeslets} the expression for the regularized fundamental solution for an axisymmetric ring of concentrated forces, the regularized ringlet, was derived. In the following, we demonstrate the validity of the method through application to simple cases of motion. 

The first two cases concern the translation and rotation of the unit sphere in a Stokesian fluid, treated independently in Sections \ref{sec:singSolComp} and \ref{sec:rotSphere} respectively. In Section \ref{sec:toroidalSwimmer} a more complicated example is considered: the propulsion of `Purcell's toroidal swimmer' \cite{purcell1977life,leshansky2008surface}, powered by tank treading of the torus surface. In considering these different cases, it is shown that the method of regularized ringlets can be used to model the surface motion of axisymmetric bodies in each principal direction $\hat{\mathbf{r}}, \hat{\bm{{\theta}}}, \hat{\mathbf{z}}$ in a cylindrical coordinate system.

\subsection{Translating unit sphere} \label{sec:singSolComp}
The validity of the regularized ringlet method is illustrated by solving the resistance problem for the translating unit sphere. Given a prescribed surface velocity ($-\hat{\mathbf{z}}$), Equation \eqref{ringsol5} yields a Fredholm first kind integral equation for the unknown force distribution \cite{smith2018nearest}. The method of regularized ringlets (implemented here via Matlab) can be used to solve this problem.

The sphere is parametrized in the $rz$ plane by $\mathbf{p} = \cos\varphi\,\hat{\mathbf{r}} + \sin\varphi\,\hat{\mathbf{z}}$ for $\varphi \in [-\pi/2,\pi/2]$, then discretized as \begin{equation} \varphi_n = \pi \frac{n-1/2}{N}-\frac{\pi}{2} \quad \text{for } n=1,\ldots,N.\end{equation} The velocity boundary condition $\mathbf{u}=-\hat{\mathbf{z}}$ is prescribed at each $\mathbf{x}_n:=\mathbf{p}(\varphi_n)$, and the resulting linear system is solved to yield the required force densities $\mathbf{g}^\ell$ at each of these locations. The drag is then calculated as \begin{equation} \sum_{n=1}^N \int_{\theta=0}^{2\pi} g_z^\ell(\mathbf{x}_n) \, r(\mathbf{x}_n)d\theta = 2\pi\sum_{n=1}^N r(\mathbf{x}_n)g_z^\ell(\mathbf{x}_n), \end{equation} which is compared with the Stokes law value of $-6\pi$. The relative errors are given for various values of $N$ and regularization parameter $\varepsilon$ in Table \ref{tab:traSph1} alongside the condition number of the resistance matrix $\mathsf{R}^\varepsilon$ in Table \ref{tab:traSph2}.

\begin{table}[ht]
	\centering
	\begin{ruledtabular}
	\begin{tabular}{lrrrrr}
		$\varepsilon$ & \multicolumn{1}{c}{$N=25$} & \multicolumn{1}{c}{51} & \multicolumn{1}{c}{101} & \multicolumn{1}{c}{201} & \multicolumn{1}{c}{401} \\ \cmidrule{1-1} \cmidrule{2-6}
		0.01  & $-1.4689\cdot 10^{-2}$ & $-2.0609\cdot 10^{-3}$ & $1.6439\cdot 10^{-3}$  & $2.4053\cdot 10^{-3}$ & $2.5104\cdot 10^{-3}$  \\
		0.005 & $-2.4754\cdot 10^{-2}$ & $-7.2086\cdot 10^{-3}$ & $-1.1116\cdot 10^{-3}$ & $7.6816\cdot 10^{-4}$  & $1.2056\cdot 10^{-3}$  \\
		0.001 & $-4.7242\cdot 10^{-2}$ & $-1.8774\cdot 10^{-2}$ & $-7.0948\cdot 10^{-3}$ & $-2.3160\cdot 10^{-3}$ & $-5.1183\cdot 10^{-4}$ \\ 
	\end{tabular}
\end{ruledtabular}
	\caption{Relative errors in the drag calculation for the resistance problem on the translating unit sphere.}
	\label{tab:traSph1}
\end{table}

\begin{table}[ht]
	\centering
		\begin{ruledtabular}
	\begin{tabular}{lrrrrr}
		$\varepsilon$ & \multicolumn{1}{c}{$N=25$} & \multicolumn{1}{c}{51} & \multicolumn{1}{c}{101} & \multicolumn{1}{c}{201} & \multicolumn{1}{c}{401} \\ \cmidrule{1-1} \cmidrule{2-6}
		0.01          & $4.6282\cdot 10^{1}$ & $1.5857\cdot 10^2$ & $7.1602\cdot 10^2$ & $7.0167\cdot 10^3$ & $4.0767\cdot 10^5$ \\
		0.005         & $3.3089\cdot 10^1$ & $9.6186\cdot 10^1$ & $3.1303\cdot 10^2$ & $1.4181\cdot 10^3$ & $1.3947\cdot 10^4$ \\
		0.001         & $1.9973\cdot 10^1$ & $4.9332\cdot 10^1$ & $1.2386\cdot 10^2$ & $5.7653\cdot 10^2$ & $1.0449\cdot 10^3$ \\ 
	\end{tabular}
\end{ruledtabular}
	\caption{Condition numbers of the resistance matrix $\mathsf{R}^\varepsilon$ for the resistance problem on the translating unit sphere.}
	\label{tab:traSph2}
\end{table}

For given $N$, excessively small $\varepsilon$ results in the drag error becoming non--monotonic. For given $\varepsilon$, increasing $N$ eventually ceases to result in a further reduction in the relative error. This is often the case with regularized Stokeslet methods (eg. see Figure \ref{fig:cc1}, Figure \ref{fig:cc2}, and references Cortez \emph{et al.} \cite{cortez2005method}, Gallagher \emph{et al.} \cite{gallagher2018sharp}). 

A thorough comparison of our results for the translating unit sphere with those of Cortez \emph{et al.} \cite{cortez2005method} can be found in Appendix \ref{sec:translatingSphere}, in which it is found that using just $N$ ringlets yields consistently more accurate solutions than using $3N^2$ regularized Stokeslets in a standard $3$D discretization of the sphere surface. A comparison to results obtainable using the axisymmetric method of fundamental solutions (a singular Stokeslet method) is given in Appendix \ref{sec:MFS}, in which the relative error in the fluid velocity is also discussed. In general, although the singular method can be tuned to give smaller relative errors in either the fluid velocity or total drag separately, it cannot do so simultaneously; regularized ringlets display more satisfactory convergence properties and are the more effective method to minimize errors in both fluid velocity and total drag. 
\FloatBarrier
\subsection{Rotating unit sphere} \label{sec:rotSphere}
The solution for the steady motion of a Stokesian fluid surrounding a solid sphere rotating uniformly about a central axis is well known and can be found in e.g. \emph{Hydrodynamics} by Lamb \cite[Chapter~XI]{lamb1945hydrodynamics}. If the sphere rotates around its $z$ axis in an $(r,\theta,z)$ cylindrical coordinate system with an angular velocity $\boldsymbol{{\Omega}} = \omega_0 \hat{\mathbf{e}}_z$, the resulting angular velocity of the fluid is given by $\boldsymbol{{\omega}} = (a/\gamma)^3\omega_0\hat{\mathbf{e}}_\theta$ where $\gamma = \sqrt{r^2 + z^2}$. This can be written in terms of the linear velocity (more readily usable in the Stokeslet formulae) over the entire domain as, \begin{equation} \label{eq:rotVel}  \mathbf{u} = \begin{cases} r\left(\frac{a}{\gamma}\right)^3\omega_0\hat{\mathbf{e}}_\theta & \forall \gamma \ge a, \\ r\omega_0\hat{\mathbf{e}}_\theta & \forall \gamma < a,\end{cases}\end{equation} where $\gamma \ge a$ corresponds to the surrounding fluid velocity and $\gamma < a$ to the solid body rotation of the sphere respectively. The zero Reynolds number torque on this sphere is given by, \begin{equation} \mathbf{T}=-8\pi\mu a^3\boldsymbol{{\Omega}},\end{equation} the derivation of which can be found in \cite{rubinow1961transverse}. This torque is associated with a drag force per unit area on the surface of the sphere given by $-3\mu\omega_0(r/a)\hat{\mathbf{e}}_\theta$, as detailed in Appendix \ref{sec:rotatingSphereForce}. The force per unit length used in the method of regularized ringlets thus takes form \begin{equation} g^l_\theta = \left(\frac{3\pi\mu}{N}\right)\left(\frac{r}{a}\right)\omega_0,\end{equation}
although for the resistance problem this is not prescribed. 

The sphere surface is again parametrized in the $rz$ plane by $\mathbf{p}=\cos\varphi\,\hat{\mathbf{r}}+\sin\varphi\,\hat{\mathbf{z}}$ for $\varphi\in[-\pi/2,\pi/2]$, discretized using $N$ ringlets at locations $\mathbf{x}_n$. 
Letting $\omega_0=-1$, the velocity $\mathbf{u}(\mathbf{x}_n)=-r(\mathbf{x}_n)\hat{\bm{{\theta}}}$ is prescribed at each $\mathbf{x}_n$ and the resistance matrix is constructed to yield the required force densities $\mathbf{g}^\ell$ at each of these locations. The torque is then calculated as
\begin{equation} \sum_{n=1}^N \int_{\theta=0}^{2\pi} g_\theta^\ell(\mathbf{x}_n) \, r^2(\mathbf{x}_n)d\theta = 2\pi\sum_{n=1}^N r^2(\mathbf{x}_n)g_\theta^\ell(\mathbf{x}_n), \end{equation} in which $r(\mathbf{x}_n)$ is squared since the torque is a moment, the product of distance and force. Comparing to the value $-8\pi$, relative errors are given for various values of $N$ and regularization parameter $\varepsilon$ in Table \ref{tab:rotSph1} alongside the condition number of the resistance matrix $\mathsf{R}_\theta^\varepsilon$ in Table \ref{tab:rotSph2}.
\begin{table}[ht]
	\centering
	\begin{ruledtabular}
	\begin{tabular}{lrrrrr}
		$\varepsilon$ & \multicolumn{1}{c}{$N=25$} & \multicolumn{1}{c}{51} & \multicolumn{1}{c}{101} & \multicolumn{1}{c}{201} & \multicolumn{1}{c}{401} \\ \cmidrule{1-1} \cmidrule{2-6}
		0.01          & $-6.6820\cdot 10^{-2}$ & $-1.3919\cdot 10^{-2}$ & $3.1012\cdot 10^{-3}$  & $7.1409\cdot 10^{-3}$  & $7.5502\cdot 10^{-3}$  \\
		0.005         & $-1.0168\cdot 10^{-1}$ & $-3.3360\cdot 10^{-2}$ & $-7.2206\cdot 10^{-3}$ & $1.5183\cdot 10^{-3}$  & $3.5556\cdot 10^{-3}$  \\
		0.001         & $-1.7339\cdot 10^{-1}$ & $-7.5656\cdot 10^{-2}$ & $-3.0238\cdot 10^{-2}$ & $-1.0422\cdot 10^{-2}$ & $-2.6879\cdot 10^{-3}$ \\ 
	\end{tabular}
\end{ruledtabular}
	\caption{Relative errors in the drag calculation for the resistance problem on the rotating unit sphere.}
\label{tab:rotSph1}
\end{table}

\begin{table}[ht]
	\centering
	\begin{ruledtabular}
	\begin{tabular}{lrrrrr}
		$\varepsilon$ & \multicolumn{1}{c}{$N=25$} & \multicolumn{1}{c}{51} & \multicolumn{1}{c}{101} & \multicolumn{1}{c}{201} & \multicolumn{1}{c}{401} \\ \cmidrule{1-1} \cmidrule{2-6}
		0.01          & $7.7965$ & $2.1678\cdot 10^1$ & $7.1174\cdot 10^1$ & $3.4925\cdot 10^2$ & $4.2212\cdot 10^3$ \\
		0.005         & $6.2313$ & $1.5492\cdot 10^1$ & $4.2389\cdot 10^1$ & $1.4119\cdot 10^2$ & $6.9613\cdot 10^2$ \\
		0.001         & $4.4051$ & \multicolumn{1}{l}{$9.5164$} & $2.1792\cdot 10^1$ & $5.4064\cdot 10^1$ & $1.4690\cdot 10^2$ \\ 
	\end{tabular}
\end{ruledtabular}
	\caption{Condition numbers of the resistance matrix $\mathsf{R}_\theta^\varepsilon$ for the resistance problem on the rotating unit sphere.}
\label{tab:rotSph2}
\end{table}

These results are similar to those of the translating unit sphere, although for given $\varepsilon, N$ the relative error in the drag is generally slightly larger and the condition number of the resistance matrix slightly smaller than for the results using the same $\varepsilon,N$ on the translating unit sphere. 
\FloatBarrier
\subsection{Purcell's toroidal swimmer}
\label{sec:toroidalSwimmer}
\begin{figure}[ht]
	\centering
	\begin{tikzpicture}
	
	\draw (-5,0) -- (5,0);
	\draw (0,-3) -- (0,5);
	
	\draw[dashed] (-2,0) circle (1);
	\draw[dashed] (2,0) circle (1);
	
	\draw[-{Latex[length=5mm, width=2mm]}] (0,4.5) -- (0,5);
	\draw[-{Latex[length=5mm, width=2mm]}] (4.5,0) -- (5,0);
	
	\draw[-{Latex[length=3mm, width=2mm]}] (3.38581929877,-0.57402514854) arc (-22.5:22.5:1.5);
	\draw[-{Latex[length=3mm, width=2mm]}] (-3.38581929877,-0.57402514854) arc (202.5:157.5:1.5);
	
	\draw[dashed] (2,0) -- (2,-2.5);
	\draw[{Latex[length=2mm, width=2mm]}-{Latex[length=2mm, width=2mm]}] (0,-2) -- (2,-2);
	\draw[{Latex[length=2mm, width=2mm]}-{Latex[length=2mm, width=2mm]}] (2,0) -- (1.393,-0.707);

	\draw[dashed] (2,0) -- (1.55,.8);
	\draw[thick,-{Latex[length=2mm, width=2mm]}] (3,0) arc (0:120:1);
	\draw (2.25,0) arc (0:120:.25);
	\node at (2.25,.4) {$\eta$};

	\node at (1.55,-.25) {$a$};
	\node at (1,-2.5) {$b$};
	\node at (4,0.5) {$\mathbf{u}^{(s)}$};
	\node at (-4,0.5) {$\mathbf{u}^{(s)}$};
	\node at (1.5,1.25) {$s$};
	
	\fill (3,0) circle (.1);

	\node at (5,-.25) {$r$};
	\node at (-.25,5) {$z$};
	
	\draw[-{Latex[length=3mm, width=2mm]}] (2.5,2.5) -- (2.5,4);
	\node at (3,3.25) {$-\mathbf{U}$};
	
	\draw[-{Latex[length=3mm, width=2mm]}] (-2.5,2.5) -- (-2.5,4);
	\node at (-3.1,3.25) {$-\mathbf{U}$};
	\end{tikzpicture}
	\caption{The geometry of the toroidal swimmer, whose cross section in the $rz$ plane is given by the dashed lines. Rotation around the $z$ axis produces the complete torus. The torus moves with velocity $\mathbf{U}$ in the direction opposing outer surface motion (such that in the given frame of reference in which the torus remains stationary, the surrounding fluid appears to move with velocity $-\mathbf{U}$). Redrawn from Leshansky \emph{et al.} \cite{leshansky2008surface}.}
	\label{fig:torusGeo}
\end{figure}
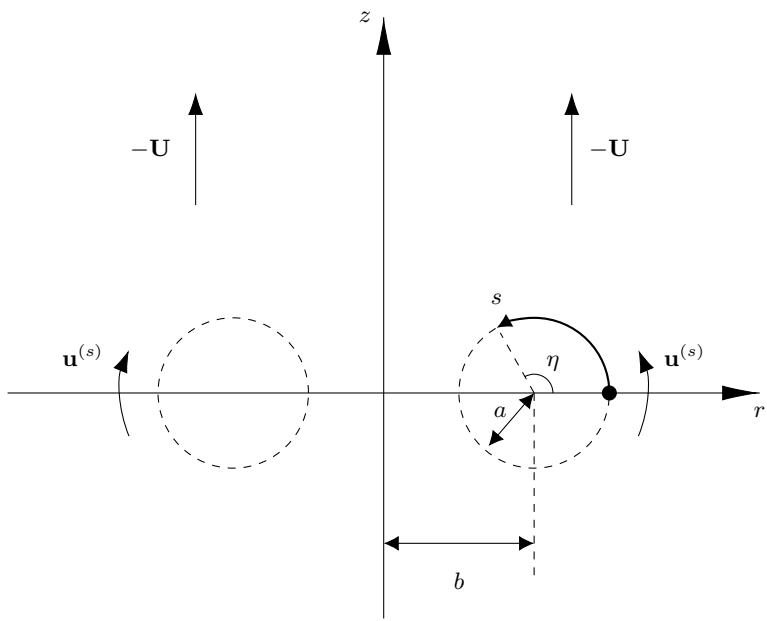

The torus is the simplest geometry capable of describing self--propelled organisms \cite{thaokar2007hydrodynamics}. Purcell's toroidal swimmer \cite{purcell1977life,taylor1952analysis} describes one such organism, the geometry of which can be seen in Figure \ref{fig:torusGeo}. Inward rotation of the torus surface produces a net force in the direction of motion of the outermost surface (against which the torus is propelled). The magnitude of this net force (and resultant propulsion speed of the torus) is dependent on the speed with which the surface of the torus rotates as well as the slenderness of the torus. It has been suggested that this mechanism could describe how a DNA miniplasmid could be turned into a self--propelled nanomachine \cite{kulic2005twirling}. Three modes of locomotion are considered by Leshansky and Kenneth \cite{leshansky2008surface}, corresponding to tank treading of (a) an incompressible surface, in which the tangential surface velocity is largest on the inner surface, (b) a weakly compressible surface, in which the tangential surface velocity is constant, (c) a highly compressible surface, in which the tangential surface velocity is largest on the outer surface. In the following, we restrict ourselves to looking at the case of constant tangential surface velocity. 

The torus geometry is reduced to a single slenderness parameter $s_0 = b/a$, where $b$ and $a$ refer to the major and minor radii of the torus respectively. The torus surface is parametrized in terms of angle $\eta \in [0,2\pi)$ in the $rz$ plane such that $ds = a\, d\eta$ and $\eta=0$ corresponds to the outermost radial point on the surface of the torus, traversed in an anticlockwise direction. For the free swimming torus, the rigid body translation $\mathbf{U}$ and rotation $\mathbf{u}^{(s)}$ of the torus surface $\partial D$ are related to the force per unit area $\mathbf g^a$ exerted by the torus on the surrounding fluid by, \begin{equation} U_\alpha(\mathbf{x}_0) + u^{(s)}_\alpha(\mathbf{x}_0) =  \frac{1}{8\pi\mu}\int_{s=0}^\ell  R_{\alpha\beta}^\varepsilon(\mathbf{x}_0,\mathbf{x}) g^a_\beta(\mathbf{s}) \, ds \quad \forall \mathbf{x}_0 \in \partial D\end{equation} subject to the condition of zero net force in the $\hat{\mathbf{z}}$ direction, \begin{equation} \int_{\partial D} \mathbf g_z^a(\mathbf{x}) \, dS(\mathbf{x}) = 0, \end{equation} where $s$ is the arclength parameterization of the cross section of the torus surface $\partial D$ in the $rz$ plane. We note that the additional free swimming conditions of zero net force in $\hat{\mathbf{r}}$ and zero total moment (as outlined by Phan--Thien \emph{et al.} \cite{phan1987boundary}) are automatically satisfied by axisymmetry and $g_\theta^a \equiv 0$ respectively. 

\begin{figure}[ht]
\centering
\subfloat[Broadwise translation of the torus.]{\includegraphics{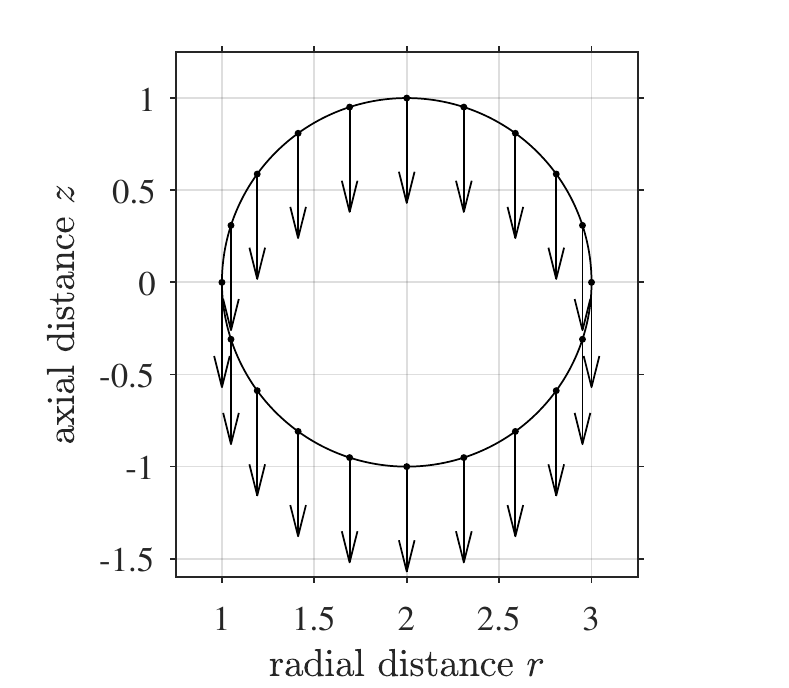} \label{fig:tga}}
\subfloat[Associated force distribution.]{\includegraphics{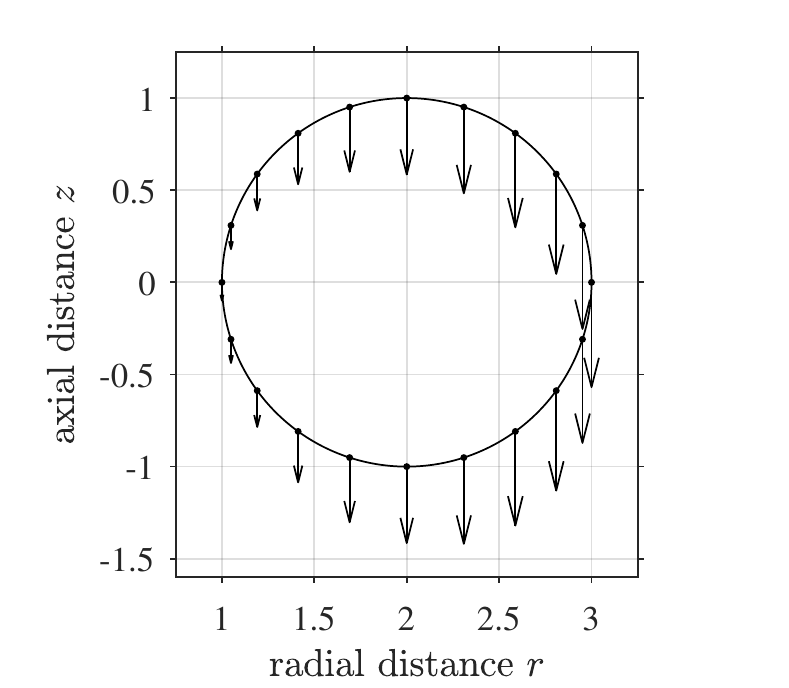} \label{fig:tgb}}
\caption{Surface velocity and associated force distribution of a toroidal glider with slenderness parameter $s_0 = 2$.}
\label{fig:torusGlider}
\end{figure}

\begin{figure}[ht]
\centering
\subfloat[Anticlockwise surface rotation of the torus.]{\includegraphics{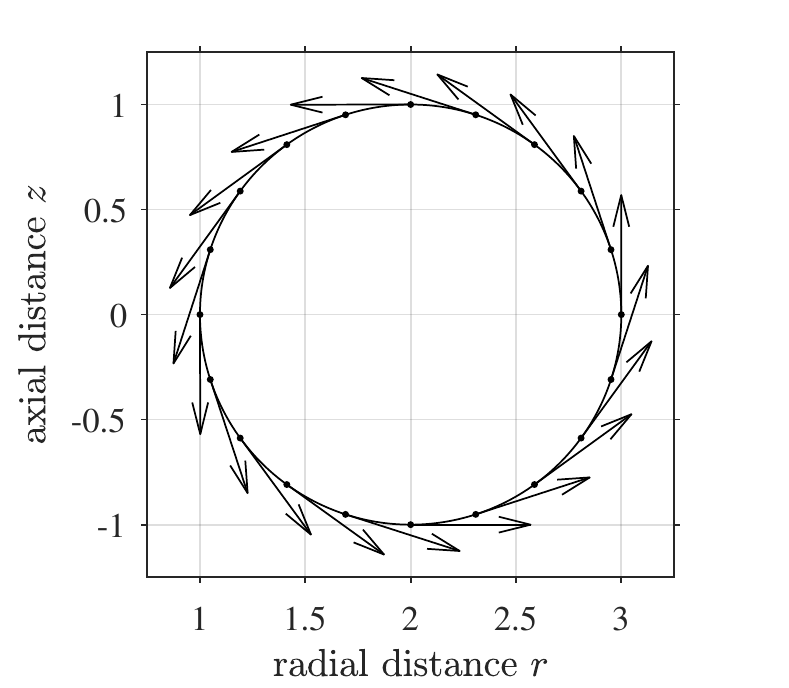} \label{fig:tpa}}
\subfloat[Associated force distribution.]{\includegraphics{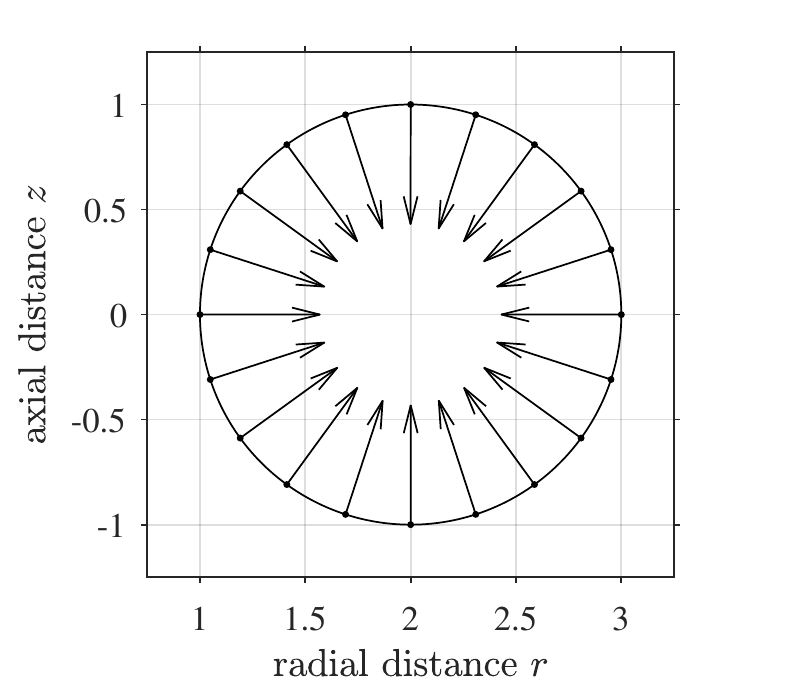} \label{fig:tpb}}
\caption{Surface velocity and associated force distribution of an anchored toroidal pump with slenderness parameter $s_0 = 2$.}
\label{fig:torusPump}
\end{figure}

The propulsion speed $U$ of the rotating torus for any given value of $s_0$ and rotation speed $u^{(s)}$ can be determined by considering two separate situations: one in which motion is purely translational in the $\hat{\mathbf{z}}$ direction (the toroidal glider with $\mathbf{U}=\hat{\mathbf{z}}, \mathbf{u}^{(s)}\equiv\mathbf{0}$) and one in which motion is purely rotational (the anchored toroidal pump with $\mathbf{U}\equiv\mathbf{0}, \mathbf{u}^{(s)} = \hat{\boldsymbol{{\eta}}}$ where $\hat{\boldsymbol{{\eta}}}$ is the unit vector whose direction varies over $s$, pointing tangential to the surface in the anticlockwise direction at all points). The glider and pump have associated force distributions $\mathbf{g}^{\text{gld}}$ and $\mathbf{g}^{\text{pmp}}$ respectively. Using the regularized ringlet, surface motions can be prescribed (Figures \ref{fig:tga} \& \ref{fig:tpa}) in order to compute the associated force distributions (Figures \ref{fig:tgb} \& \ref{fig:tpb}) responsible for producing each motion. For the toroidal glider the radial force is assumed to be zero ($g_r^{\text{gld}}\equiv 0$), hence \begin{align} \label{eq:glider1} U_r(\mathbf{x}_0) & = \frac{1}{8\pi\mu}\int_{\eta=0}^{2\pi}  R_{rz}^\varepsilon(\mathbf{x},\mathbf{x}_0) g_z^{\text{gld}}(\mathbf{x}) \, a\, d\eta = 0 \quad \forall \, \mathbf{x}_0 \in \partial D, \\ U_z(\mathbf{x}_0) & = \frac{1}{8\pi\mu}\int_{\eta=0}^{2\pi}  R_{zz}^\varepsilon(\mathbf{x},\mathbf{x}_0) g_z^{\text{gld}}(\mathbf{x}) \, a\, d\eta = 1 \quad \forall \, \mathbf{x}_0 \in \partial D, \label{eq:glider2}\end{align} which, after solving for the unknown force distribution $\mathbf g_z^{\text{gld}}$, yields the net axial force in the $\hat{\mathbf{z}}$ direction \begin{equation} G^{\text{gld}} = \int_{\partial D}  g_z^{\text{gld}}(\mathbf{x}) \, dS(\mathbf{x}),\end{equation} which is nonzero. For the anchored toroidal pump the surface velocity is given by $\mathbf{u}^{(s)}=\hat{\boldsymbol{{\eta}}}=(-\sin\eta,\cos\eta)$ in $(r,z)$ coordinates. It follows that \begin{align}\label{eq:pump1} u_r^{(s)}(\mathbf{x}_0)&=\frac{1}{8\pi\mu}\int_{\eta=0}^{2\pi} R_{r\beta}^\varepsilon(\mathbf{x},\mathbf{x}_0)g_\beta^{\text{pmp}}(\eta)\,a\,d\eta = -\sin\eta, \quad \forall \, \mathbf{x}_0 \in \partial D, \\ u_z^{(s)}(\mathbf{x}_0)&=\frac{1}{8\pi\mu}\int_{\eta=0}^{2\pi} R_{z\beta}^\varepsilon(\mathbf{x},\mathbf{x}_0) g_\beta^{\text{pmp}}(\eta)\,a\,d\eta =\cos\eta, \quad \forall \, \mathbf{x}_0 \in \partial D, \label{eq:pump2}\end{align} from which it can be determined that the net axial $\hat{\mathbf{z}}$ force is given by \begin{equation} G^{\text{pmp}} = \int_{\partial D} g_z^{\text{pmp}}(\eta) \, dS(\mathbf{x}).\end{equation} By linearity of the Stokes flow equations we may subtract the gliding solution from the pump solution to rewrite the system in the form \begin{align}\label{eq:swimmer1} u_r^{(s)}(\mathbf{x}_0)&=\frac{1}{8\pi\mu}\int_{\eta=0}^{2\pi} R_{r\beta}^\varepsilon(\mathbf{x},\mathbf{x}_0)g_\beta^a(\eta)\,a\,d\eta, \quad \forall \, \mathbf{x}_0 \in \partial D, \\ -\frac{G^{\text{pmp}}}{G^{\text{gld}}} + u_z^{(s)}(\mathbf{x}_0)&=\frac{1}{8\pi\mu}\int_{\eta=0}^{2\pi} R_{z\beta}^\varepsilon(\mathbf{x},\mathbf{x}_0)g_\beta^a(\eta)\,a\,d\eta, \quad \forall \, \mathbf{x}_0 \in \partial D,\label{eq:swimmer2}\end{align} where $\mathbf{g}^a = \mathbf{g}^{\text{pmp}} - (G^{\text{pmp}}/G^{\text{gld}})g^{\text{gld}}_z\hat{\mathbf{z}}$ and the net force is equal to \begin{equation} \int_{\partial D}g^a_z(\eta) \, dS(\mathbf{x}) = 0,\end{equation} as required for the free--swimmer. The propulsion velocity of the swimming torus is thus given by $\mathbf{U} = - (G^{\text{pmp}}/G^{\text{gld}})\hat{\mathbf{z}}$, opposing the direction of outer surface motion. The speed $U=|\mathbf{U}|$ is dependent on both the rotation speed $u^{(s)}:=|\mathbf{u}^{(s)}|$ and the slenderness ratio $s_0$, so both $G^{\text{gld}}$ and $G^{\text{pmp}}$ must be recomputed whenever one of these parameters is changed. However, Leshansky and Kenneth \cite{leshansky2008surface} were able to show that the propulsion speed depends linearly on the rotation speed, such that by considering the scaled propulsion speed $U/u^{(s)}$ it is only necessary vary $s_0$ to be able to consider all possible propulsion speeds resulting from a given constant rotational surface velocity. 

\begin{figure}[ht]
	\centering
	\includegraphics{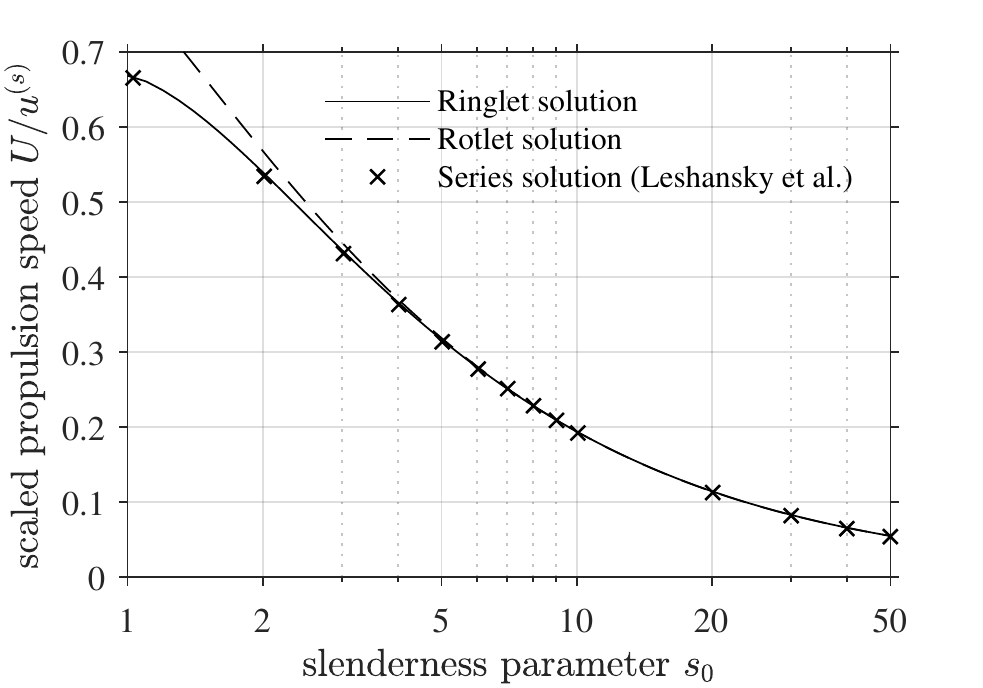}
	\caption{Scaled propulsion speed $U/u^{(s)}$ versus slenderness parameter $s_0$ using different numerical schemes. \textbf{Solid line} shows regularized ringlet solution using $N=100$ rings (method derived in this paper). \textbf{Dashed line} shows rotlet solution as detailed in \cite{leshansky2008surface}. \textbf{Crosses} denote values of the exact series solution, obtained from Figure 7 of Leshansky \emph{et al.} \cite{leshansky2008surface} using MathWorks' grabit function \cite{grabit} in Matlab.}
	\label{fig:scaledPropVel}
\end{figure}

Our results for the scaled propulsion velocity found using a discretization of the torus surface using $N=100$ regularized ringlets are compared with those obtained by Leshansky and Kenneth \cite{leshansky2008surface}, who tackled the same problem using a line distribution of rotlets at the torus centreline (inaccurate as $s_0 \rightarrow 1$) and an exact series solution via expansion in toroidal harmonics (Figure \ref{fig:scaledPropVel}). Of particular note, in the limit as $s_0 \rightarrow 1$ it is found that using $N=100$ rings with $\varepsilon=0.01$ in the regularized ringlet method yields a scaled propulsion velocity of $0.6684$, representing just a $0.513\%$ error when compared to the series solution value of $0.665.$ This error can be reduced to $<0.1\%$ by using $N=1000$ rings, at which point the value of the scaled propulsion velocity as calculated by the method of regularized ringlets is $0.6656$. This is a significant improvement over the solution found using a line distribution of rotlets, in which the error is  $>1\%$ for all $s_0 \le 6$. 

In addition to giving  values for the scaled propulsion speed that are in excellent agreement with the series solution of Leshansky and Kenneth \cite{leshansky2008surface}, the regularized ringlet solution also provides the force required at each point on the torus surface to produce the swimming motion. The series and rotlet solutions do not yield this information, with the propulsion speed instead being calculated according to the net drag force on the toroidal glider in these methods. This, in combination with Figures \ref{fig:tga} \& \ref{fig:tgb}, highlights why the centreline rotlet solution is inaccurate for small values of $s_0$; as slenderness decreases, the difference between the magnitude of the force required at the inner- and outer-most surfaces to produce rigid body translation of the torus grows large. In Figure \ref{fig:tgb} (in which $s_0=2$), the drag force on the outer surface is approximately $5.7$ times larger than that on the inner surface. The constant centreline force associated with the rotlet cannot account for this discrepancy, whereas the full discretization of the torus surface using regularized ringlets can. 

\begin{figure}[ht]
\centering
\subfloat[\raggedright{Streamlines of fluid velocity.}]{\includegraphics[trim=.5cm 1cm .5cm 1cm,clip]{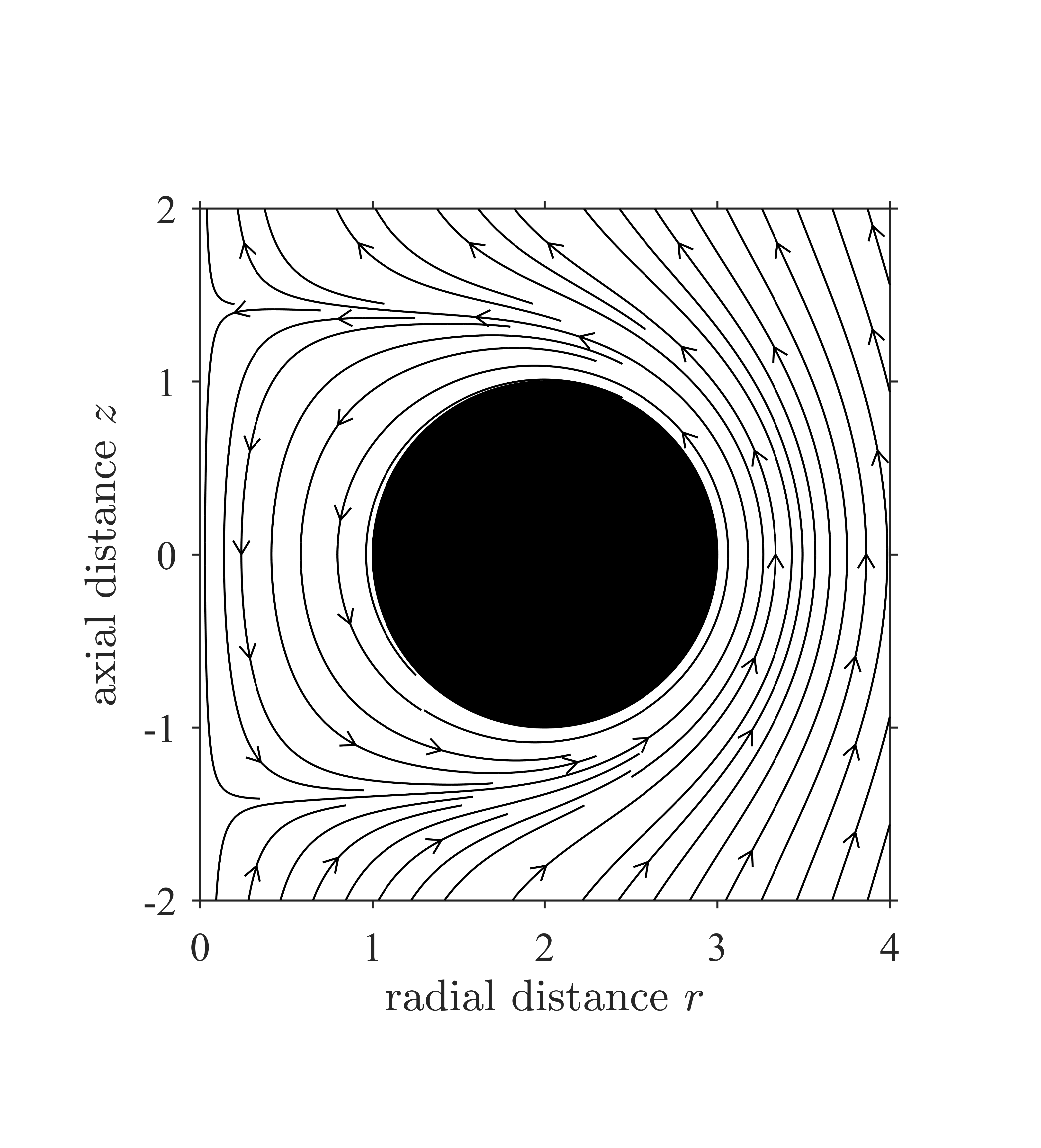} \label{fig:torusStreamlines}} \quad
\subfloat[\raggedright{Magnitude of fluid velocity.}]{\includegraphics[trim=.5cm 1cm .5cm 1cm,clip]{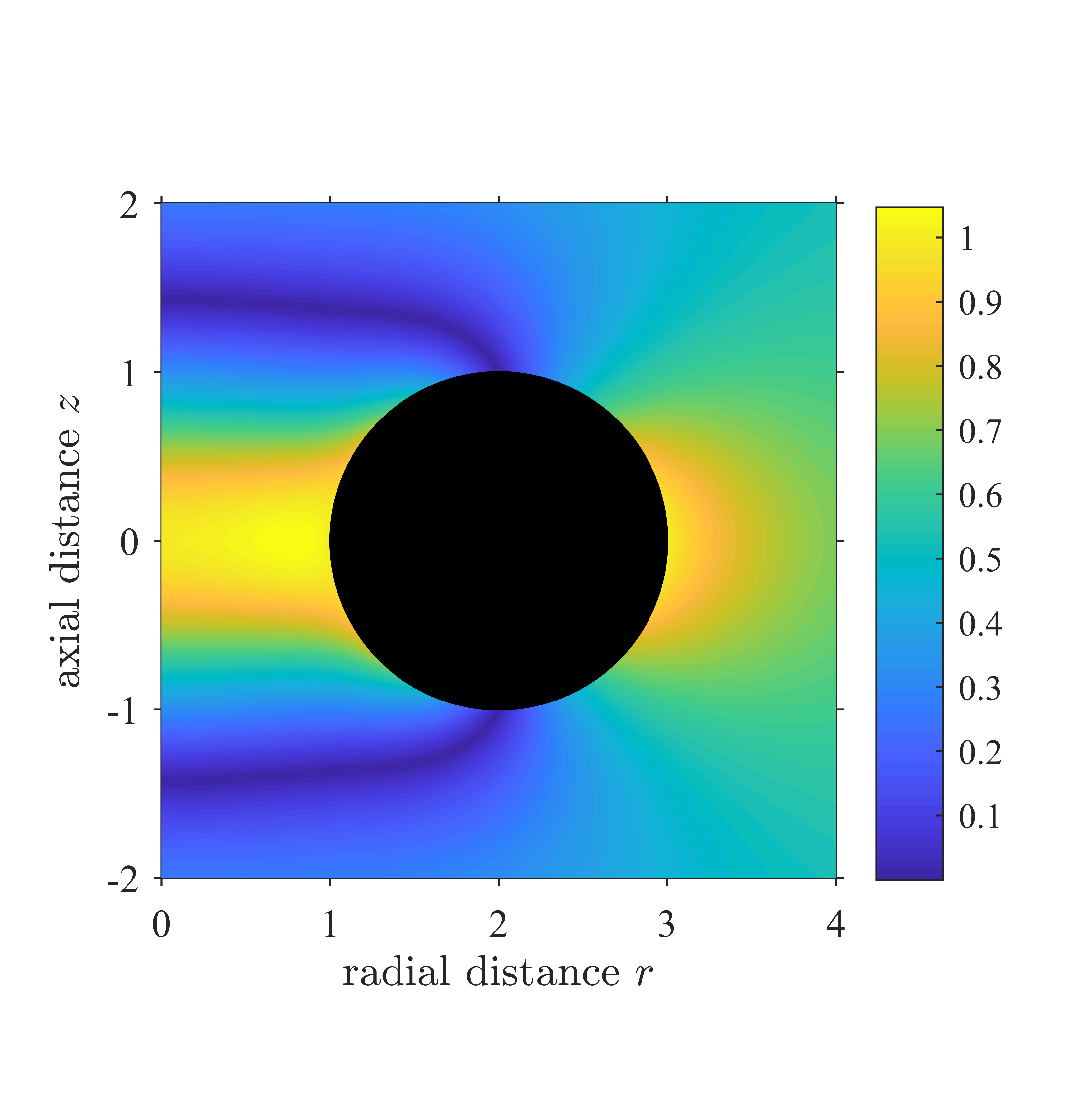} \label{fig:torusFlowVel}}
\caption{Streamlines and magnitude of fluid velocity in the region surrounding the force--free toroidal swimmer with slenderness parameter $s_0=2$, undergoing uniform counterclockwise surface rotation with unit angular velocity. The torus propagates in direction $-\hat{\mathbf{z}}$ and the frame of reference moves with the swimmer.} 
\end{figure}

Figures \ref{fig:torusStreamlines} \& \ref{fig:torusFlowVel} show the streamlines and magnitude of the fluid velocity in a region near the force--free toroidal swimmer with slenderness parameter $s_0=2$, undergoing uniform counterclockwise surface rotation with unit angular velocity. This results in propagation of the torus in the direction $-\hat{\mathbf{z}}$. Fluid passing through the central hole of the torus is caught in closed streamlines, in agreement with the results of Leshansky and Kenneth \cite{leshansky2008surface}.

%


\FloatBarrier
\section{Cytosolic flow in the pollen tube - an illustrative example of the benefits of regularization} \label{sec:pollenTube}
In Section \ref{sec:examples}, the method of regularized ringlets was applied to simple cases of motion to illustrate validity and applicability. These canonical examples demonstrate the use of ringlets in situations with moving boundaries. However, perhaps the most important feature of the regularized method is the ability to place source points directly in the fluid, in which case the regularized ring represents a curve in $3$D space. The regularization parameter $\varepsilon$ can then be used to control the spreading of the force (the size of the region over which it is applied). This is useful in situations in which the exact location at which a force is applied in a fluid is unknown or when a force is applied over a large area. In the following, these features of regularization are explored in more detail by considering the example of fluid flow in the angiosperm pollen tube. A more extensive study of the biomechanics of pollen tube growth (with a particular focus on the transport and distribution of elements of the cytoplasm) will follow in a future paper.

\subsection{A brief overview of the pollen tube} \label{sec:tubeOverview}
The pollen tube, a cellular protuberance originating from the pollen grain, is a vital component of the fertilisation process in plants. Responsible for the delivery of sperm cells from the pollen grain to the ovule, the tube may have to grow over $30\,$cm in length at speeds in excess of \SI{1}{cm.h^{-1}} \cite{bove2008magnitude,booy1992analysis}. This rapid elongation of the tube occurs via tip growth, with expansion localised close to the apical region of the cell \cite{fayant2010finite}. To prevent rupture, new cell membrane and cell wall material must be targeted to the sites of highest expansion. Spherical vesicles in the cytosolic region act as the delivery vectors for this material \cite{chebli2013transport} and are carried by motor myosin proteins along actin filaments in a process referred to as actomyosin transport. These actin filaments, arranged in bundles in the periphery and centre of the tube, are oriented in such a way that `full' peripheral vesicles travel towards the growing apex where they secrete their contents into the wall. In order to maintain the correct ratio of structural components, `empty' vesicles (comprised mostly of membrane) are also secreted by the wall back into the cytoplasm, where it is hypothesised that they are picked up by myosin on the central actin bundle and travel away from the apex. The combined movement of these vesicles induces a flow in the cytosol, known as `cytoplasmic streaming' or `cyclosis', which further aids in the cycling of vesicles towards and away from the apical region. To our knowledge, no comprehensive attempt has been made to produce a complete model of cytosolic flow in the pollen tube in a manner deriving directly from physical principles. In this section, we show how this is easily achievable using the method of regularized ringlets. 

\subsection{Mathematical model}
\subsubsection{Geometry and boundary conditions} \label{sec:tubeGeo}
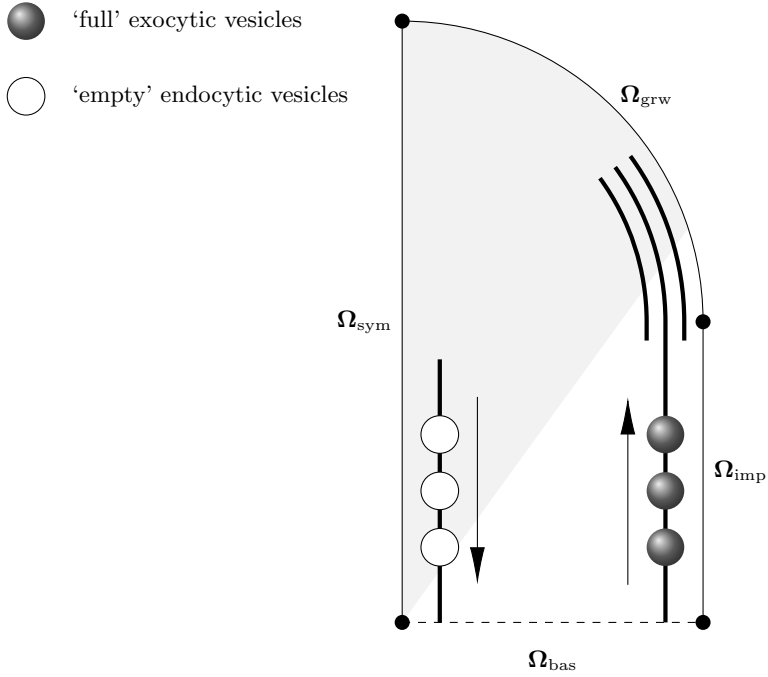
\begin{figure}[ht]
	\begin{tikzpicture}
		\fill[black!5] (3.8,5.24) arc (18:90:4) -- (0,0);
		
		\draw (4,0) -- (4,4) arc (0:90:4) -- (0,0);
		\draw[dashed] (0,0) -- (4,0);
		\draw[ultra thick] (3.25,3.75) -- (3.25,4) arc (0:36:3.25);
		\draw[ultra thick] (3.5,0) -- (3.5,4) arc (0:36:3.5);
		\draw[ultra thick] (3.75,3.75) -- (3.75,4) arc (0:36:3.75);
		
		\draw[ultra thick] (0.5,0) -- (0.5,3.5);
		
		\shade[ball color=black!50] (3.5,1) circle (.25);
		\shade[ball color=black!50] (3.5,1.75) circle (.25);
		\shade[ball color=black!50] (3.5,2.5) circle (.25);
		\draw[-{Latex[length=5mm, width=2mm]}] (3,.5) -- (3,3);
		
		\fill[white] (.5,1) circle (.25);
		\fill[white] (.5,1.75) circle (.25);
		\fill[white] (.5,2.5) circle (.25);
		\draw (.5,1) circle (.25);
		\draw (.5,1.75) circle (.25);
		\draw (.5,2.5) circle (.25);
		\draw[-{Latex[length=5mm, width=2mm]}] (1,3) -- (1,.5);
		
		\fill (0,8) circle (.1);
		\fill (0,0) circle (.1);
		\fill (4,0) circle (.1);
		\fill (4,4) circle (.1);
		
		\node at (-.5,4) {$\boldsymbol{{\Omega}}_{\text{sym}}$};
		\node at (3.25,7) {$\boldsymbol{{\Omega}}_{\text{grw}}$};
		\node at (2,-.5) {$\boldsymbol{{\Omega}}_{\text{bas}}$};
		\node at (4.5,2) {$\boldsymbol{{\Omega}}_{\text{imp}}$};	
		
		\fill[white] (-5,7) circle (.25);
		\draw (-5,7) circle (.25);		
		
		\shade[ball color=black!50] (-5,8) circle (.25);
		
		\node[right] at (-4.5,7) {`empty' endocytic vesicles};
		\node[right] at (-4.5,8) {`full' exocytic vesicles};

		\node at (9.25,8.25) {};
		
	\end{tikzpicture}
		\caption{Suggested mechanism for transport of vesicles in the pollen tube, showing actin bundles (thick interior lines) and dense, apical actin fringe running parallel to the peripheral wall, as well as pooling of apical vesicles and shape of the `inverted vesicle cone' (shaded area). Direction of vesicle movement along actin bundles is given by arrows. The location of each of the four boundaries of the domain are also shown. Image not drawn to scale.}
		\label{fig:dom_geo}
	\end{figure}

The typical geometry of the tube (in $2$D cylindrical coordinates) along with the two vesicle populations and actin bundles can be seen in Figure \ref{fig:dom_geo}, in which the static cylindrical shank is joined to a growing apical hemispherical cap. The boundary of the domain is split into four sections: the symmetry boundary $\boldsymbol{{\Omega}}_{\textrm{sym}}$ running down the centreline of the tube, the static, impermeable peripheral wall $\boldsymbol{{\Omega}}_{\textrm{imp}}$, the growing apical hemisphere $\boldsymbol{{\Omega}}_{\textrm{grw}}$ and the artificial basal boundary $\boldsymbol{{\Omega}}_{\textrm{bas}}$  (where the computational domain is truncated).

The Reynolds number for fluid flow in the pollen tube can be evaluated using the typical flow speed ($\approx \SI{1}{\micro m.s^{-1}}$), tube radius ($\approx \SI{8.13}{\micro m}$) and the kinematic viscosity of water ($\approx \SI{e6}{\micro m^2.s^{-1}}$) to find $\text{Re} \approx \SI{e-5}{}$, firmly in the regime of Stokes flow. Although pollen tube growth is typically oscillatory, acceleration and deceleration are small compared to growth speed itself. This enables the use of the steady Stokes equations.

Since the cytosolic flow is induced by the actomyosin transport of vesicles along cytoskeletal actin bundles, forces must be applied inside the computational domain. Singular stokeslet methods typically do not allow for this unless the surfaces of individual vesicles are discretized, but their small size ($\approx \SI{100}{nm}$ radius) compared to the typical length scale of the problem and their large number density make this impractical. One solution is to use the method of regularized ringlets instead, placing rings in series along the centreline of the peripheral actin bundle with the parameter $\varepsilon$ being used to control the bundle thickness. 

\begin{figure}[ht]
	\centering
	\includegraphics{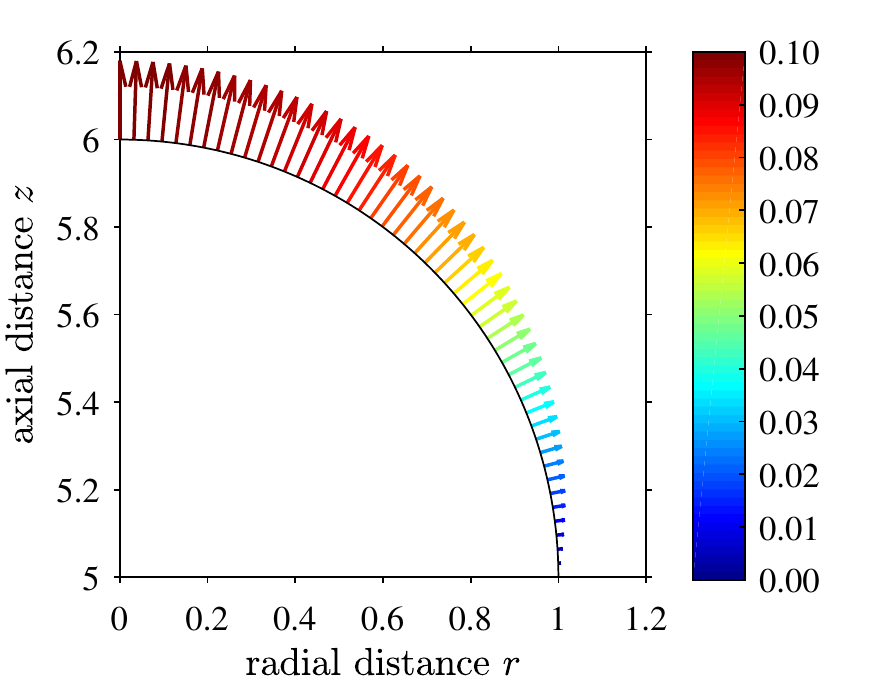}
	\caption{Growth velocity of the apical boundary, normal to the cell surface and varying from a maximum at the extreme apex to zero at the point where the apical hemisphere joins the adjacent shank.  Velocities are scaled by the speed of vesicles on actin (\SI{1}{\micro m.s^{-1}}), with lengths scaled by the radius of the tube (\SI{8.13}{\micro m}).}
	\label{fig:apexFlow3}
\end{figure}

The boundary conditions for the fluid velocity $\mathbf{u}$ are given by,
\begin{equation}
\begin{array}{lll}
\mathbf{u}=\mathbf{u}_g \text{ on } \boldsymbol{{\Omega}}_{\text{grw}}, &
\mathbf{u}=\mathbf{0} \text{ on } \boldsymbol{{\Omega}}_{\text{imp}}, &
u_r=\partial u_z / \partial r =0 \text{ on } \boldsymbol{{\Omega}}_{\text{sym}}, 
\end{array}
\end{equation}
in accordance with the assumptions of tip growth and axisymmetry. No restriction is placed on $\mathbf{u}$ on the artificial boundary $\boldsymbol{{\Omega}}_{\textrm{bas}}$. For the growth velocity $\mathbf{u}_g$ in the hemispherical apex, the normal displacement growth assumption of Dumais \cite{dumais2006anisotropic} is employed to define \begin{equation}\mathbf{u}_g = \mathbf{u}_g(\varphi) \coloneqq v_g\sin\varphi(\cos\varphi,\sin\varphi),\end{equation} where $\varphi$ is the angle between the outward-pointing surface normal and the positive $r$ axis (varying from $0$ at the point where the hemisphere joins the shank to $\frac{\pi}{2}$ at the extreme apex) and $v_g$ is the growth speed of the tube. An example of how the growth of the boundary varies over the apical hemisphere can be seen in Figure \ref{fig:apexFlow3}, where the maximum growth speed $v_g$ is equal to $0.1 \, \mu$m s$^{-1}$ and the wall velocity in the adjacent shank (not pictured) is equal to zero. 

\subsubsection{Actin bundle model} \label{sec:epsSelection}
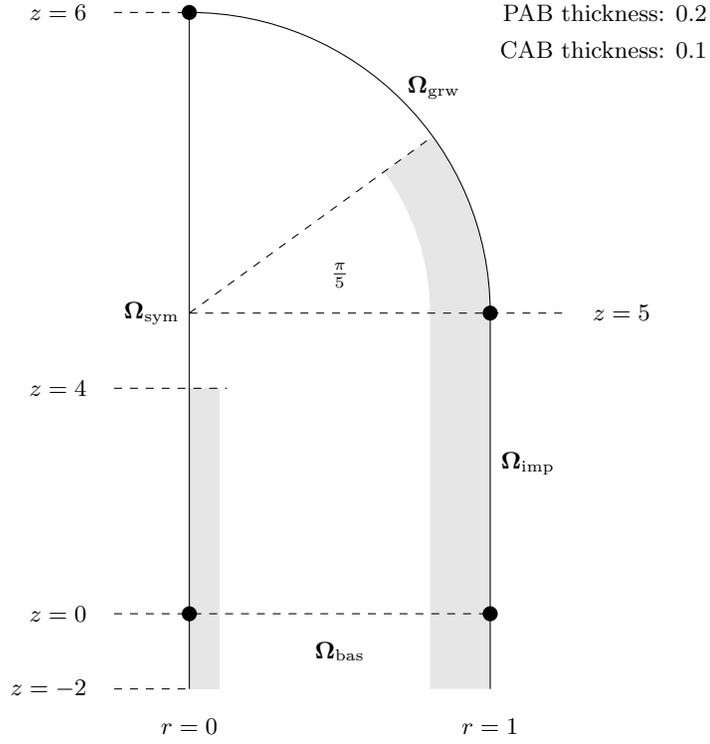
\begin{figure}[ht]
	\begin{tikzpicture}
	\fill[black!10] (3.2,-1) -- (4,-1) -- (4,4) arc (0:36:4) -- (2.589,5.881) arc (36:0:3.2);
	\fill[black!10] (0,-1) -- (0,3) -- (.4,3) -- (.4,-1);
	
	\draw (4,0) -- (4,4) arc (0:90:4) -- (0,0);
	\draw[dashed] (-1,0) -- (4,0);
	
	\draw[dashed] (0,4) -- (5,4);
	\draw[dashed] (0,4) -- (3.236,6.351);
	
	\draw[dashed] (-1,3) -- (.5,3);
	\draw[dashed] (-1,-1) -- (0,-1);
	\draw[dashed] (-1,8) -- (0,8);
%


	\fill (0,8) circle (.1);
	\fill (0,0) circle (.1);
	\fill (4,0) circle (.1);
	\fill (4,4) circle (.1);
	
	\node at (-.5,4) {$\boldsymbol{{\Omega}}_{\text{sym}}$};
	\node at (3.25,7) {$\boldsymbol{{\Omega}}_{\text{grw}}$};
	\node at (2,-.5) {$\boldsymbol{{\Omega}}_{\text{bas}}$};
	\node at (4.5,2) {$\boldsymbol{{\Omega}}_{\text{imp}}$};	
	
	\draw (0,0) -- (0,-1);
	\draw (4,0) -- (4,-1);
	
	\node[left] at (-1.25,-1) {$z=-2$};
	\node[left] at (-1.25,0) {$z=0$};
	\node[left] at (-1.25,3) {$z=4$};
	\node[right] at (5.25,4) {$z=5$};
	\node[left] at (-1.25,8) {$z=6$};
	

\node at (2,4.5) {$\frac{\pi}{5}$};

\node at (0,-1.5) {$r=0$};
\node at (4,-1.5) {$r=1$};
	
\node[left] at (7,8) {PAB thickness: 0.2};
\node[left] at (7,7.5) {CAB thickness: 0.1};

\node at (-3,8) {};
	
	\end{tikzpicture}
	\caption{Geometric elements of the pollen tube model. PAB and CAB refer to peripheral actin bundle and central actin bundle respectively, with CAB thickness half that of PAB by axisymmetry. Image not drawn to scale in $\hat{z}$.}
	\label{fig:tubeGeo}
\end{figure}

A section of the pollen tube is modelled ranging from the extreme apex to a point equal to six tube radii distal where the domain is artificially truncated. In dimensionless values, the extreme apex is thus given by $(r,z) = (0,6)$ with $z=0$ being the distal truncation line. The central line of axisymmetry is given by $r=0$, with $r=1$ denoting the peripheral boundary in the shank. The hemispherical apex is the upper-right quarter circle of radius $1$, centred at $(0,5)$. The peripheral actin bundle (PAB) is considered adjacent to the pollen tube wall, with its width equal to one fifth of the pollen tube radius ($0.2$) and extending at an angle $\varphi = \pi/5$ into the apical hemisphere (see Figure \ref{fig:tubeGeo}) in accordance with the confocal microscopy imaging of Lovy-Wheeler \emph{et al.} \cite{lovy2005enhanced}. The central actin bundle (CAB) is located further away from the apex than the PAB, and is modelled with a reduced thickness (half that of the PAB) in cylindrical coordinates to account for the central axisymmetry of the tube.

\begin{figure}[ht]
	\centering
	\includegraphics{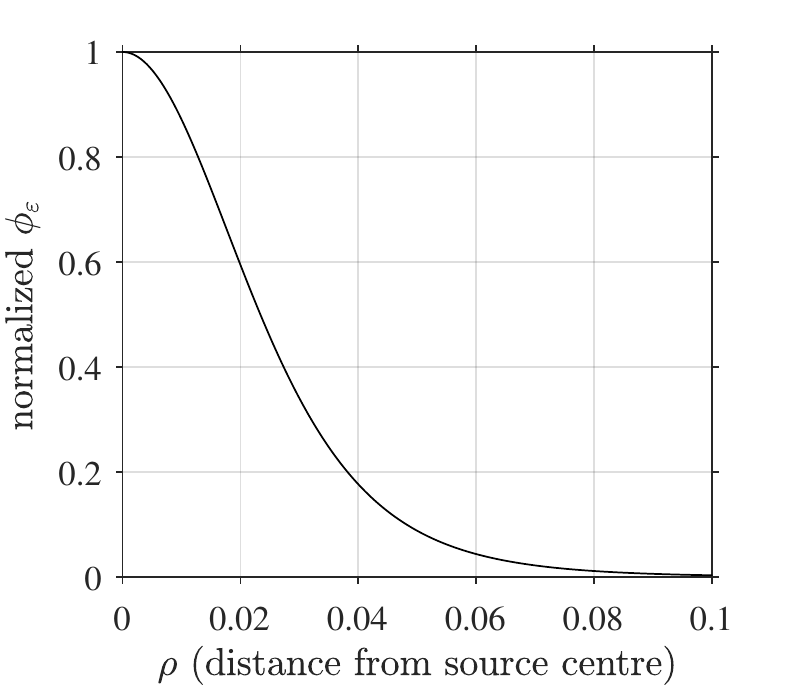}
	\caption{Controlled spreading of force distribution using $\phi_\varepsilon$ with $\varepsilon=0.05$}.
	\label{fig:epsSelect}
\end{figure}

By placing rings of regularized Stokeslets in series along the centre line of the PAB, it is possible to carefully select an appropriate value for the regularization parameter $\varepsilon$ such that the region over which the majority of the force distribution is applied is roughly the same as the bundle thickness. The same procedure is employed for the CAB, only with individual regularized Stokeslets rather than ringlets since the centreline coincides with $r=0$. Using $\varepsilon = 0.05$ it is found that $\approx93\%$ of the total applied force is contained within a region of radius $0.1$ which corresponds well to the approximate thickness of the actin bundles. This is shown in Figure \ref{fig:epsSelect} where the maximal value of $\phi_\varepsilon$ has been scaled to $1$ for the sake of clarity. Smaller values of $\varepsilon$ result in an even larger percentage of the total force being contained within $0 \le \rho \le 0.1$ but are increasingly skewed towards $\rho=0$. 

We note that our modeling ignores the presence of a third f-actin structure, the short actin bundles observed in the extreme apex \cite{fu2001rop}; these bundles are transient and significantly less dense than the thick peripheral and central actin bundles. Further, due to their proximity to the growing tube wall, any effect that these short actin bundles may have on the fluid velocity is likely insignificant compared to the effect of the growth of the wall.

\subsubsection{Frames of reference}
In each of the velocity profiles to be produced, cytosolic velocities are prescribed along the actin bundles in accordance with values taken from the STICS analysis of Bove \emph{et al.} \cite{bove2008magnitude}. This is done under the assumption that the large-scale directed movement of vesicles must be a consequence of cytosolic flow (induced by the smaller-scale movement of individual vesicles along actin). The fluid velocity is defined in terms of two different coordinate systems: the static lab frame where $\mathbf{u}_L=(u_r,u_z)$ and the moving tip frame with $\mathbf{u}_T=(u_r,u_z-v_g)$. The lab frame represents the fluid velocity relative to a stationary observation point (in which $\mathbf{u}_L\cdot\hat{\mathbf{n}} \neq 0$ on $\boldsymbol{{\Omega}}_{\text{grw}}$ due to boundary growth), whereas the tip frame represents the fluid velocity relative to the growing tip (resulting in a static domain with $\mathbf{u}_T\cdot\hat{\mathbf{n}}$ at all boundary points except on $\boldsymbol{{\Omega}}_{\text{bas}}$). The steady geometry of the domain for the advancing tip frame means that $\mathbf{u}_T$ satisfies the steady Stokes equations, from which it follows that $\mathbf{u}_L$ is also a solution (since the two differ by the uniform flow field $(0,v_g)$ only).

\subsubsection{Two model scenarios for comparison}
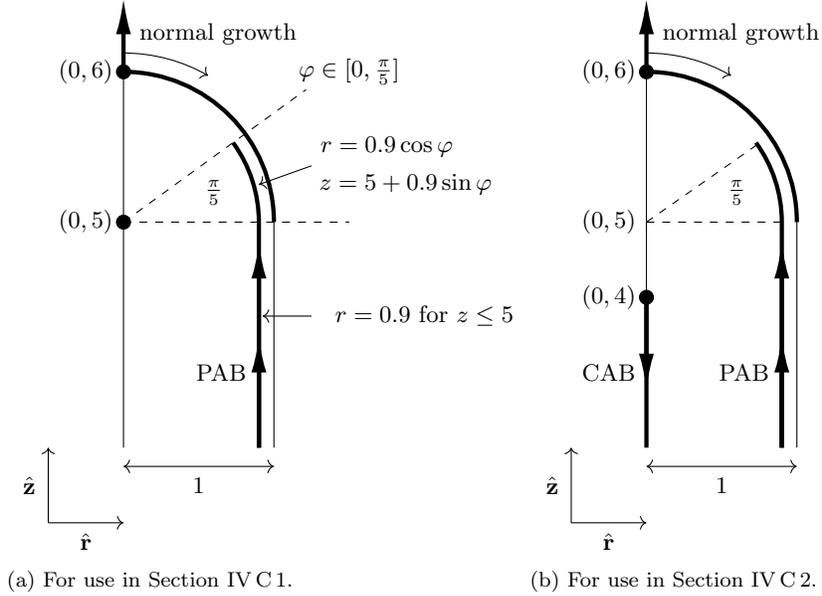
\begin{figure}[ht]
	\centering
\subfloat[\raggedright{For use in Section \ref{sec:centralBund}.}]{
	\begin{tikzpicture}
	\draw (2,0) -- (2,3);
	\draw[ultra thick] (2,3) arc (0:90:2);
	\draw (0,5) -- (0,0);
	\draw[ultra thick] (1.8,0) -- (1.8,3) arc (0:36:1.8);
	\draw[ultra thick, -{Latex[length=5mm, width=2mm]}] (1.8,0) -- (1.8,1.5);
	\draw[ultra thick, -{Latex[length=5mm, width=2mm]}] (1.8,0) -- (1.8,2.75);
	\draw[ultra thick, -{Latex[length=5mm, width=2mm]}] (0,5) -- (0,6);
	\draw[->] (0,5.25) arc (90:60:2.25);
	\node[right] at (0.1,5.5) {normal growth};
	\draw [dashed] (0,3) -- (3,3);
	\draw[dashed] (0,3) -- (2.427,4.763);
	\node at (1.2,3.4) {$\frac{\pi}{5}$};
	\fill (0,5) circle (.1);
	\node at (-.5,5) {$(0,6)$};
	\draw [<->] (0,-.25) -- (2,-.25);
	\node at (1,-.5) {$1$};
	\draw [->] (2.5,1.75) -- (1.85,1.75);
	\node at (4,1.75) {$r=0.9$ for $z\le 5$};
	\draw [->] (2.5,3.75) -- (1.8,3.5);
	\node at (3.5,4) {$r = 0.9 \cos\varphi$};
	\node at (3.75,3.5) {$z = 5 + 0.9\sin\varphi$};
	\node at (3,5) {$\varphi \in [0,\frac{\pi}{5}]$};
	\fill (0,3) circle (.1);
	\node at (-.5,3) {$(0,5)$};
	\draw [->] (-1,-1) -- (0,-1);
	\draw [->] (-1,-1) -- (-1,0);
	\node at (-.5,-1.25) {$\hat{\mathbf{r}}$};
	\node at (-1.25,-.5) {$\hat{\mathbf{z}}$};
	\node at (1.3,1) {PAB};
	\end{tikzpicture}
	\label{fig:sch1}
}
\subfloat[\raggedright{For use in Section \ref{sec:growthSpeed}.}]{
 \begin{tikzpicture}
	\draw (2,0) -- (2,3) arc (0:90:2) -- (0,0);
	\draw[ultra thick] (1.8,0) -- (1.8,3) arc (0:36:1.8);
	\draw [dashed] (0,3) -- (1.8,3);
	\draw[dashed] (0,3) -- (1.5,4.05);
	\node at (1.2,3.4) {$\frac{\pi}{5}$};
	\fill (0,5) circle (.1);
	\node at (-.5,5) {$(0,6)$};
	\draw [<->] (0,-.25) -- (2,-.25);
	\node at (1,-.5) {$1$};
	\draw[ultra thick, -{Latex[length=5mm, width=2mm]}] (1.8,0) -- (1.8,1.5);
	\draw[ultra thick, -{Latex[length=5mm, width=2mm]}] (1.8,0) -- (1.8,2.75);
	\draw[ultra thick, -{Latex[length=5mm, width=2mm]}] (0,2) -- (0,.75);
	\draw[ultra thick, -{Latex[length=5mm, width=2mm]}] (0,5) -- (0,6);
	\draw[->] (0,5.25) arc (90:60:2.25);
	\node[right] at (0.1,5.5) {normal growth};
	\draw [->] (-1,-1) -- (0,-1);
	\draw [->] (-1,-1) -- (-1,0);
	\node at (-.5,-1.25) {$\hat{\mathbf{r}}$};
	\node at (-1.25,-.5) {$\hat{\mathbf{z}}$};
	\fill (0,2) circle (.1);
	\node at (-.5,2) {$(0,4)$};
	\node at (-.5,3) {$(0,5)$};
	\draw[ultra thick] (0,2) -- (0,0);
	\draw[ultra thick] (2,3) arc (0:90:2);
	\node at (-.5,1) {CAB};
	\node at (1.3,1) {PAB};
	\end{tikzpicture}
	\label{fig:sch2}
	}
\caption{Dimensionless schematic diagrams for situations considered in \textbf{(a)} Section \ref{sec:centralBund}, and \textbf{(b)} Section \ref{sec:growthSpeed}. Thick lines denote locations on which velocity is prescribed in each case. PAB and CAB refer to peripheral actin bundle and central actin bundle respectively.}
\end{figure}

In Section \ref{sec:centralBund}, we investigate the role of the CAB in vesicle transport. Fluid velocity is prescribed on the PAB only (see Figure \ref{fig:sch1}) in order to compare the resultant central fluid velocity to experimental results. The centreline of the PAB is modelled using the union of the straight line extending from $(r,z) = (0.9,-2)$ to $(0.9,5)$ and the curve $(r,z) = (0.9\cos\varphi,5+0.9\sin\varphi)$ for $\varphi \in [0,\pi/5]$. The regularized ringlet placement is extended to $z=-2$ to ensure the velocity profile at $z=0$ is consistent with the rest of the tube. The fluid velocity on the straight line is given by $(u_r,u_z) = (0,0.5)$ with the fluid velocity on the curve being tangential and of constant magnitude, that is, $(u_r,u_z) = 0.5(\sin\varphi,\cos\varphi)$ for $\varphi\in[0,\pi/5]$. On the peripheral wall (again extended to $z=-2$), the velocity is $\mathbf{0}$ in the shank and prescribed according to the growth speed $v_g$ of the tube in the apical hemisphere using the velocity function $\mathbf{u}_g(\varphi) =v_g\sin\varphi(\cos\varphi,\sin\varphi)$ for $\varphi \in [0,\pi/2]$. 

In Section \ref{sec:growthSpeed}, the velocity of the growing boundary is varied in order to determine the effect of the growth speed of the tube on the cytosolic velocity. Fluid velocity is prescribed on each of the PAB, the CAB, and the growing boundary accordingly (see Figure \ref{fig:sch2}). Fluid velocity along the PAB is kept constant at \SI{0.5}{\micro m.s^{-1}} based on the observation that cytoplasmic streaming rates are typically independent of pollen tube growth rates \cite{vidali2001actin}. Fluid velocity on the centreline of the central actin bundle, given by $r=0$ for $-2 \le z \le 4$, is also prescribed. It is assumed that vesicles are sufficiently closely packed on this bundle that the fluid velocity at its centre can be approximated by the speed of vesicles on actin (\SI{1}{\micro m.s^{-1}}), giving $(u_r,u_z) = (0,-1)$. Since these node locations are at $r=0$, standard regularized Stokeslets must be used here.

A regularisation parameter of $\varepsilon=0.05$ is chosen in all cases and the ringlets are linearly spaced a distance approximately 0.025 apart, resulting in a smooth velocity profile. The flow velocity for $z < 3$ (not pictured) always matches the flow velocity at $z=3$ almost exactly, with no further change occurring in the $\hat{\mathbf{z}}$ direction.

\FloatBarrier
\subsection{Results} 
\subsubsection{Role of the central actin bundle}\label{sec:centralBund}
The CAB has long been hypothesized to aid in the removal of vesicles from the apical region, but direct observation of vesicle transport along actin is often hindered by the small size of vesicles (typically below the resolution limit of conventional confocal microscopes). Higher resolution imaging methods such as evanescent wave microscopy have been used to observe long-range vesicle movement (presumably a result of actomyosin transport) in the periphery of the tube \cite{wang2006imaging}, but the limited penetration depths available in these methods ($\le \SI{400}{nm}$) do not allow for imaging of the central region. STICS analysis suggests a way of determining whether the central bundle participates in vesicle transport. Directed vesicle movement in the periphery of the tube is not seen to exceed speeds of $\approx\SI{0.5}{\micro m.s^{-1}}$, smaller than the $\approx \SI{0.8}{\micro m.s^{-1}}$ observed in the centre \cite{bove2008magnitude}. By using the method of regularized ringlets and prescribing fluid velocity only along, 
\begin{itemize} 
\item the PAB, where $|\mathbf{u}_T| = \SI{0.5}{\micro m.s^{-1}}$,
\item the tube boundary with apical growth speed $v_g = \SI{0.1}{\micro m.s^{-1}}$,
\end{itemize}
the resulting fluid velocity in the centre should provide further insight insight into whether the central bundle participates in vesicle transport. 

\begin{figure}[ht]
	\centering
\subfloat[Lab frame.]{\includegraphics[trim = 0 0 1cm 0,clip]{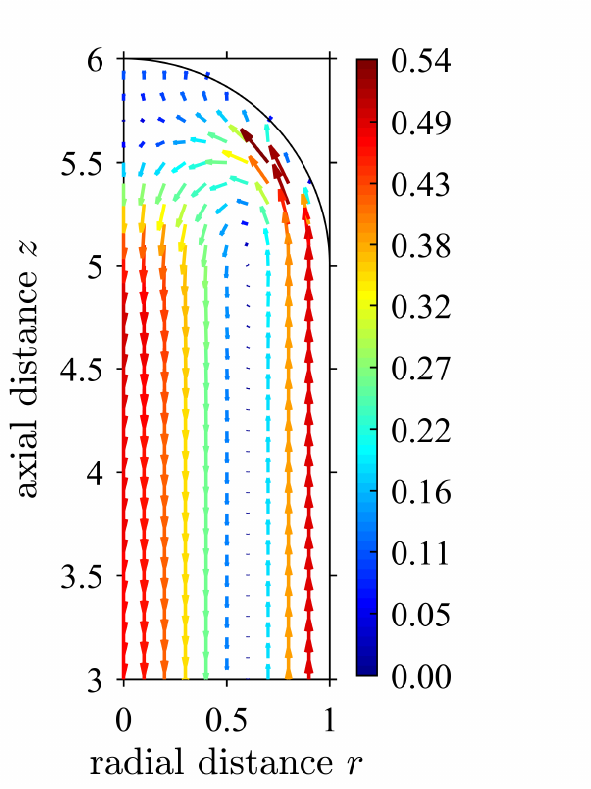} \label{fig:pFL}}\qquad
\subfloat[Tip frame.]{\includegraphics[trim = 0 0 1cm 0,clip]{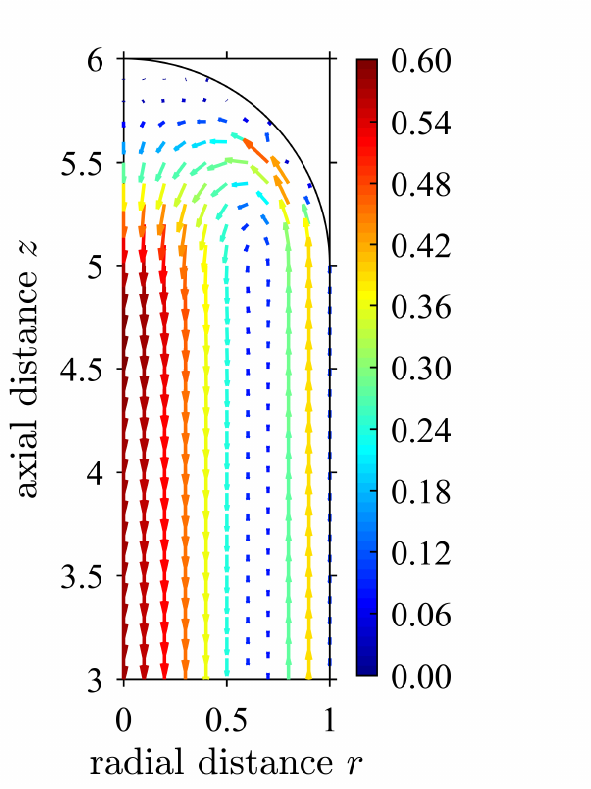} \label{fig:pFT}}
\caption{Dimensionless magnitude and direction of apical cytosolic flow in (a) \textbf{lab frame}, and (b) \textbf{tip frame}, calculated using prescribed velocity of magnitude \SI{0.5}{\micro m.s^{-1}} on the peripheral actin bundle, as well as a prescribed normal velocity with maximum magnitude equal to the growth speed \SI{0.1}{\micro m.s^{-1}} at the apical boundary. Velocities are scaled by the speed of vesicles on actin (\SI{1}{\micro m.s^{-1}}), with lengths scaled by the radius of the tube (\SI{8.13}{\micro m}).}
\label{fig:centralBundleTest}
\end{figure}

The results of this investigation can be seen in Figure \ref{fig:centralBundleTest}, with both lab (Figure \ref{fig:pFL}) and tip (Figure \ref{fig:pFT}) frames shown. The shape of the velocity profile in Figure \ref{fig:pFL} is in excellent agreement with the STICS analysis of Bove \emph{et al.} \cite{bove2008magnitude}, with a wider band of basal flow through the center than apical flow in the periphery. However, despite an arguably exaggerated prescribed peripheral fluid velocity of \SI{0.5}{\micro m.s^{-1}}, fluid velocity in the centre does not achieve speeds of \SI{0.8}{\micro m.s^{-1}}. This is a strong indication that the CAB must also participate in the transport of vesicles, particularly considering that that our current implementation of the Stokes equations does not account for variations in local fluid viscosity (known to be larger in the presence of filamentous actin networks \cite{wong2004anomalous}, reducing the fluid velocity induced by any given force). The increased flow speed through the centre of the tube can be easily accounted for by inclusion of additional drag from actomyosin vesicle transport (known to reach speeds of up to \SI{2}{\micro m.s^{-1}} \cite{vidali2001actin,de1997quantitative}) along the CAB, with the largest velocities being observed at the very centre as a result of the reduced cytosolic volume in this region. 

\FloatBarrier

\subsubsection{The influence of growth speed} \label{sec:growthSpeed}
We now turn our attention to the effect of tube growth speed on the cytosolic velocity profile. A further three new velocity profiles are produced, based on three different growth speeds for the tube ($0,0.1, \text{ and } 0.2 \, \SI{}{\micro m.s^{-1}}$). These speeds (approximately) correspond to that of a static tube, the typical growth rate cited in Bove \emph{et al.} \cite{bove2008magnitude}, and the average growth rate measured by Vidali \emph{et al.} \cite{vidali2001actin} for the \emph{Lilium longiflorum} species. Fluid velocity is prescribed along 
\begin{itemize} \item the PAB ($|\mathbf{u}_T|=\SI{.5}{\micro m.s^{-1}}$), 
\item the CAB ($u_z=-\SI{1}{\micro m.s^{-1}}$), 
\item and the tube boundary ($v_g = 0,0.1,\SI{0.2}{\micro m.s^{-1}}$),
\end{itemize}
in each case.

\begin{figure}[ht]
	\centering
	\subfloat[\SI{0}{\micro m.s^{-1}}.]{\includegraphics[trim = 0 0 1cm 0,clip]{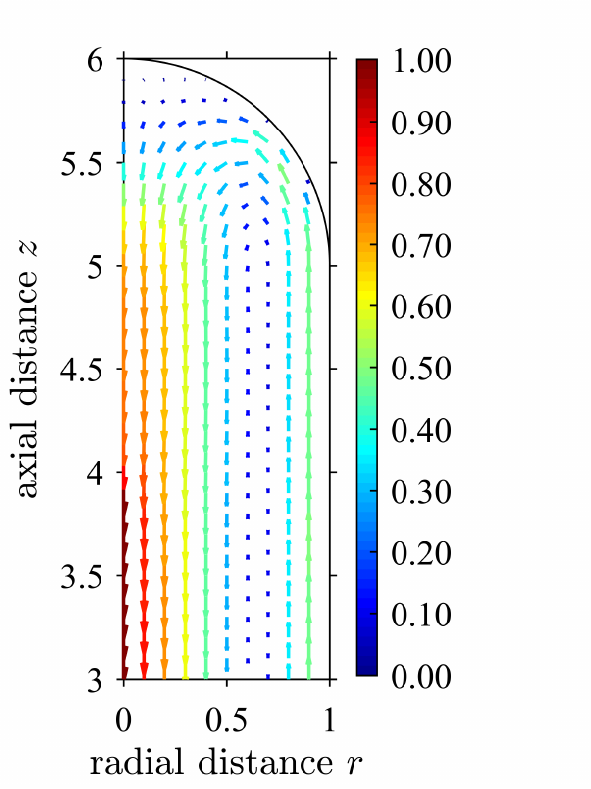} \label{fig:staticGrowth}}\quad
	\subfloat[\SI{0.1}{\micro m.s^{-1}}.]{\includegraphics[trim = 0 0 1cm 0,clip]{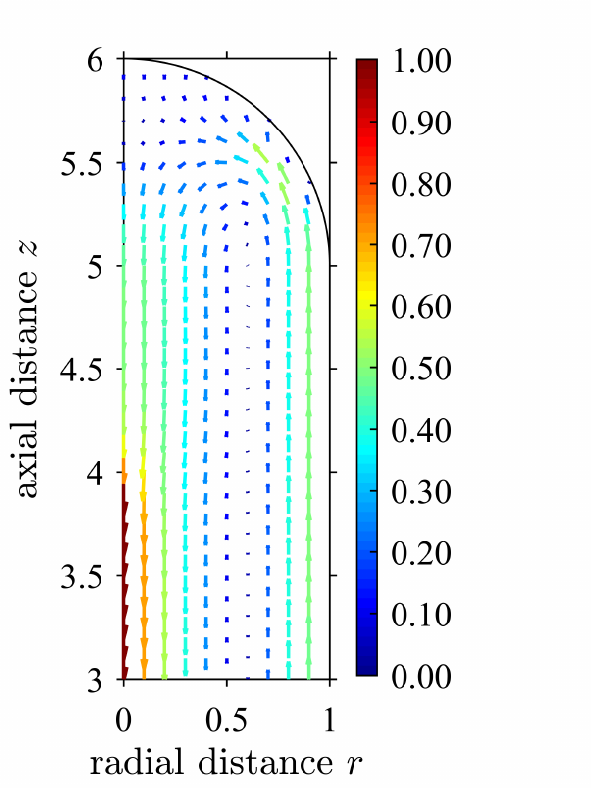} \label{fig:apexFlow4}}\quad
	\subfloat[\SI{0.2}{\micro m.s^{-1}}.]{\includegraphics[trim = 0 0 1cm 0,clip]{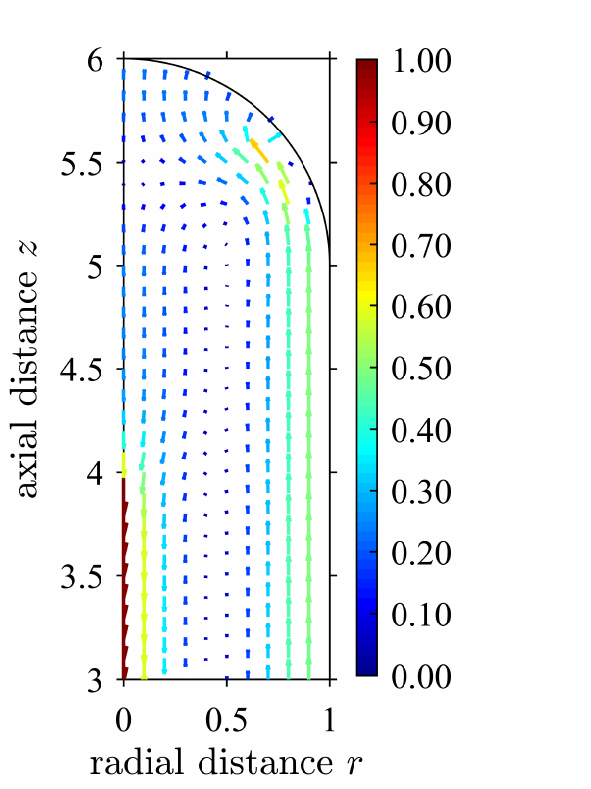} \label{fig:apexFlow5}} \\ \vspace{.5cm}
	\textbf{Lab frame} \\ \vspace{.5cm} \rule{0.5\textwidth}{.4pt}
	
	\subfloat[\SI{0}{\micro m.s^{-1}}.]{\includegraphics[trim = 0 0 1cm 0,clip]{staticGrowth} \label{fig:staticGrowth2}}\quad
	\subfloat[\SI{0.1}{\micro m.s^{-1}}.]{\includegraphics[trim = 0 0 1cm 0,clip]{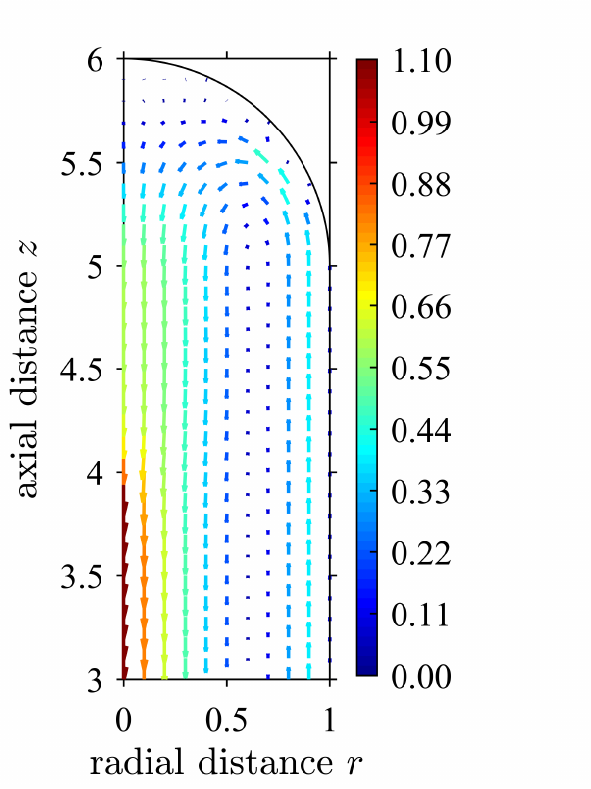} \label{fig:apexFlow6}}\quad
	\subfloat[\SI{0.2}{\micro m.s^{-1}}.]{\includegraphics[trim = 0 0 1cm 0,clip]{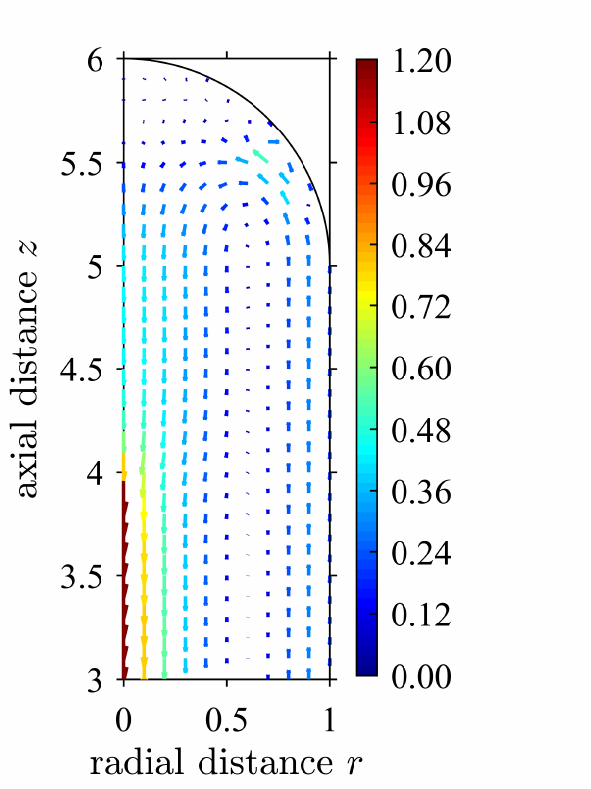} \label{fig:apexFlow7}} \\ \vspace{.5cm}
	\textbf{Tip frame}
	\caption{Dimensionless magnitude and direction of apical cytosolic flow in \textbf{lab} (top row) and \textbf{tip} (bottom row) frames, calculated using prescribed velocities of magnitude \SI{0.5}{\micro m.s^{-1}} and \SI{1}{\micro m.s^{-1}} on the peripheral and central actin bundles respectively, as well as a prescribed normal velocity with maximum magnitude equal to the growth speed (a,d) \SI{0}{\micro m.s^{-1}}, (b,e) \SI{0.1}{\micro m.s^{-1}}, (c,f) \SI{0.2}{\micro m.s^{-1}}, at the apical boundary. Velocities are scaled by the speed of vesicles on actin (\SI{1}{\micro m.s^{-1}}), with lengths scaled by the radius of the tube (\SI{8.13}{\micro m}).}
	\label{fig:prescribedVelocityTipFrame}
\end{figure}

Figures \ref{fig:staticGrowth} -- \ref{fig:apexFlow5} show the velocity profiles for each of these three tube growth speeds in the \textbf{lab frame}. Significant differences can be seen between the three profiles, both in the magnitude and direction of the fluid velocity. In particular, at larger growth speeds there is a wider band of cytosolic flow in the positive $\hat{\mathbf{z}}$ direction in the peripheral region, and the central band of basal flow is both narrower and of a reduced magnitude. This is an expected consequence of mass conservation, since in a tube with a faster growth speed more fluid must flow towards the apical region to fill the increasing space.

Figures \ref{fig:staticGrowth2} -- \ref{fig:apexFlow7} show the velocity profiles for each of these three tube growth speeds in the \textbf{tip frame}. Here, the growth speed of the tube has been subtracted from the $\hat{\mathbf{z}}$ component of the fluid velocity for each corresponding velocity profile in the lab frame. Interestingly, the differences between the velocity profiles are significantly less pronounced in this frame of reference. Differences in the magnitude of the fluid velocity still persist, but the overall shape of the profiles bear a striking similarity. The persistent shape of the velocity profiles seen in Figures \ref{fig:staticGrowth2} -- \ref{fig:apexFlow7} could help explain the observed similarities in the distribution of the apical vesicle population and other elements of the cytoplasm across multiple pollen tube species and throughout different phases of oscillatory growth. 

\FloatBarrier
\section{Conclusions and future work} \label{sec:conclusion}
In this paper, the regularized fundamental solution for the velocity (single layer potential) and stress (double layer potential) due to an axisymmetric ring of smoothed point forces (the `regularized ringlet') was derived, expanding on the work of Cortez and colleagues \cite{cortez2001method,cortez2005method}. The velocity solution, written in the form of complete elliptic integrals of the first and second kind, tends to the singular solution of Pozrikidis \cite{pozrikidis1992boundary} in the limit as the regularization parameter $\varepsilon$ tends to $0$. 

For the resistance problem on the translating and rotating unit sphere, the method of regularized ringlets was shown to produce small relative errors in drag calculations and small condition numbers for the underlying resistance matrices. Further testing in the case of the translating unit sphere shows that the regularized ringlet performs favourably compared to the traditional method of regularized Stokeslets both in terms of accuracy (Appendix \ref{sec:translatingSphere}) and speed (Appendix \ref{sec:accuracyApplications}) and that it possesses more satisfactory convergence properties than the singular ringlet (Appendix \ref{sec:MFS}). 

The applicability of the regularized ringlet to fluid flow problems involving body surface motion in the $\hat{\mathbf{r}}$ direction (perpendicular to the line of axisymmetry) was established in the example of Purcell's toroidal swimmer. Using the regularized ringlet method, we were able to reproduce Leshansky and Kenneth's \cite{leshansky2008surface} results for the scaled propulsion velocity of the toroidal swimmer, propelled by surface tank treading against the direction of motion of its outer surface. Our results show a significant improvement over the asymptotic solution found by integrating a centre-line distribution of rotlets in the limit as slenderness decreases ($s_0 \rightarrow 0$). The use of regularized ringlets also yields the drag force at all points on the torus surface for the toroidal swimmer, information that is not readily available using the series or rotlet solutions.

In order to elucidate the further benefits of regularization, we studied the case of cytosolic flow in the growing pollen tube. Here the inducing drag force is the result of the directed actomyosin transport of vesicles inside the fluid, making application of singular Stokeslet methods particularly challenging. Since the area over which the force is applied is also relatively large (and somewhat indeterminate), being able to use the regularization parameter $\varepsilon$ to control the spreading of the force is also vital to the solution. 

Using the regularized ringlet solution for the steady Stokes equations, we were able to show that it is highly likely that the central actin bundle plays a role in the removal of vesicles from the apical pool. This has long been hypothesised, with vesicle movement through the central region seen to reach speeds of up to \SI{2}{\micro m.s^{-1}} \cite{bove2008magnitude} (similar to those of actomyosin transport), but direct evidence has been lacking. Fluid velocity profiles based on drag induced by actomyosin transport of vesicles along the peripheral bundle alone shows that cytosolic flow in the central region does not reach these speeds, strongly suggesting that actomyosin transport takes place on the central actin bundle. 

During further investigation, we were able to show that the shape of the cytosolic velocity profile relative to the moving tip is largely independent of the growth speed of the tube (under the assumption that the actin profile is able to keep up with the advancing tip). This provides some insight into how the tube is able to consistently orient its internal architecture across multiple pollen tube species and in spite of changes in growth speed during different phases of oscillatory growth. In a future paper, we will continue our work on the pollen tube by using the cytosolic velocity profiles we have produced in tandem with an advection-diffusion-reaction equation for the spatio-temporal distribution of vesicles in the pollen tube. Investigating the conditions necessary to produce the `inverted vesicle cone' almost universally observed in the apex of the angiosperm pollen tube yields new results concerning the appropriate values for different parameters pertaining to tube growth. We believe that the regularized Stokeslet ring method will be of benefit to further studies in the fluid dynamics of axisymmetric growth and beyond.

\begin{acknowledgments}
The authors acknowledge funding from the Engineering and Physical Sciences Research Council (EPSRC) in the form of a Doctoral Training Award (EP/M508202/1 to JT) and grants (EP/N021096/1 to DJS;  EP/M00015X/1 to RJD).

The authors would also like to thank Prof Anja Geitmann and Dr Youssef Chebli (both of the Department of Plant Science, Faculty of Agriculture and Environmental Sciences, McGill University, Sainte-Anne-de-Bellevue, Canada) for their valuable insights on pollen tube physiology and growth. 
\end{acknowledgments}

\appendix

\section{Regularized Stokeslets in cylindrical coordinates for ringlet evaluation}\label{sec:IntKer}
The expressions $S_{ij}^\varepsilon(\mathbf{x}_0,\mathbf{x})$ for fluid point $\mathbf{x}_0=(r_0,0,z_0)$ and ring point $\mathbf{x}_n=(r_n,\theta,z_n)$ in cylindrical coordinates are
\begin{align}
S_{11}^\varepsilon &= \frac{1}{\hat{r}_\varepsilon^3}\left[2(r_0-r_n\cos\theta)^2 + (r_n\sin\theta)^2 + (z_0-z_n)^2 + 2\varepsilon^2\right], \nonumber \\
S_{12}^\varepsilon &= \frac{1}{\hat{r}_\varepsilon^3}\left[-(r_0-r_n\cos\theta)(r_n\sin\theta)\right],
\nonumber \\
S_{13}^\varepsilon &= \frac{1}{\hat{r}_\varepsilon^3}\left[(r_0-r_n\cos\theta)(z_0-z_n)\right],
\nonumber \\
S_{21}^\varepsilon &= \frac{1}{\hat{r}_\varepsilon^3}\left[-(r_0-r_n\cos\theta)(r_n\sin\theta)\right],
\nonumber \\
S_{22}^\varepsilon &= \frac{1}{\hat{r}_\varepsilon^3}\left[(r_0-r_n\cos\theta)^2 + 2(r_n\sin\theta)^2 + (z_0-z_n)^2 + 2\varepsilon^2 \right],
\nonumber \\
S_{23}^\varepsilon &= \frac{1}{\hat{r}_\varepsilon^3}\left[-(r_n\sin\theta)(z_0-z_n)\right],
\nonumber \\
S_{31}^\varepsilon &= \frac{1}{\hat{r}_\varepsilon^3}\left[(r_0-r_n\cos\theta)(z_0-z_n)\right],
\nonumber \\
S_{32}^\varepsilon &= \frac{1}{\hat{r}_\varepsilon^3}\left[-(r_n\sin\theta)(z_0-z_n)\right],
\nonumber \\
S_{33}^\varepsilon &= \frac{1}{\hat{r}_\varepsilon^3}\left[(r_0-r_n\cos\theta)^2 + (r_n\sin\theta)^2 + 2(z_0-z_n)^2 + 2\varepsilon^2\right], \label{eq:ringlets} \end{align} in which $\hat{r}_\varepsilon = ((r_0-r_n\cos\theta)^2 + (r_n\sin\theta)^2 + (z_0-z_n)^2 + \varepsilon^2)^{1/2}.$

\FloatBarrier
\section{Evaluating the limiting behaviour of $R_{\alpha\beta}^\varepsilon$} \label{app:limits}
To understand the behaviour of $R_{\alpha\beta}^\varepsilon$ as $r_n,r_0\rightarrow 0$, it is easiest to consider $R_{\alpha\beta}^\varepsilon$ in the form
\begin{align} \label{eq:R1}
R_{rr}^\varepsilon(\mathbf{x}_0,\mathbf{x}_n) &= r_n(-r_0r_nI_0+(2\tau-(z_0-z_n)^2)I_1 - 3r_0r_nI_2), \\ R_{rz}^\varepsilon(\mathbf{x}_0,\mathbf{x}_n) &= r_n(z_0-z_n)(r_0I_0-r_nI_1), \\ R_{zr}^\varepsilon(\mathbf{x}_0,\mathbf{x}_n) &= r_n(z_0-z_n)(-r_nI_0+r_0I_1), \\ R_{zz}^\varepsilon(\mathbf{x}_0,\mathbf{x}_n) &= r_n((\tau + (z_0-z_n)^2 + \varepsilon^2)I_0 - 2r_0r_nI_1), \\ R_{\theta\theta}^\varepsilon(\mathbf{x}_0,\mathbf{x}_n) &= r_n(r_0r_nI_0 + (\tau+\varepsilon^2)I_1 - 3r_0r_nI_2),  \label{eq:R2} \end{align}
in which
\begin{align}
I_0 &= \frac{4k^3}{(4r_0r_n)^{3/2}}\left(\frac{1}{1-k^2}E\right) , \\
I_1 &= \frac{4k^3}{(4r_0r_n)^{3/2}}\left(\frac{2-k^2}{k^2(1-k^2)}E - \frac{2}{k^2}F\right), \\
I_2 &= \frac{4k^3}{(4r_0r_n)^{3/2}}\left(\frac{k^4-8k^2+8}{k^4(1-k^2)}E-\frac{4(2-k^2)}{k^4}F\right),
\end{align}
with $\tau = r_0^2 + r_n^2 + (z_0-z_n)^2 + \varepsilon^2$ and $k^2 := 4r_0r_n/(\tau+2r_0r_n)$. Evaluating $\lim_{k\rightarrow 0}I_n$ for each $n\in\{0,1,2\}$ and substituting these back into Equations \eqref{eq:R1} -- \eqref{eq:R2} yields the desired results for $\lim_{r_n\rightarrow 0}R_{\alpha\beta}^\varepsilon$ and $\lim_{r_0\rightarrow 0}R_{\alpha\beta}^\varepsilon$.

The first step in evaluating these limits is to observe that from the definition of $k$, it follows that $k\rightarrow 0$ as either $r_0\rightarrow 0$ or $r_n\rightarrow 0$ (or both). Further, noting that
\begin{equation}
\lim_{k\rightarrow 0} \frac{4k^3}{(4r_0r_n)^{3/2}} = \lim_{k\rightarrow 0}\frac{4\left(\frac{4r_0r_n}{\tau+2r_0r_n}\right)^{3/2}}{(4r_0r_n)^{3/2}}= \frac{4}{\tau^{3/2}},\end{equation} in which the value of $\tau$ as $k\rightarrow 0$ depends on whether $r_0\rightarrow 0$ or $r_n\rightarrow 0$ (or both), it is observed that $I_0,I_1,I_2$ contain common finite term outside of the larger brackets. Evaluating the remaining parts of $I_0,I_1,I_2$ in the limit as $k\rightarrow 0$ requires employing the power series expansions of the complete elliptic integrals \cite{gradshteyn2014table}, such that 
\begin{align}
F(k) &= \frac{\pi}{2}\left(1 + \frac{1}{4}k^2 + \frac{9}{64}k^4 + \ldots \right), \\
E(k) &= \frac{\pi}{2}\left(1 - \frac{1}{4}k^2 - \frac{9}{64}k^4 + \ldots \right).
\end{align}

Letting $E \sim \pi/2$ it follows that
\begin{equation} \lim_{k\rightarrow 0}\left(\frac{1}{1-k^2}E\right) = \frac{\pi}{2}.\end{equation} Similarly, letting $E \sim \frac{\pi}{2}(1-\frac{1}{4}k^2)$ and $F \sim \frac{\pi}{2}(1+\frac{1}{4}k^2)$ it can be found that 
\begin{equation}\lim_{k\rightarrow 0}\left(\frac{2-k^2}{k^2(1-k^2)}E - \frac{2}{k^2}F\right) = 0.\end{equation}
Finally, using $E \sim \frac{\pi}{2}(1-\frac{1}{4}k^2 - \frac{9}{64}k^4)$ and $F \sim \frac{\pi}{2}(1+\frac{1}{4}k^2 + \frac{9}{64}k^4)$ yields
\begin{equation} \lim_{k\rightarrow 0}\left(\frac{k^4-8k^2+8}{k^4(1-k^2)}E-\frac{4(2-k^2)}{k^4}F\right) = -\frac{\pi}{8}.\end{equation}
Compiling all of the above gives
\begin{equation} \begin{array}{llll} I_0\rightarrow 2\pi/\tau^{3/2}, & I_1\rightarrow 0, & I_2\rightarrow -\pi/2\tau^{3/2}, & \text{as } k\rightarrow 0, \end{array}\end{equation}
which upon substitution into the expressions for $R_{\alpha\beta}^\varepsilon$ yields 
\begin{align} \lim_{r_n\rightarrow 0}R_{\alpha\beta}^\varepsilon & \equiv 0, \\
\lim_{r_0\rightarrow 0} R_{rr}^\varepsilon & = \lim_{r_0\rightarrow 0} R_{rz}^\varepsilon = \lim_{r_0\rightarrow 0} R_{\theta\theta}^\varepsilon  = 0, \\
\lim_{r_0\rightarrow 0} R_{zr}^\varepsilon & = -2\pi r_n^2(z_0-z_n)/\tau^{3/2},\\
\lim_{r_0\rightarrow 0} R_{zz}^\varepsilon & = 2\pi r_n(\tau+(z_0-z_n)^2+\varepsilon^2)/\tau^{3/2}. \end{align}

\section{Fluid flow induced by unit ringlet forces} \label{sec:streamlines}
\begin{figure}[ht]
	\centering
	\subfloat[\raggedright{Unit force in direction $\hat{\mathbf{z}}$.}]{\includegraphics{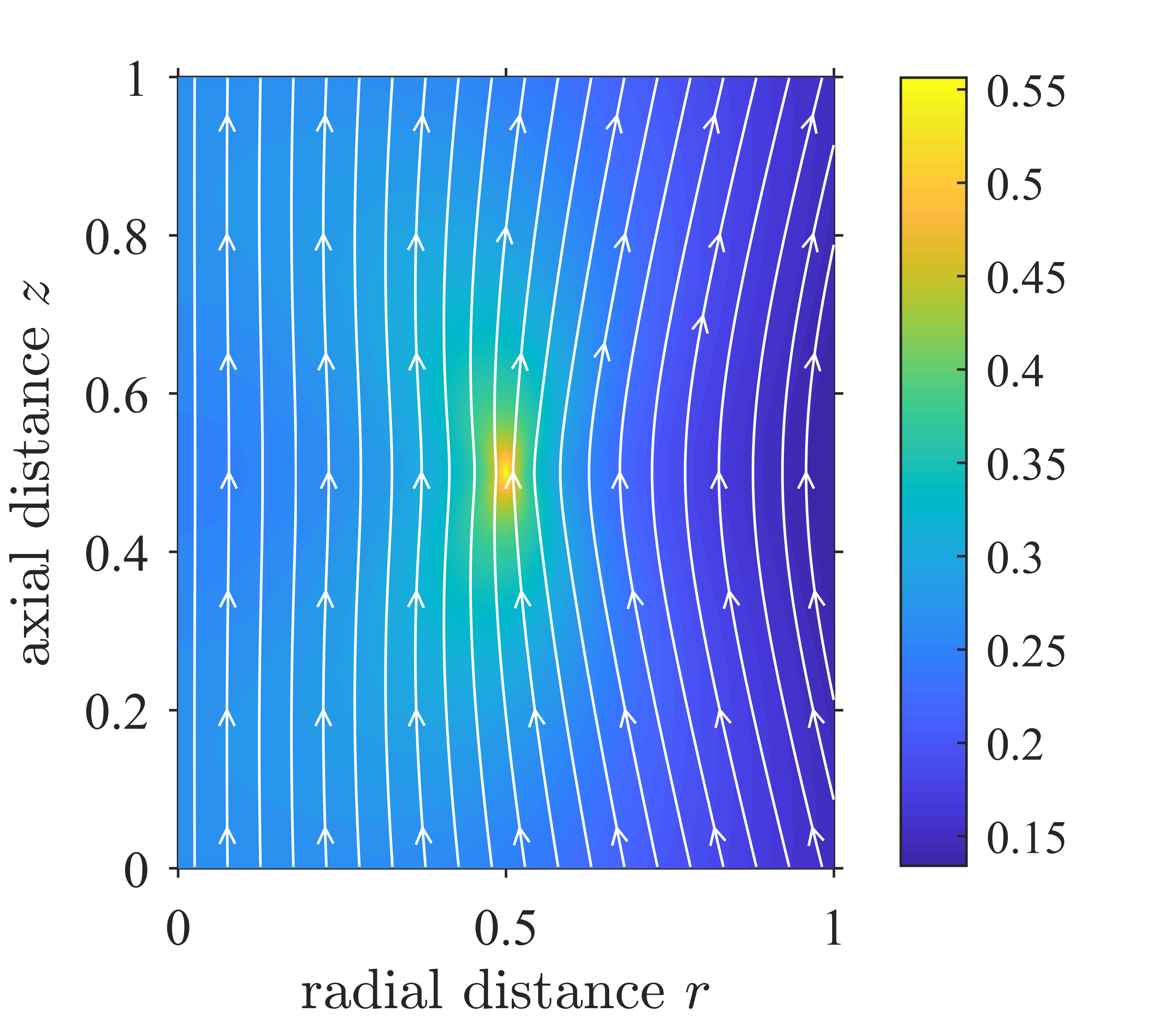}}\qquad
	\subfloat[\raggedright{Unit force in direction $\hat{\mathbf{z}}$ with wall at $r=1$.}]{\includegraphics{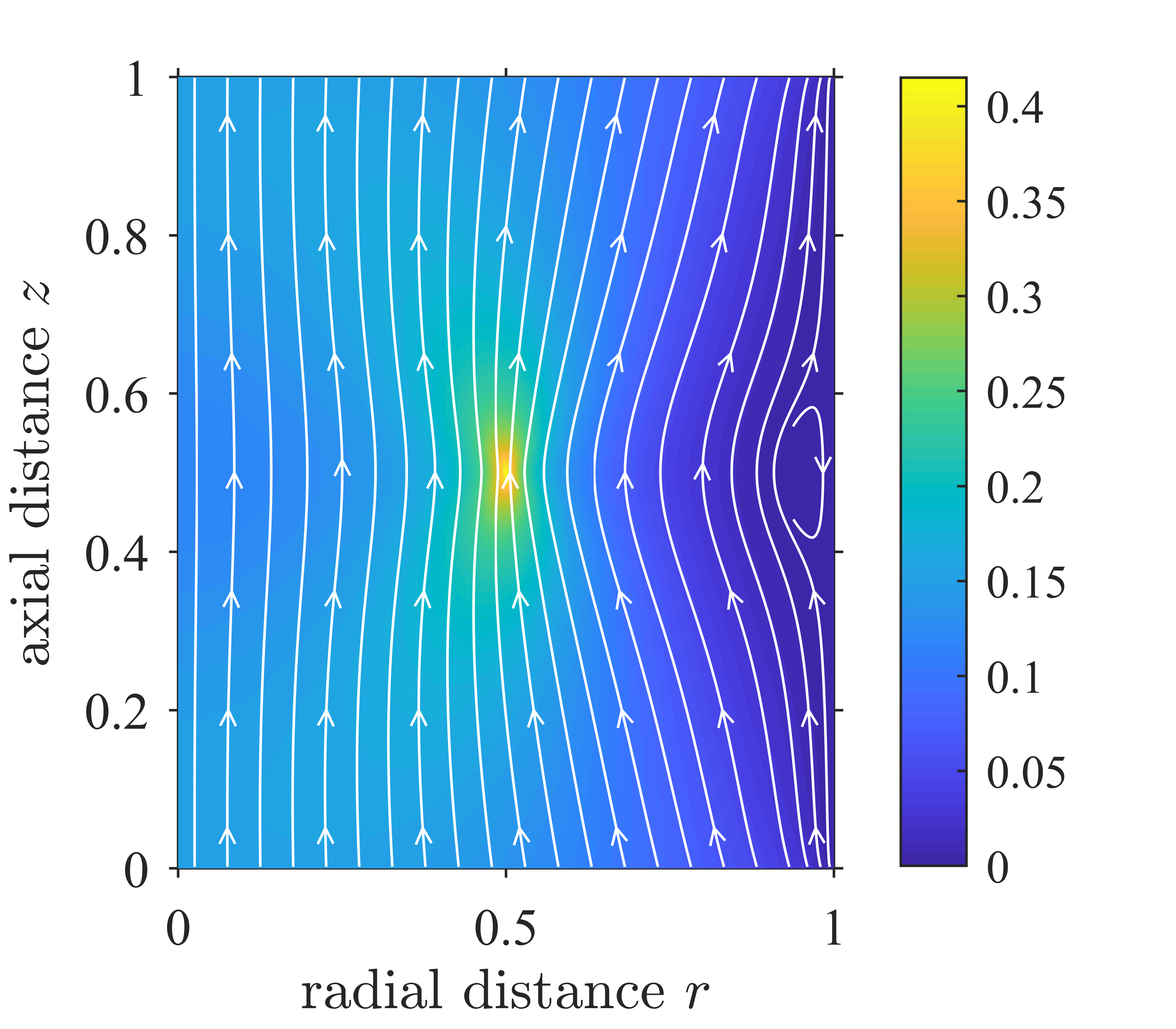}} \\
	\subfloat[\raggedright{Unit force in direction $\hat{\mathbf{r}}$.}]{\includegraphics{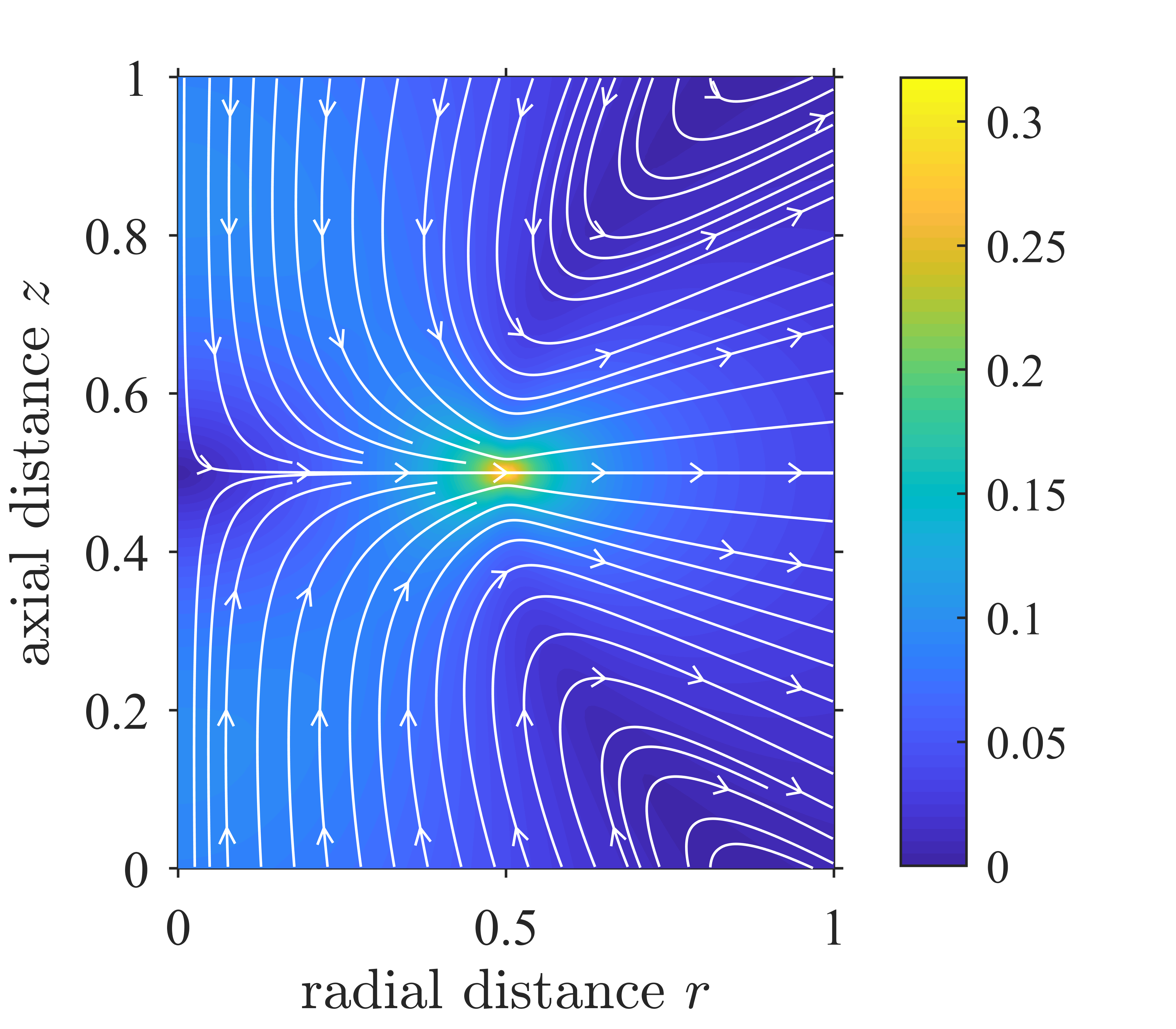}}\qquad
	\subfloat[\raggedright{Unit force in direction $\hat{\mathbf{r}}$ with wall at $r=1$.}]{\includegraphics{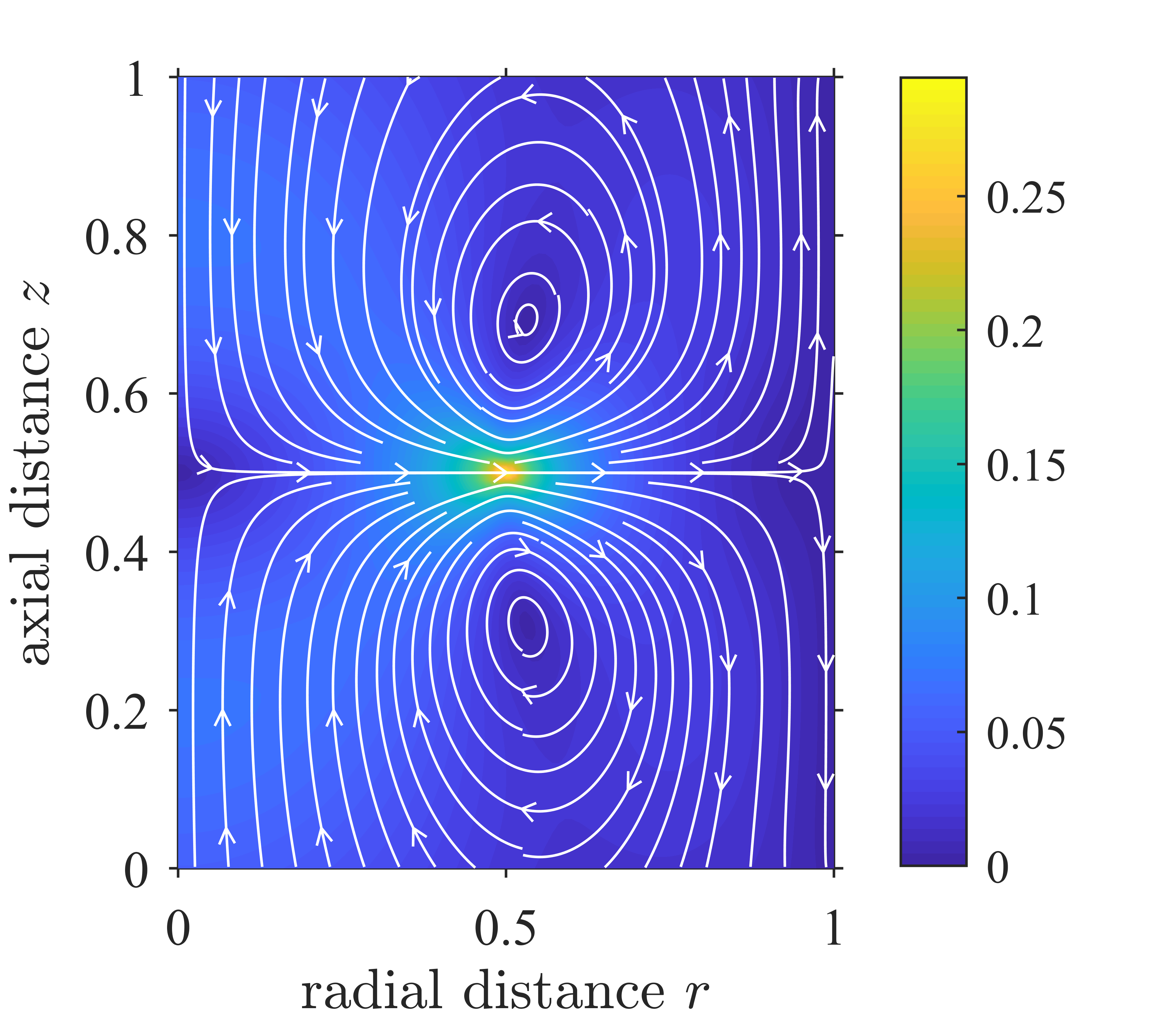}}
		\caption{Fluid flow induced by unit force on ringlet placed at $(r,z) = (0.5,0.5)$. (a) and (b) show axial force, with (c) and (d) showing radial force. A no--slip bounding wall is present at $r=1$ in cases (b) and (d).}
	\label{fig:streamlines}
\end{figure}

Figure \ref{fig:streamlines} shows the fluid velocity induced by a ringlet located at $(r,z)=(0.5,0.5)$ with associated unit force in $\hat{\mathbf{r}}$ and $\hat{\mathbf{z}}$ directions, both under free--space conditions ((a) \& (c)) and in the presence of a cylindrical wall at $r=1$ ((b) \& (d), with the wall also represented by Stokes ringlets). In the presence of the bounding wall, closed streamlines and toroidal eddies are observed.

\FloatBarrier
\section{Evaluating the double layer potential} \label{sec:DLP}
Recall the form of the double layer potential
\begin{equation} \label{eq:doubleLayerPot} \text{(DLP)}_i = \frac{1}{8\pi}\int_{\partial D} u_j(\mathbf{x})T_{ijk}^\varepsilon(\mathbf{x}_0,\mathbf{x})n_k(\mathbf{x}) \, dS(\mathbf{x}),\end{equation} 
in which the stress tensor $T_{ijk}^\varepsilon$ is given by
\begin{equation} T_{ijk}^\varepsilon(\mathbf{x}_0,\mathbf{x}) = -6\frac{(x_{0,i}-x_i)(x_{0,j}-x_j)(x_{0,k}-x_k)}{(|\mathbf{x}_0-\mathbf{x}|^2 + \varepsilon^2)^{5/2}} - 3\varepsilon^2 \frac{(x_{0,i}-x_i)\delta_{jk} + (x_{0,j}-x_j)\delta_{ik} + (x_{0,k}-x_k)\delta_{ij}}{(|\mathbf{x}_0-\mathbf{x}|^2 + \varepsilon^2)^{5/2}}.\end{equation}

As with the Stokeslet $S_{ij}^\varepsilon$ in the SLP, the stress tensor $T_{ijk}^\varepsilon$ can be expressed in cylindrical coordinates via
\begin{align} (x_{0,1}-x_1) & = r_0-r_n\cos\theta, \\ (x_{0,2}-x_2) &= -r_n\sin\theta, \\ (x_{0,3}-x_3) &= z_0-z_n, \\ |\mathbf{x}_0-\mathbf{x}|^2 &= (r_0-r_n\cos\theta)^2+(r_n\sin\theta)^2 + (z_0-z_n)^2.
\end{align} 
and the Cartesian and cylindrical forms of the flow vector $\mathbf{u}$ are related via $u_j = \Theta_{j\beta}u_\beta$ in which
\begin{equation}\Theta(\theta) = \left(\begin{array}{ccc} \cos\theta & -\sin\theta & 0 \\ \sin\theta & \cos\theta & 0 \\ 0 & 0 & 1 \end{array} \right), \end{equation}
Recalling that the azimuthal component of the normal to an axisymmetric body is zero, the transformation from polar to Cartesian coordinates is $n_i = \Psi_{i\alpha}n_\alpha$ where 
\begin{equation}\Psi(\theta) = \left(\begin{array}{ccc} \cos\theta & 0 & 0 \\ \sin\theta & 0 & 0 \\ 0 & 0 & 1 \end{array} \right). \end{equation} Let \begin{equation} Q_{\alpha\beta\gamma} = r \delta_{\alpha i} \int_0^{2\pi} \Theta_{j\beta }T_{ijk}^\varepsilon\Psi_{k \gamma} \, d\theta,\end{equation} such that \begin{equation} \text{(DLP)}_\alpha = \frac{1}{8\pi} \int_0^L Q_{\alpha\beta\gamma}u_{\beta}n_\gamma\, ds.\end{equation}
For fixed $\alpha$, nonzero elements of $Q_{\alpha\beta\gamma}$ are given by
\begin{equation}
\begin{array}{ll} Q_{\alpha 1 1} & = r_n \delta_{\alpha i} \int_0^{2\pi} \left(T^\varepsilon_{i 1 1}\cos^2\theta + T^\varepsilon_{i 1 2}\sin\theta\cos\theta + T^\varepsilon_{i 2 1}\sin\theta\cos\theta + T^\varepsilon_{i 2 2}\sin^2\theta\right) d\theta, \\
Q_{\alpha 1 3} & = r_n\delta_{\alpha i}\int_0^{2\pi} \left(T^\varepsilon_{i 1 3}\cos\theta + T^\varepsilon_{i 2 3}\sin\theta\right)d\theta, \\
Q_{\alpha 2 1} & = r_n\delta_{\alpha i} \int_0^{2\pi} \left(- T^\varepsilon_{i 1 1} \sin\theta\cos\theta - T^\varepsilon_{i 1 2}\sin^2\theta + T^\varepsilon_{i 2 1}\cos^2\theta + T^\varepsilon_{i 2 2}\sin\theta\cos\theta\right)d\theta, \\
Q_{\alpha 2 3} & = r_n\delta_{\alpha i}\int_0^{2\pi} \left(-T^\varepsilon_{i 1 3}\sin\theta + T^\varepsilon_{i 2 3}\cos\theta\right)d\theta, \\
Q_{\alpha 3 1} & = r_n\delta_{\alpha i} \int_0^{2\pi} \left(T^\varepsilon_{i 3 1}\cos\theta + T^\varepsilon_{i 3 2}\sin\theta\right)d\theta, \\
Q_{\alpha 3 3} & = r_n\delta_{\alpha i} \int_0^{2\pi} \left(T^\varepsilon_{i 33}\right)d\theta,\end{array}
\end{equation}
such that each $Q_{\alpha\beta\gamma}$ is a linear sum of terms of the form
\begin{equation} \llangle \bullet \rrangle_{ijk}:=r_n\int_0^{2\pi}T_{ijk}^\varepsilon \bullet \, d\theta.\end{equation} 
Letting
\begin{equation} \label{eq:J_n} J_{m,n} := r_n\int_0^{2\pi} \frac{\sin^m\theta\cos^n\theta}{(\tau-2r_0r_n\cos\theta)^{5/2}} \,d\theta, \end{equation} in which $J_{m,n} = 0$ for $m$ odd, each of the necessary $\llangle \bullet \rrangle_{ijk}$ can be written as: 
\begin{equation}
\begin{array}{ll}
\llangle\cos^2\theta\rrangle_{111} &= -(6r_0^3+9\varepsilon^2r_0)J_{0,2} + (18r_0^2r_n + 9\varepsilon^2r_n)J_{0,3} - 18r_0r_n^2J_{0,4} + 6r_n^3J_{0,5}, \\
\llangle\sin\theta\cos\theta\rrangle_{112} &= (3\varepsilon^2 r_n + 6r_0^2r_n)J_{2,1} - 12r_0r_n^2J_{2,2} + 6r_n^3J_{2,3}, \\
\llangle\sin^2\theta\rrangle_{122} &= -3\varepsilon^2r_0J_{2,0} + 3\varepsilon^2r_nJ_{2,1} - 6r_0r_n^2J_{4,0} + 6r_n^3J_{4,1}, \\
\llangle\cos\theta\rrangle_{113} &= -3(z_0-z_n)((2r_0^2+\varepsilon^2)J_{0,1}-4r_0r_nJ_{0,2}+2r_n^2J_{0,3}), \\
\llangle\sin\theta\rrangle_{123} &= 6r_n(z_0-z_n)(r_0J_{2,0}-r_nJ_{2,1}), \\
\llangle 1\rrangle_{133} &= (6(z_0-z_n)^2 + 3\varepsilon^2)(-r_0J_{0,0}+r_nJ_{0,1}), \\ 
\\
\llangle\sin\theta\cos\theta\rrangle_{211} &= (3\varepsilon^2r_n+6r_0^2r_n)J_{2,1} - 12r_0r_n^2J_{2,2} + 6r_n^3J_{2,3}, \\
\llangle\sin^2\theta\rrangle_{212} &= -3\varepsilon^2r_0J_{2,0} + 3\varepsilon^2 r_nJ_{2,1} - 6r_0r_n^2J_{4,0} + 6r_n^3J_{4,1}, \\
\llangle\cos^2\theta\rrangle_{221} &= -3\varepsilon^2 r_0J_{0,2} + 3\varepsilon^2 r_nJ_{0,3} + 6r_n^3J_{2,3} - 6r_0r_n^2J_{2,2}, \\
\llangle\sin\theta\cos\theta\rrangle_{222} &= 9\varepsilon^2r_nJ_{2,1} + 6r_n^3J_{4,1}, \\
\llangle\sin\theta\rrangle_{213} &= 6r_n(z_0-z_n)(r_0J_{2,0}-r_nJ_{2,1}), \\
\llangle\cos\theta\rrangle_{223} &= - 3\varepsilon^2(z_0-z_n)J_{0,1} -6r_n^2(z_0-z_n)J_{2,1}, \\
\\
\llangle\cos^2\theta\rrangle_{311} &= -3(z_0-z_n)((2r_0^2+\varepsilon^2)J_{0,2}-4r_0r_nJ_{0,3}+2r_n^2J_{0,4}), \\
\llangle\sin\theta\cos\theta\rrangle_{312} &= 6r_0r_n(z_0-z_n)J_{2,1}-6r_n^2(z_0-z_n)J_{2,2}, \\
\llangle\sin^2\theta\rrangle_{322} &= -3\varepsilon^2(z_0-z_n)J_{2,0}-6r_n^2(z_0-z_n)J_{4,0},\\
\llangle\cos\theta\rrangle_{313}&=(6(z_0-z_n)^2+3\varepsilon^2)(-r_0J_{0,1}+r_nJ_{0,2}),\\
\llangle\sin\theta\rrangle_{323} &= (6r_n(z_0-z_n)^2+3r_n\varepsilon^2)J_{2,0},\\
\llangle 1\rrangle_{333} &= -(6(z_0-z_n)^3 + 9\varepsilon^2(z_0-z_n))J_{0,0}.
\end{array}
\end{equation}
Using the double angle formulae for $\sin$ and $\cos$ we are able to express $J_{m,n}$ purely in terms of even powers of $\cos$, such that
\begin{equation} \label{eq:J_int} J_{m,n} = \lambda \cdot 2^m \int_0^{\frac{\pi}{2}}  \frac{(2\cos^2\theta-1)^n(1-\cos^2\theta)^{m/2}\cos^m\theta}{(1-k^2\cos^2\theta)^{5/2}} \,d\theta \quad \text{for } m \text{ even}, \end{equation} 
where $\lambda = r_n(k/\sqrt{r_0r_n})^5/8 = 4r_n(\sqrt{\tau+2r_0r_n})^{-5}$. Note that the upper limit of integration was first reduced by application of the double angle formulae, followed by the even parity of the resulting integrand about $\pi/2$. If we further define
\[C_m = \lambda \int_0^{\frac{\pi}{2}} \frac{\cos^{2m}\theta}{(1-k^2\cos^2\theta)^{5/2}} \, d\theta, \] then expanding Equation \eqref{eq:J_int} for the relevant values of $m,n$ yields

\begin{equation}
\begin{array}{rrrrrrr}
J_{0,0} = &  +C_0, &&&&&  \\
J_{0,1} = & -C_0 & +2C_1, &&&& \\
J_{0,2} = &  +C_0 & -4C_1 & +4C_2, &&& \\
J_{0,3} = & -C_0 & +6C_1 & -12C_2 & +8C_3, && \\
J_{0,4} = &  +C_0 & -8C_1 & +24C_2 & -32C_3 & +16C_4, & \\
J_{0,5} = & -C_0 & +10C_1 & -40C_2 & +80C_3 & -80C_4 & +32C_5, \\ 
\\
J_{2,0} = && 4(+C_1 & -C_2), &&& \\
J_{2,1} = && 4(-C_1 & +3C_2 & -2C_3), && \\
J_{2,2} = && 4(+C_1 & -5C_2 & +8C_3 & -4C_4), & \\
J_{2,3} = && 4(-C_1 & +7C_2 & -18C_3 & +20C_4 & -8C_5), \\ 
\\
J_{4,0} = &&& 16(+C_2 & -2C_3 & +C_4), & \\
J_{4,1} = &&& 16(-C_2 & +4C_3 & -5C_4 & +2C_5),
\end{array}
\end{equation}

in which the integrals $C_m$ can be expressed in terms of complete elliptic integrals of the first and second kind ($F$ and $E$ respecitvely) with elliptic modulus $k$ as
\begin{align} 
C_0 & = \frac{\lambda}{3(1-k^2)}\left(-F + \frac{2(2-k^2)}{1-k^2}E\right), \nonumber \\
C_1 & = \frac{\lambda}{3k^2(1-k^2)}\left(-F + \frac{1+k^2}{1-k^2}E\right), \nonumber \\
C_2 & = \frac{\lambda}{3k^4(1-k^2)}\left((2-3k^2)F + \frac{2(2k^2-1)}{1-k^2}E\right), \\
C_3 & = \frac{\lambda}{3k^6(1-k^2)}\left((8-9k^2)F - \frac{3k^4-13k^2+8}{1-k^2}E\right), \nonumber \\
C_4 & = \frac{\lambda}{3k^8(1-k^2)}\left((16-16k^2-k^4)F - \frac{2k^6 + 4k^4 - 24k^2 + 16}{1-k^2}E\right), \nonumber \\ 
C_5 & = \frac{\lambda}{15k^{10}(1-k^2)}\left(128-120k^2-9k^4-4k^6)F - \frac{8k^8+11k^6+27k^4-184k^2+128}{1-k^2}E\right). \nonumber
\end{align}

\section{Translating unit sphere} \label{sec:translatingSphere}
Following the example of Cortez \emph{et al.} \cite{cortez2005method}, consider the translating unit sphere with velocity $\mathbf{u}=-\hat{\mathbf{z}}.$ The sphere is parametrized and discretized in the same manner as in Section \ref{sec:singSolComp}. Assuming the sphere experiences zero azimuthal spin, the fluid velocity at any point $\mathbf{x}_0$ can be approximated using the $N$ ringlets via Equation \eqref{eq:stokes_summation}. Using intervals of equal size, the quadrature weight associated with numerical integration over $\mathbf{p}$ is simply $w_n=\frac{\pi}{N}\,\forall n$. Now considering the velocity evaluated at the location of each ring in the $rz$ plane yields a system of equations which can be written in matrix form
\begin{equation}   \mathsf{U} = \frac{1}{8N\mu}\left[ \begin{array}{cc} \mathsf R_{rr}^{\varepsilon} & \mathsf R_{rz}^{\varepsilon} \\ \mathsf R_{zr}^{\varepsilon} & \mathsf R_{zz}^{\varepsilon} \end{array}\right]\mathsf{G}^{a},
\end{equation}
as outlined in Equations \eqref{eq:invertStart} - \eqref{eq:invertEnd}. By setting \begin{equation} g^a_r(\mathbf{x}_i)=0, \quad g^a_z(\mathbf{x}_i)=\frac{3\mu}{2a}U, \quad \forall i=1,...\,,N,\end{equation} where $\mu,a,U$ are the fluid viscosity, sphere radius, and sphere speed (in $z$) respectively, the classical solution for Stokes flow \begin{equation}u_r(\mathbf{x}_i)=0, \quad u_z(\mathbf{x}_i)=U, \quad \forall i=1,...\,,N,\end{equation} should follow. Note that this calculation is independent of the value of $\mu$ by cancellation (besides the implicit requirement that $\text{Re}=UL/\mu\ll 1$ for the Stokes equations to be valid), and that we use $a=1$, $U=-1$ as described in the outline of the problem. 

\begin{figure}[ht]  \centering \includegraphics{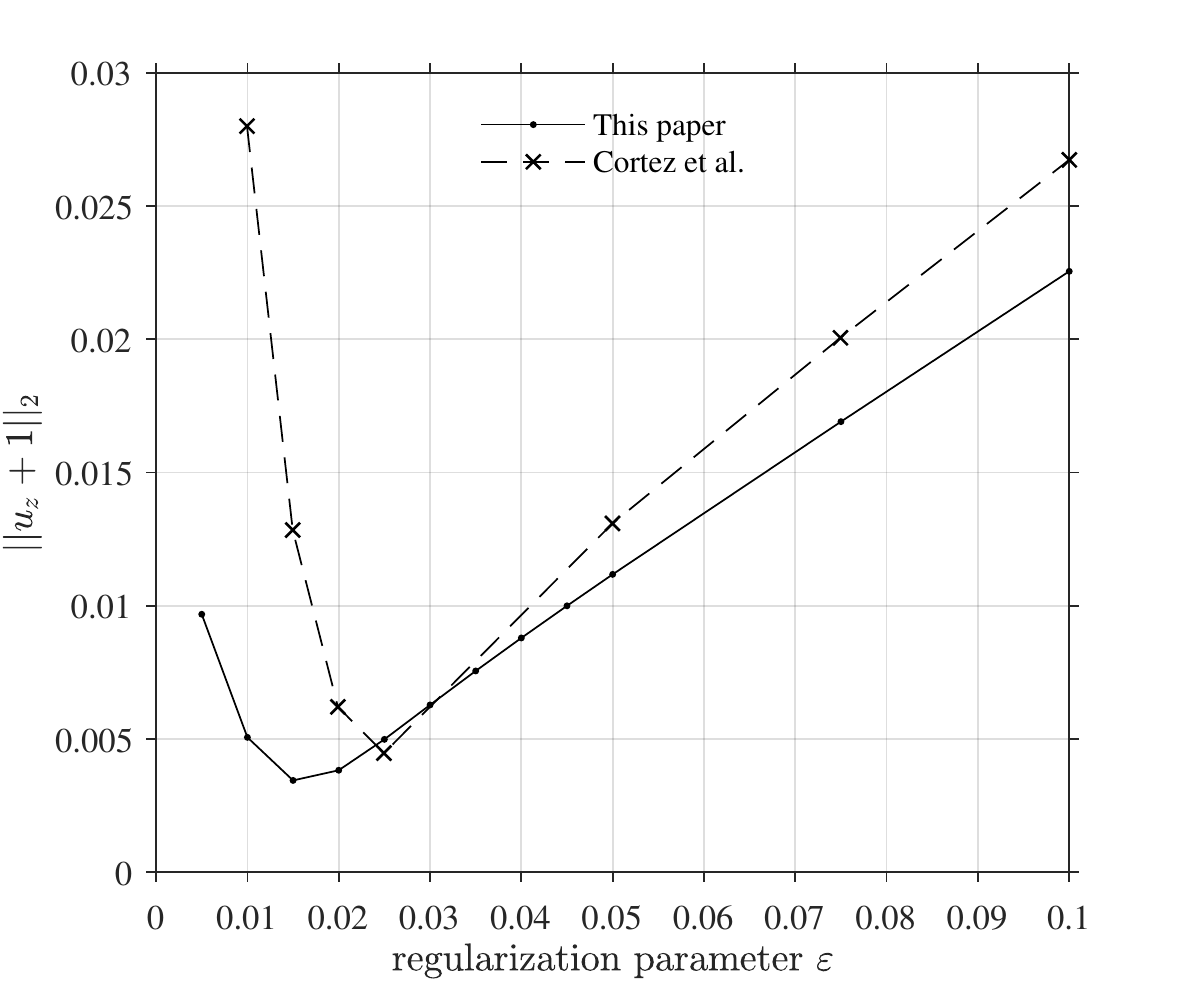}
	\caption{$\ell^2$ errors for various values of $\varepsilon$ using $N=50$ regularized Stokeslet rings in our discretization of the sphere surface (solid line), plotted against data taken from Cortez \emph{et al.} \cite{cortez2005method} (dashed line).} \label{fig:cc1} \end{figure}

The first test involves using $N=50$ regularized rings so that the grid size is given by $\pi/50\approx0.065$. This is chosen such that the minimum distance between adjacent rings in the axisymmetric discretization of the sphere surface is approximately the same as the distance between adjacent points in Cortez and colleagues'  discretization using $3N^2$ regularized Stokeslets. The regularization parameter $\varepsilon$ is varied between $0.005$ and $0.1$, and the error in the $\ell^2$ norm for the $z$ component of the flow field is recorded in each case. This error is defined as \begin{equation} |u_z+1|_2 :=\sqrt{\frac{\sum_{i=1}^N (u_z(\mathbf{x}_i)+1)^2}{N}}.\end{equation} Division by $N$ is necessary for the sake of comparison of errors with later tests where the value of $N$ will change in order to alter the grid size. Initial results using regularized ringlets are shown in Figure \ref{fig:cc1} 
and are favourable compared to those using regularized Stokeslets. The regularized ringlet method appears to be optimal for a lower value of $\varepsilon$ than the regularized Stokeslet method, with a minimal error found at $\varepsilon\approx0.015$ with our method and $\varepsilon\approx0.025$ with that of the regularized Stokeslet, as well as being slightly more accurate for almost all values of $\varepsilon$ tested. Interestingly, the magnitude of the errors using the two methods briefly appear to coincide near the point at which the regularized Stokeslet error is minimized.

\begin{figure}[ht]  \centering \includegraphics{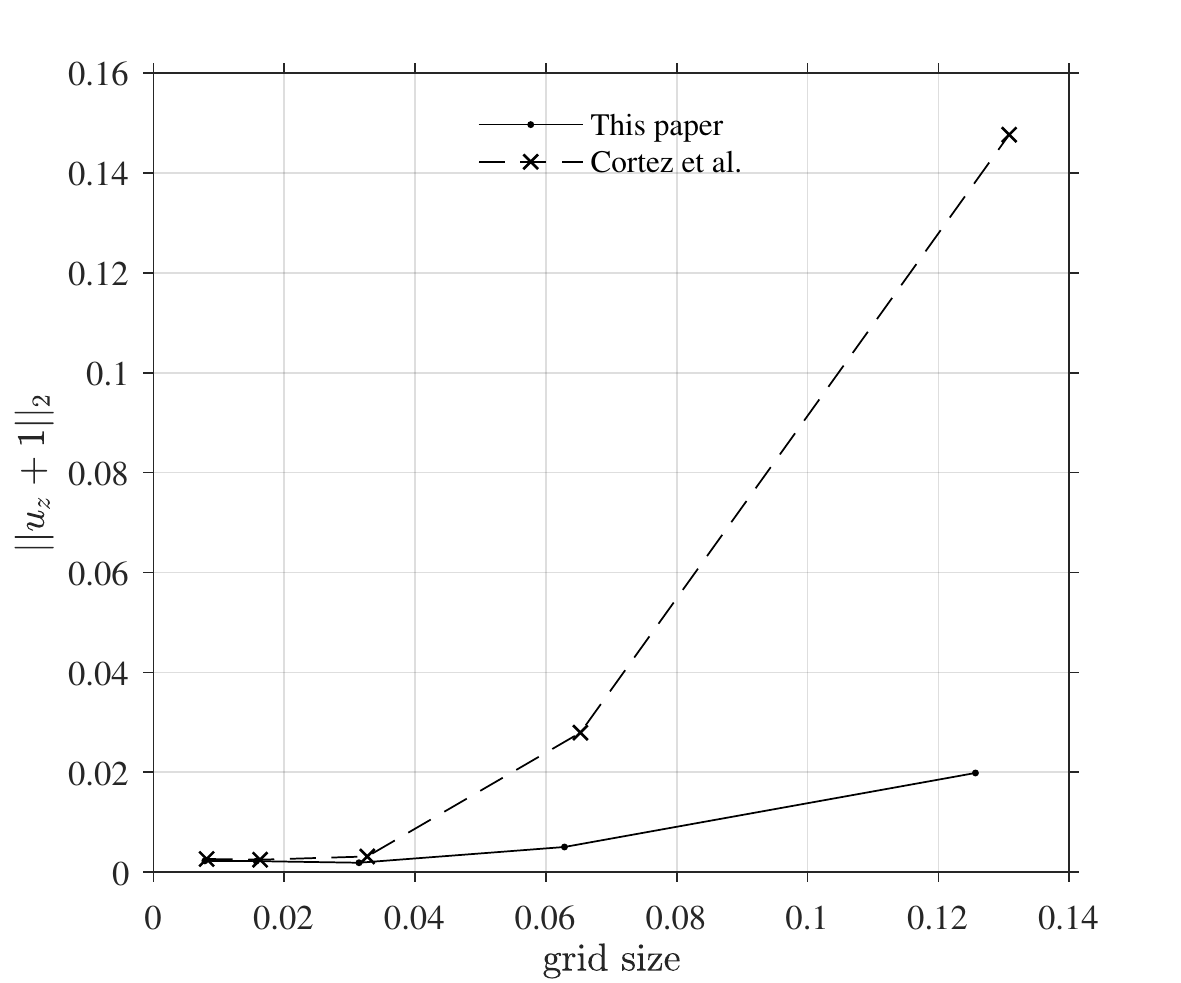}
		\caption{$\ell^2$ errors for different grid sizes ($N=25,50,100,200,400$) in our regularized ring discretization of the sphere surface using fixed $\varepsilon=0.01$ (solid line), plotted against data taken from Cortez \emph{et al.} (dashed line).} \label{fig:cc2} 
\end{figure}

The second test involves varying the grid size for a fixed value of $\varepsilon=0.01.$ Regularized ringlet results are shown in Figure \ref{fig:cc2} and again compare very favourably to results using regularized Stokeslets. For larger grid sizes our errors are significantly reduced compared to those found using regularized Stokeslets. This is at least in part a result of the ringlet method being better suited to handling the small value of $\varepsilon=0.01$. With both methods, the error eventually stops decreasing as the grid size tends towards $0$ since in this regime the regularization error dominates.

\begin{figure}[ht]
	\centering 
	\subfloat[]{\includegraphics[scale=.75]{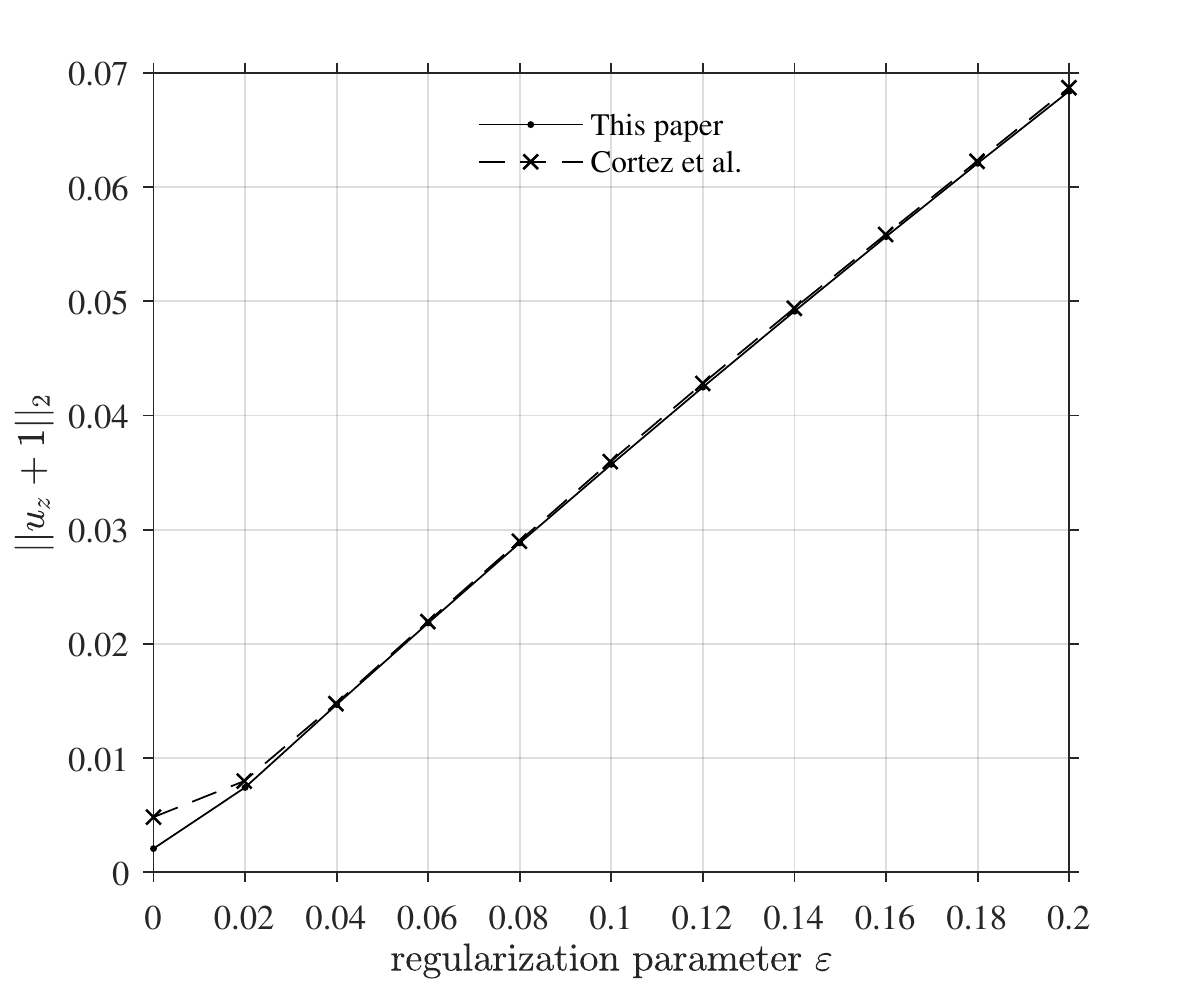} \label{fig:cc3}}
	\subfloat[]{\includegraphics[scale=.75]{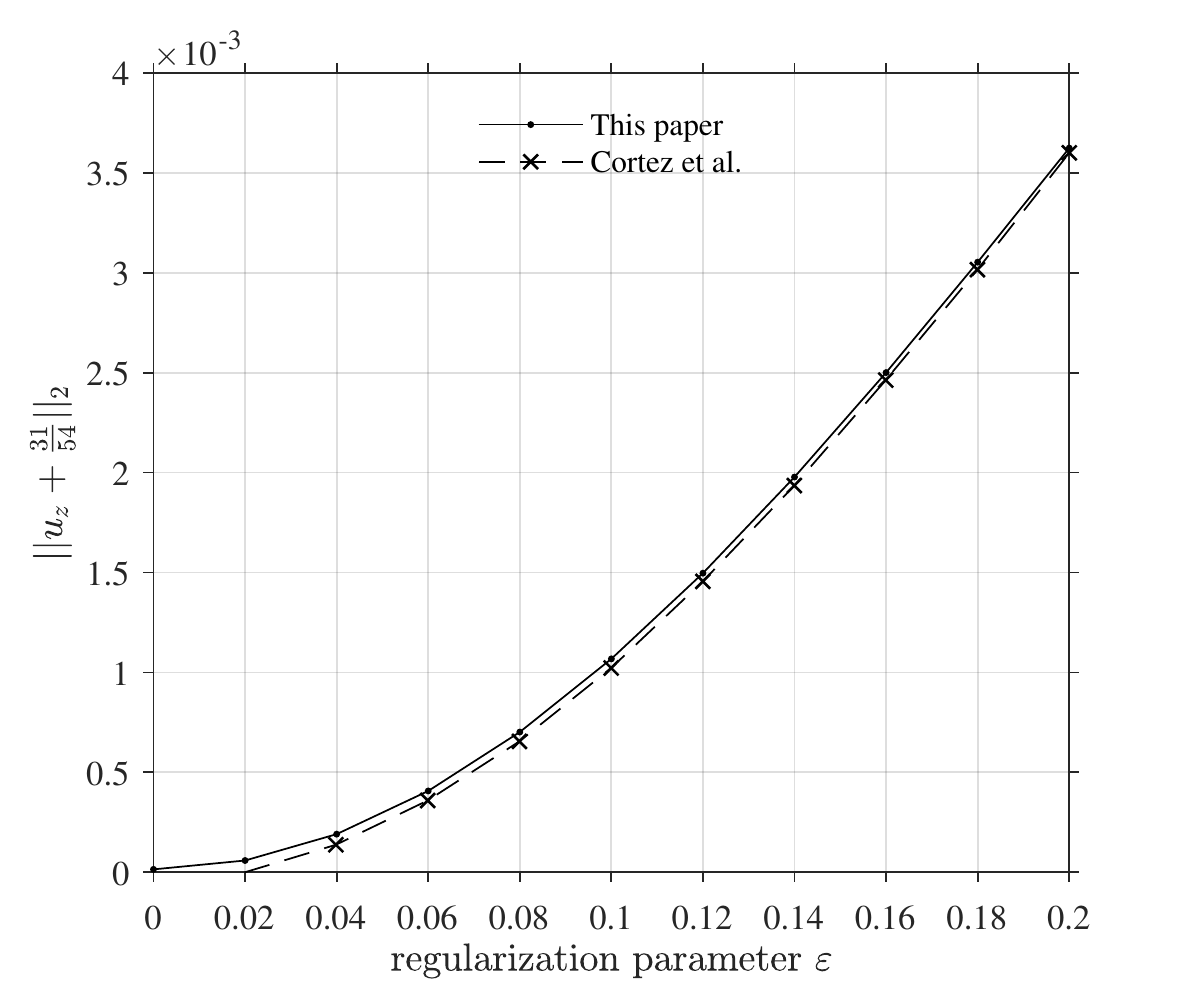} \label{fig:cc4}}
\caption{$\ell^2$ errors at surface point (a) $(r,z) = (1,0)$, (b) $(r,z) = (1.5,0)$ for various values of $\varepsilon$ using $N=124$ regularized Stokeslet rings in our discretization of the sphere surface (solid line), plotted against data taken from Cortez \emph{et al.} (dashed line).}
\end{figure}

The final test again looks at the effect of varying the value of $\varepsilon$ on the magnitude of the numerical error, this time using $N=124$ ringlets for a grid size approximately equal to $0.026$ (note that we do not use $N=125$ ringlets to avoid placing a ring at the point $(r,z)=(1,0)$, which would result in a singular velocity when $\varepsilon=0$). The velocity error is compared at two distinct points: $(r,z)=(1,0)$ lying on the surface of the sphere, and $(r,z)=(1.5,0)$ lying a distance of half the sphere radius away. Ringlet results are shown in Figures \ref{fig:cc3} \& \ref{fig:cc4}, and once more match regularized Stokeslet results very closely. The magnitude of the error is linear with respect to $\varepsilon$ on the surface of the sphere and quadratic a sufficient distance away.

\FloatBarrier

\section{Comparison to singular solutions}\label{sec:MFS}
Regularized solutions for Stokes flow have the advantage of being simple to implement and readily usable without needing to worry about the presence of singularities in the computational domain. This does not mean that singular solutions cannot be used, however; so long as the appropriate care is taken to deal with the singularities in some way, singular solutions can also yield excellent results. The method of fundamental solutions \cite{young2006method,alves2004density} is a popular choice for implementing singular Stokeslet solutions, in which a fictitious boundary is placed outside of the computational domain and adjacent to the physical boundary of the problem considered. Stokeslets (or `source points') are placed on this fictitious boundary and are associated with collocation points (typically of an equal number) on the physical boundary, with the force density for each Stokeslet being calculated using the resistance matrix such that the physical boundary conditions are satisfied. What the appropriate distance between the fictitious and physical boundaries should be is difficult to determine \emph{a priori}, and in some sense this distance can be regarded as a regularization parameter for the singular problem \cite{barrero2012method}. If the separation distance is too small, the proximity between the Stokeslet singularities and the physical boundary may lead to inaccurate solutions, whereas if the distance is too large the resistance matrix may become ill-conditioned \cite{chen2005method}. In some cases, placement of the fictitious boundary may also be constrained by the geometry of the problem itself, leading to solutions that are far from optimal. The difficulty in balancing all of these factors is one of the reasons for the popularity of regularized methods. 

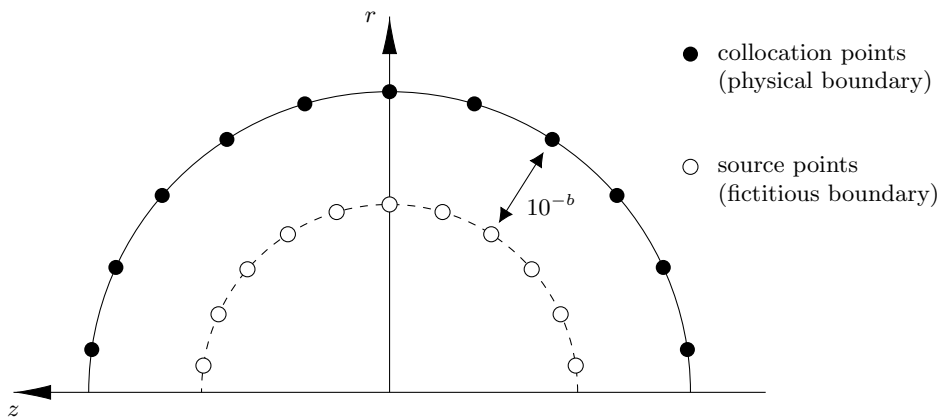
\begin{figure}[ht]
	\centering
	\begin{tikzpicture}
	
	\draw (-5,0) -- (5,0);
	\draw (0,0) -- (0,5);
	
	\draw[-{Latex[length=5mm, width=2mm]}] (0,4.5) -- (0,5);
	\draw[-{Latex[length=5mm, width=2mm]}] (-4.5,0) -- (-5,0);
	
	\draw (4,0) arc (0:180:4);
	\draw[dashed] (2.5,0) arc (0:180:2.5);
	\node at (-5,-.25) {$z$};
	\node at (-.25,5) {$r$};
	
	\foreach \Point in {(0,4),(1.1269,3.8380),(2.1626,3.3650),(3.0230,2.6194),(3.6385,1.6617),(3.9593,0.5693),(-1.1269,3.8380),(-2.1626,3.3650),(-3.0230,2.6194),(-3.6385,1.6617),(-3.9593,0.5693)}{
		\fill \Point circle (.1);
	}
	
	\foreach \P in {(0,2.5),(0.7043,2.3987),(1.3516,2.1031),(1.8894,1.6372),(2.2741,1.0385),(2.4746,0.3558),(-0.7043,2.3987),(-1.3516,2.1031),(-1.8894,1.6372),(-2.2741,1.0385),(-2.4746,0.3558)}{
		\fill[white] \P circle (.1);
		\draw \P circle (.1);
	}
	
	\draw[{Latex[length=2mm, width=2mm]}-{Latex[length=2mm, width=2mm]}] (1.45,2.25) -- (2.05,3.2);
	
	\node at (2.15,2.5) {$10^{-b}$};
	
	\fill (4,4.5) circle (.1);
	\fill[white] (4,3) circle (.1);
	\draw (4,3) circle (.1);
	
	\node[right] at (4.25,4.5) {collocation points};
	\node [below right] at (4.25,4.4) {(physical boundary)};
	\node[right] at (4.25,3) {source points};
	\node[below right] at (4.25,2.9) {(fictitious boundary)};
	\end{tikzpicture}
	\caption{Schematic diagram for implementation of the method of fundamental solutions on the unit sphere using singular rings in the $rz$ plane. Separation distance $10^{-b}$ is exaggerated for the sake of clarity.}
	\label{fig:singularSchematic}
\end{figure}

\begin{figure}[ht]  \centering \includegraphics{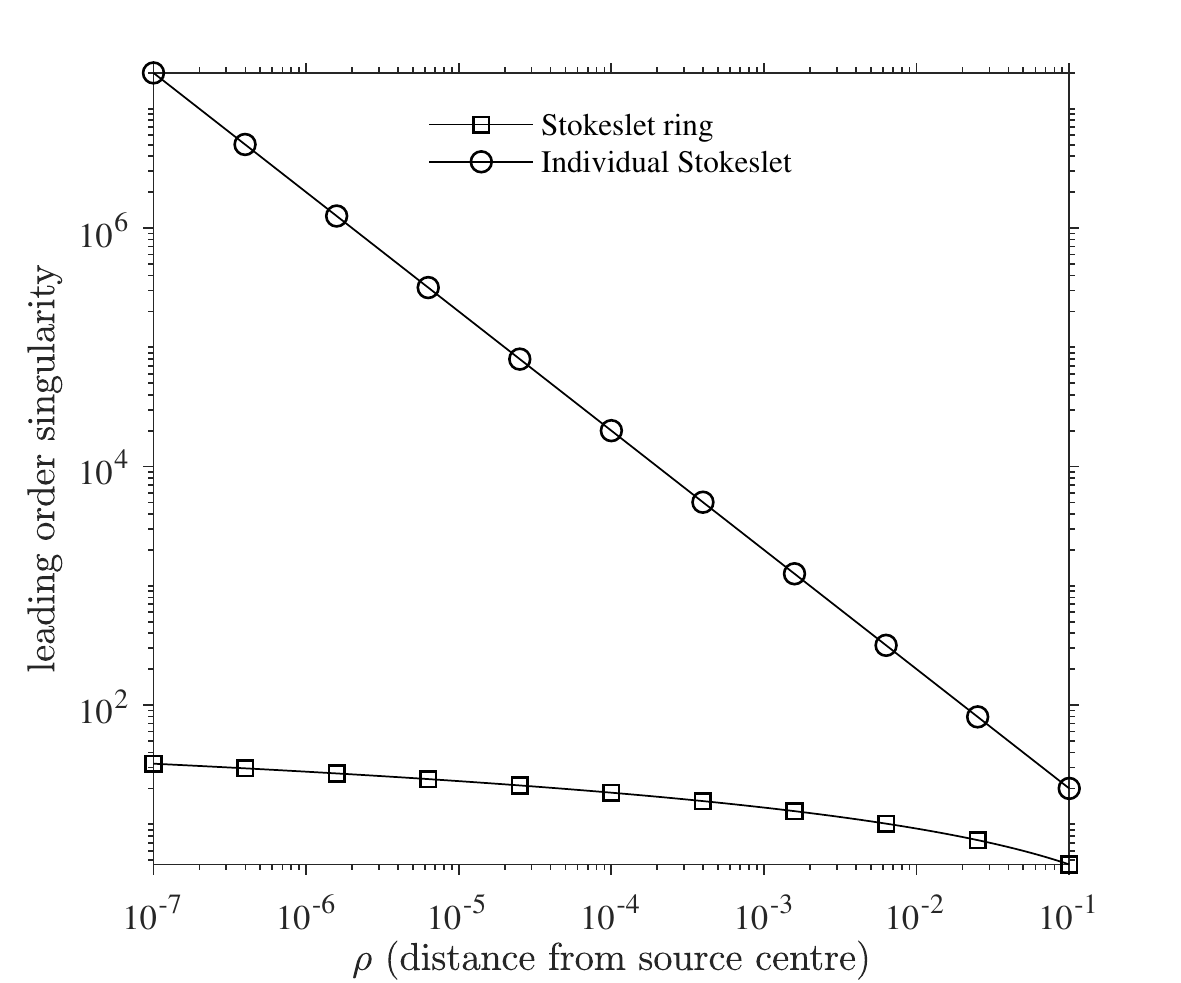}
	\caption{Comparison of the magnitude of the singularities present in the Stokeslet (squares) and the Stokeslet ring (circles) in the limit as the source and collocation point coincide.}
	\label{fig:singularSolution}
\end{figure}

In the case of the axisymmetric ring of singular Stokeslets, some interesting behaviour occurs in the limit as the source and collocation points coincide. In this limit, $k$ tends to unity from which it follows that $F \rightarrow \infty$ and $E\rightarrow 1$ in Equations \eqref{eq:poz1} -- \eqref{eq:poz2}. By employing the asymptotic expansion $F \approx -\ln \hat{r} + \ldots$ in which $\hat{r} = |\mathbf{x}_0 - \mathbf{x}_n|$, we find that $R_{\theta\theta}^0 \approx 2R_{rr}^0\approx 2R_{zz}^0 \approx - 4\ln\hat r + \ldots$ which all tend to infinity as $\hat r \rightarrow 0$ but at a significantly slower rate than the individual Stokeslet ($\approx \hat{r}^{-1}$). The matrix elements $R_{rz}^0$ and $R_{zr}^0$ are similarly divergent but typically take values in the range $[-1,1]$. Figure \ref{fig:singularSolution} shows a comparision of the magnitude of the singularities as $\hat r \rightarrow 0$ for $R_{rr}^0,R_{zz}^0 \approx -2\ln\hat r$ (red circles) versus $S_{ij}^0\approx2\hat{r}^{-1}$ for $i=j$ (blue squares). A log-log plot must be employed due to the speed with which the Stokeslet singularity increases for small $\hat r$. 

As a result of the slow rate with which the singularity in the axisymmetric Stokeslet ring approaches infinity, the method of fundamental solutions can be employed with a separation distance between the fictious and physical boundaries that is several orders of magnitude smaller than is possible using individual Stokeslets. This allows for the discovery of accurate solutions with a well-conditioned underlying resistance matrix. We illustrate this by once again conducting Matlab simulations for the resistance problem on the translating unit sphere (velocity $-\hat{\mathbf{z}}$) using both regularized and singular Stokeslet rings.

As before, the sphere is parametrized in $2$D using the arc segment $\mathbf{p} = \cos\varphi\,\hat{\mathbf{r}} + \sin\varphi\,\hat{\mathbf{z}}$ for $\varphi \in [-\pi/2,\pi/2]$ with $N=400$ points placed along its length at locations $\mathbf{x}_n$. For the method regularized ringlets, these $\mathbf{x}_n$ denote ringlet locations. The velocity $\mathbf{u}(\mathbf{x}_n)=-\hat{\mathbf{z}}$ is prescribed at all ringlet locations and the resistance matrix is constructed to yield the required force densities $\mathbf{g}^\ell(\mathbf{x}_n)$. The condition number of the resistance matrix $\mathsf{R}^\varepsilon$ is determined using the "cond" function in Matlab. The analytical solution for the drag on the unit sphere with unit velocity $-\hat{\mathbf{z}}$ is known to be equal to $(3/2)\hat{\mathbf{z}}$, so that the mean relative error for the calculation of the force term resulting from the resistance matrix is given by
\begin{equation}e(g^a_z) = \frac{1}{N}\sum_{n=1}^N \left|\frac{ g^a_z(\mathbf{x}_n) + 3/2}{-3/2}\right|,\end{equation} in which we recall that $\mathbf{g}^a$ is the force per unit area exerted by the sphere on the fluid and so has opposite sign to the drag. The fluid velocity at the near-- and far--field locations $(r,z) = (0,1.1), (0,1.5)$ is evaluated using the force densities $\mathbf{g}^a$ according to Equation \eqref{eq:stokes_summation}. The relative error in these velocities is given by
\begin{equation}e(1.1) = \left|\frac{u_z(0,1.1) + 1315/1331}{-1315/1331}\right|, \quad e(1.5) = \left|\frac{u_z(0,1.5) + 23/27}{-23/27}\right|,\end{equation}
based on the analytical solution given by Cortez \emph{et al.} \cite{cortez2005method}. Table \ref{tab:MFS} shows a summary of the results obtained for each of these four metrics for various values of regularisation parameter $\varepsilon$.

\begin{table}[ht]
	\centering 
	\begin{ruledtabular}
	\begin{tabular}{lllll}
		$\varepsilon$ & $e(g^a_z)$     & $e(1.1)$      & $e(1.5)$      & Cond \\ \midrule
		$.1$    & $5.3$  & $4.0\cdot 10^{-3}$ & $1.7\cdot 10^{-2}$ & $1.1\cdot 10^{19}$ \\
		$.05$   & $1.8\cdot 10^{-2}$ & $2.5\cdot 10^{-3}$  & $8.4\cdot 10^{-3}$  & $8.6\cdot 10^{15}$ \\
		$.025$  & $9.7\cdot 10^{-3}$ & $1.4\cdot 10^{-4}$  & $4.1\cdot 10^{-3}$  & $2.9\cdot 10^{10}$ \\
		$.01$   & $4.3\cdot 10^{-3}$ & $5.8\cdot 10^{-4}$  & $1.6\cdot 10^{-3}$  & $4.2\cdot 10^{5}$ \\
		$.005$  & $2.5\cdot 10^{-3}$ & $2.2\cdot 10^{-4}$  & $7.6\cdot 10^{-4}$  & $1.4\cdot 10^{4}$ \\
		$.0025$ & $2.1\cdot 10^{-3}$ & $2.3\cdot 10^{-4}$  & $1.1\cdot 10^{-4}$  & $2.8\cdot 10^{3}$ \\
		$.001$  & $2.5\cdot 10^{-3}$ & $1.0\cdot 10^{-3}$  & $7.4\cdot 10^{-4}$  & $1.0\cdot 10^{3}$ \\ 
	\end{tabular}
\end{ruledtabular}
	\caption{Accuracy of the method of regularized ringlets for various values of the regularization parameter $\varepsilon$ in the resistance problem on the translating unit sphere. Columns refer to relative errors in total drag on the sphere $e( g_z^a)$, fluid velocity at the point $(r,z)=(0,1.1)$ denoted $e(1.1)$ and analogous error at $(0,1.5)$, and finally the condition number of the underlying resistance matrix.}
	\label{tab:MFS}
\end{table}

For the method of fundamental solutions, let $\mathbf{x}_n$ denote collocation points on the sphere surface. $N$ source points are placed at positions $\mathbf{x}_s = (1-10^{-b})\cdot\mathbf{x}_c$ where $b$ is a parameter that represents the separation distance between the physical and fictitious boundaries (Figure \ref{fig:singularSchematic}). The velocity $\mathbf{u}(\mathbf{x}_n)=-\hat{\mathbf{z}}$ is prescribed at all collocation points and the resistance matrix is constructed to yield the required force densities $\mathbf{g}^\ell(\mathbf{x}_s)$ at each source point. The accuracy and applicability of the method is measured using the same four metrics as for the method of regularized ringlets. Table \ref{tab:MFS2} shows a summary of results for various values of separation distance $10^{-b}$, analogous to the results in the regularized case from Table \ref{tab:MFS}.

\begin{table}[ht]
	\centering 
	\begin{ruledtabular}
	\begin{tabular}{lllll}
$b$ & $e(g^a_z)$     & $e(1.1)$      & $e(1.5)$      & Cond \\ \midrule
$1$ & $4.8\cdot 10^{1}$  & $1.5\cdot 10^{-13}$ & $3.5\cdot 10^{-14}$ & $1.1\cdot 10^{19}$ \\
$2$ & $1.6\cdot 10^{-2}$ & $1.6\cdot 10^{-6}$  & $3.8\cdot 10^{-7}$  & $9.5\cdot 10^{6}$ \\
$3$ & $4.3\cdot 10^{-3}$ & $9.5\cdot 10^{-4}$  & $6.3\cdot 10^{-4}$  & $2.5\cdot 10^{3}$ \\
$4$ & $5.4\cdot 10^{-4}$ & $3.0\cdot 10^{-3}$  & $2.5\cdot 10^{-3}$  & $4.4\cdot 10^{2}$ \\
$5$ & $7.3\cdot 10^{-3}$ & $5.1\cdot 10^{-3}$  & $4.6\cdot 10^{-3}$  & $2.4\cdot 10^{2}$ \\
$6$ & $9.4\cdot 10^{-3}$ & $7.2\cdot 10^{-3}$  & $6.7\cdot 10^{-3}$  & $1.6\cdot 10^{2}$ \\
$7$ & $1.1\cdot 10^{-2}$ & $9.3\cdot 10^{-3}$  & $8.8\cdot 10^{-3}$  & $1.2\cdot 10^{2}$ \\ 
	\end{tabular}
\end{ruledtabular}
	\caption{Accuracy of the method of fundamental solutions using singular ringlets for various values of the separation distance $10^{-b}$ in the resistance problem on the translating unit sphere. Columns refer to relative errors in total drag on the sphere $e(g_z^a)$, fluid velocity at the point $(r,z)=(0,1.1)$ denoted $e(1.1)$ and analogous error at $(0,1.5)$, and finally the condition number of the underlying resistance matrix.}
	\label{tab:MFS2}
\end{table}

Similar results are achievable using both methods, although the singular method generally yields more well-conditioned resistance matrices and can produce relative errors of a smaller magnitude for the drag calculation and fluid velocities. It should however be noted that it is not possible to minimize each error in the singular calculation concurrently (as the smallest values for $e(1.1)$ and $e(1.5)$ are generally associated with larger values for the condition number and $e(g_z^a)$). Despite axisymmetry enabling a drastic reduction of the separation distance for the singular problem, the question of what distance is considered `optimal' still persists. This is not an issue for the regularized ringlet, in which case each error achieves a minimal value for similar values of $\varepsilon$ ($\approx 0.0025$ to $0.005$). For excessively small $\varepsilon$ ($< 0.05$), the error becomes non--monotonic as is often the case with regularized Stokeslet methods (eg. see Figure \ref{fig:cc1}, Cortez \emph{et al.} \cite{cortez2005method}, Gallagher \emph{et al.} \cite{gallagher2018sharp}). The same is also true for small $b$ in the singular case. For both regularized and singular methods, reduction of regularization parameter $\varepsilon$ or $b$ always appears to result in a reduction of the condition number of the resistance matrix.

Although the singular method can be tuned to give smaller relative errors in either the fluid velocity or total drag separately, it cannot do so simultaneously; regularized ringlets display more satisfactory convergence properties and are the more effective method to minimize errors in both fluid velocity and total drag. 

\FloatBarrier
\section{Computational speed} \label{sec:accuracyApplications}
Table \ref{tab:speedComp} shows a comparison of the computational times $t_R$ and $t_S$ (measured in seconds),  associated with constructing the $2N \times 2N$ ringlet matrix $R^\varepsilon$ (in the irrotational fluid case) and the $3N \times 3N$ Stokeslet matrix $S^\varepsilon$ respectively. The increase in computational time for computing $R^\varepsilon$ is a result of needing to compute the complete elliptic integrals $F(k)$ and $E(k)$ for all combinations of ring locations $(\mathbf{x}_m,\mathbf{x}_n) \, \forall \, m,n \in{1,...,N}$, which requires the construction of two further $2N \times 2N$ matrices $E$ and $F$. The computational time needed for this isolated operation, $t_e$ is also listed in Table \ref{tab:speedComp}. The total additional time needed to construct $R^\varepsilon$ is modest, typically between $10\%$ and $20\%$ of the time needed for $S^\varepsilon$. 

The computational time associated with $F$ and $E$ (and by extension, $R^\varepsilon$) can be reduced by evaluating $F(k)$ and $E(k)$ to a lower degree of accuracy; the Matlab function "ellipke(k,TOL)" calculates $F(k)$ and $E(k)$ to the accuracy defined by TOL, which has a default value of $2^{-52} \approx 2.2 \cdot 10^{-16}$ (double-precision accuracy). This is a far greater accuracy than we are typically able to achieve using regularized Stokeslet methods, suggesting a larger value of TOL will suffice. 

\begin{table}[ht]
	\centering
	\begin{ruledtabular}
		\begin{tabular}{lllllll}
			
			& \multicolumn{6}{l}{number of nodes $N$} \\
			& $1000$   & $2000$   & $3000$   & $4000$   & $5000$   & $6000$   \\ \midrule
			$t_e$ & $0.1029$ & $0.4100$ & $0.9208$ & $1.6228$ & $2.5303$ & $3.6553$ \\
			$t_R$ & $0.2863$ & $1.1182$ & $2.4406$ & $4.4124$ & $6.9682$ & $9.9643$ \\ 
			$t_S$          & $0.2380$ & $0.9396$ & $2.1676$ & $3.7562$ & $6.0355$ & $8.9524$ \\ 
		\end{tabular}
	\end{ruledtabular}
	\caption{Comparison of computational time (in seconds) for evaluating elliptic integrals as well as constructing ring matrix $\mathsf R^\varepsilon$ and Stokeslet matrix $\mathsf S^\varepsilon$ using varying numbers of nodes.}
	\label{tab:speedComp}
\end{table}

We note that by consideration of the size of the matrices involved and the typically small value of $t_e$, we should hypothetically be able to achieve $t_R = (4/9)t_S + t_e < t_S$ for any given value of $N$. In practice, this is not the case. The elements $S_{ij}^\varepsilon$ for $i,j \in \{1,2,3\}$ share a common form that enables them to be encoded in matrix form $\mathsf S^\varepsilon$ very efficiently. The same is not true of the elements $R_{\alpha\beta}^\varepsilon$ for $\alpha,\beta \in \{r,z\},$ hence why $t_R > t_S \, \forall \, N$ in Table \ref{tab:speedComp}.

Inclusion of fluid rotation (such that an additional $N \times N$ matrix must be constructed for $\mathsf R_{\theta}^{\varepsilon}$ incurs an additional cost of small value, approximately equal to $(t_R - t_e)/4$ for any given value of $N$. We divide by $4$ under the assumption that the cost associated with constructing the additional $N \times N$ matrix $\mathsf R_{\theta}^{\varepsilon}$ is one quarter of the cost associated with constructing the irrotational $2N \times 2N$ matrix $\mathsf R^\varepsilon$. 

The true value of working with ringlets in $2$D can be seen by by evaluating the cost of inverting the matrices $\mathsf R_{\theta}^{\varepsilon}$ and $\mathsf R^\varepsilon$ versus $\mathsf S^\varepsilon$ (as is necessary in the resistance problem for evaluating the force associated with a given boundary velocity). These matrices are of size $N\times N$, $2N \times 2N$ and $3N \times 3N$ respectively. We investigate the cost of solving a linear system $\mathbf{X} = \mathbf{A}\setminus\mathbf{b}$ in which $\mathbf{A}$ and $\mathbf{b}$ are a $M \times M$ matrix and a $M \times 1$ vector of normally distributed random data respectively, with $M \, \in \, \{N,2N,3N\}.$ We denote the time taken to solve this system by $t_N,t_{2N},$ and $t_{3N}$ in each case. The results are given in Table \ref{tab:invertSpeed}, in which it is clear to see that $t_{3N}$ is significantly larger than the sum of $t_N$ and $t_{2N}$ for all values of $N$ tested. In practice, this means that using a $3$D Stokeslet implementation with $\mathsf S^\varepsilon$ for solving an axisymmetric resistance problem will always be significantly more costly than our $2$D ringlet implementation with $\mathsf R^\varepsilon, \mathsf R_{\theta}^{\varepsilon}.$

\begin{table}[ht]
	\centering
	\begin{ruledtabular}
		\begin{tabular}{lllllll}
			
			& \multicolumn{6}{l}{number of nodes $N$} \\
			& $1000$   & $2000$   & $3000$   & $4000$    & $5000$    & $6000$    \\ \midrule
			$t_N$    & $0.0178$ & $0.1119$ & $0.2986$ & $0.6377$  & $1.2263$  & $2.0053$  \\ 
			$t_{2N}$ & $0.1018$ & $0.7139$ & $2.0614$ & $5.1004$  & $9.3121$  & $15.7495$ \\ 
			$t_{3N}$  & $0.2978$ & $2.3084$ & $6.7434$ & $15.1372$ & $30.0496$ & $50.7495$ \\ 
		\end{tabular}
	\end{ruledtabular}
	\caption{Computational time (in seconds) associated with solving the linear system $\mathbf{X} = \mathbf{A}\setminus\mathbf{b}$ for varying sizes of matrix $\mathbf{A}$ and vector $\mathbf{b}$.}
	\label{tab:invertSpeed}
\end{table}

As well as producing smaller resistance matrices with a reduced associated computational cost for \emph{a given number of nodes}, our axisymmetric ringlet method also \emph{requires far fewer nodes} in order to achieve the same level of accuracy as the traditional regularized Stokeslet method. In Appendix \ref{sec:translatingSphere} it was shown that for the case of the translating unit sphere the axisymmetric discretization of the sphere surface with $N$ rings produces results that are consistently more accurate than the traditional $3$D patch discretization using $3N^2$ nodes. To put this into context, the result using $1000$ ringlets in $2$D (computational time $\approx \SI{0.1}{s}$) thus corresponds to using $1,500,000$ regularized Stokeslets in $3$D. A lower bound estimate for the associated computational time (using $t_{3N} \propto 3N$ with data from Table \ref{tab:invertSpeed}) for method of regularized Stokeslets using this number of nodes yields a time in excess of $500$ days. Assuming each element of the resulting matrix $\mathsf S^\varepsilon$ is stored as a double-precision floating-point number, this would also require $\approx \SI{1}{PB}$ of storage.

\FloatBarrier
\section{Force calculation for the rotating sphere} \label{sec:rotatingSphereForce}
The leading order expression (in terms of Re) for the torque on the rotating sphere is given by, \begin{equation} \mathbf{T}=-8\pi\mu a^3\boldsymbol{{\Omega}},\end{equation} the derivation of which can be found in \cite{rubinow1961transverse}. This torque is associated with a drag force per unit area on the surface of the sphere given by $\mathbf{f}=-3\mu\omega_0(r/a)\hat{\mathbf{e}}_\theta$, which can be verified by considering the identity, \begin{equation} \mathbf{T} = \iint_S \mathbf{x} \times \mathbf{f} \, dS, \end{equation} where $S$ denotes the sphere surface. Multiplication by $\hat{\mathbf{e}}_z$ yields, \begin{equation} -8\pi\mu a^3\omega_0 = \hat{\mathbf{e}}_z \cdot \iint_S \mathbf{x} \times \mathbf{f} \, dS. \end{equation} In cylindrical coordinates, we have \begin{equation} \begin{split} \hat{\mathbf{e}}_z\cdot(\mathbf{x}\times\mathbf{f}) & = \hat{\mathbf{e}}_z\cdot((r\hat{\mathbf{e}}_r+\theta\hat{\mathbf{e}}_\theta+z\hat{\mathbf{e}}_z)\times(-3\mu\omega_0(r/a)\hat{\mathbf{e}}_\theta)) \\ &= -3\mu\omega_0(r/a)\hat{\mathbf{e}}_z\cdot(r\hat{\mathbf{e}}_z-z\hat{\mathbf{e}}_r) \\ &= -3\mu\omega_0(r^2/a).\end{split}\end{equation} Converting to a spherical system $(r,\theta,\varphi)$ in which $\theta$ denotes the azimuthal angle and $\varphi$ the polar angle, we substitute $r^2=a^2\sin^2\varphi$ and $dS = a^2\sin\varphi\,d\theta d\varphi$ such that, \begin{equation} \begin{split} \hat{\mathbf{e}}_z \cdot \iint_S \mathbf{x} \times \mathbf{f} \, dS &= -3\mu a^3\omega_0 \int_{\theta=0}^{2\pi}\left(\int_{\varphi=0}^\pi \sin^3\varphi \, d\varphi\right) d\theta, \\ &= -6\pi\mu a^3 \omega_0 \int_{\varphi=0}^\pi \sin^3\varphi \, d\varphi, \\ &= -6\pi\mu a^3 \omega_0 \cdot \frac{4}{3} = -8\pi\mu a^3 \omega_0, \end{split} \end{equation} as required. 

\end{document}